\documentclass[fleqn,usenatbib]{mnras}

\usepackage{newtxtext,newtxmath}

\usepackage[T1]{fontenc}

\DeclareRobustCommand{\VAN}[3]{#2}
\let\VANthebibliography\thebibliography
\def\thebibliography{\DeclareRobustCommand{\VAN}[3]{##3}\VANthebibliography}

\usepackage{graphicx}	
\usepackage{amsmath}	
\usepackage{fix-cm}
\usepackage{subcaption}
\usepackage{float}
\usepackage{placeins}
\usepackage{MnSymbol}
\usepackage{enumerate}
\usepackage{threeparttable}
\usepackage{booktabs,caption}
\usepackage{xspace}
\usepackage{adjustbox}  
\usepackage{changepage} 
\usepackage{multirow}
\usepackage{array}

\newcommand\orcid[1]{\href{http://orcid.org/#1}{\adjustbox{trim={-.15\width} {0\height} {-.15\width} {0\height},clip}{\includegraphics[height=10pt]{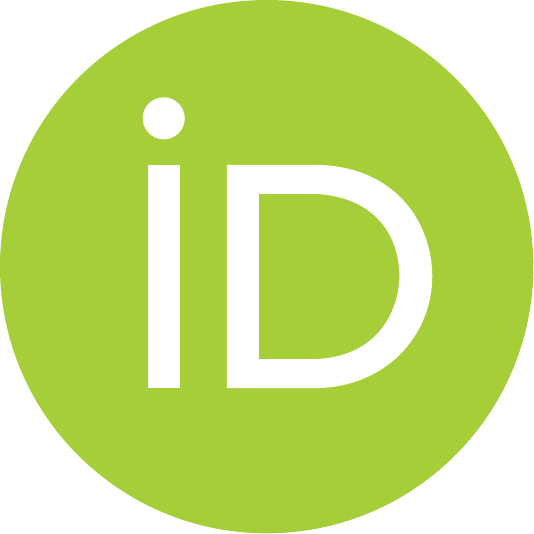}}}}

\usepackage{silence}
\title[Clumpy fraction of galaxies]{The clumpy fraction of galaxies at $z=1-9$ with JADES: evidence for a mass-dependent transition in clump origin}

\author[Puskás et al.]{Dávid Puskás\orcid{0000-0001-8630-2031},$^{1,2,3}$\thanks{E-mail: dpuskas@arizona.edu}
Sandro Tacchella\orcid{0000-0002-8224-4505},$^{1,2}$
Phoebe Chapman,$^{1,2}$
Floor van Donkelaar\orcid{0000-0002-7235-9747},$^{2,4}$\newauthor
Andrew J.\ Bunker\orcid{0000-0002-8651-9879},$^{5}$
Courtney Carreira\orcid{0000-0001-6301-3667},$^{6}$
A. Lola Danhaive\orcid{0000-0002-9708-9958},$^{1,2}$
Francesco D'Eugenio\orcid{0000-0003-2388-8172},$^{1,2}$\newauthor
Qiao Duan\orcid{0009-0009-8105-4564},$^{1,2}$
Patricia Iglesias-Navarro\orcid{0009-0009-8959-2404},$^{7,8}$
Amanda Stoffers\orcid{0009-0009-5464-3558},$^{1,2}$
Hannah \"Ubler\orcid{0000-0003-4891-0794},$^{9}$\newauthor
Yongda Zhu\orcid{0000-0003-3307-7525},$^{3}$
Mirko Curti\orcid{0000-0002-2678-2560},$^{10}$
Emma Curtis-Lake\orcid{0000-0002-9551-0534},$^{11}$
Kevin Hainline\orcid{0000-0003-4565-8239},$^{3}$
Eiichi Egami\orcid{0000-0003-1344-9475},$^{3}$\newauthor
Daniel J.\ Eisenstein\orcid{0000-0002-2929-3121},$^{12}$
Zhiyuan Ji\orcid{0000-0001-7673-2257},$^{3}$
Benjamin D.\ Johnson\orcid{0000-0002-9280-7594},$^{12}$
Roberto Maiolino\orcid{0000-0002-4985-3819},$^{1,2,13}$\newauthor
Marcia Rieke\orcid{0000-0002-7893-6170},$^{3}$
Brant Robertson\orcid{0000-0002-4271-0364},$^{6}$ and
Christina C. Williams\orcid{0000-0003-2919-7495}$^{14}$
\\
\\
$^{1}$Cavendish Laboratory, University of Cambridge, 19 JJ Thomson Avenue, Cambridge, CB3 0HE, UK\\
$^{2}$Kavli Institute for Cosmology, Madingley Road, Cambridge, CB3 0HA, UK\\
$^{3}$Steward Observatory, University of Arizona, 933 N. Cherry Avenue, Tucson, AZ 85721, USA\\
$^{4}$Institute of Astronomy, University of Cambridge, Madingley Road, Cambridge CB3 0HA, UK\\
$^{5}$Department of Physics, University of Oxford, Denys Wilkinson Building, Keble Road, Oxford OX1 3RH, UK\\
$^{6}$Department of Astronomy and Astrophysics University of California, Santa Cruz, 1156 High Street, Santa Cruz, CA 95064, USA\\
$^{7}$Instituto de Astrofísica de Canarias, C/ Vía Láctea s/n, 38205 La Laguna, Tenerife, Spain\\
$^{8}$Departamento de Astrofísica, Universidad de La Laguna, 38200 La Laguna, Tenerife, Spain\\
$^{9}$Max-Planck-Institut f\"ur extraterrestrische Physik (MPE), Gie{\ss}enbachstra{\ss}e 1, 85748 Garching, Germany\\
$^{10}$INAF, Osservatorio di Astrofisica e Scienza dello Spazio, Via P. Gobetti 93/3, I-40129 Bologna, Italy\\
$^{11}$Centre for Astrophysics Research, Department of Physics, Astronomy and Mathematics, University of Hertfordshire, Hatfield AL10 9AB, UK\\
$^{12}$Center for Astrophysics $|$ Harvard \& Smithsonian, 60 Garden St., Cambridge MA 02138 USA\\
$^{13}$Department of Physics and Astronomy, University College London, Gower Street, London WC1E 6BT, UK\\
$^{14}$NSF National Optical-Infrared Astronomy Research Laboratory, 950 North Cherry Avenue, Tucson, AZ 85719, USA
}

\date{Accepted XXX. Received YYY; in original form ZZZ}

\pubyear{\the\year{}}

\begin{document}
\label{firstpage}
\pagerange{\pageref{firstpage}--\pageref{lastpage}}
\maketitle

\begin{abstract}
We measure the fraction of clumpy galaxies ($f_{\rm clumpy}$) over the redshift range $z=1$--9 and stellar masses $\log_{10}(M_\star/{\rm M_\odot}) \geq 8$ using \textit{JWST}/NIRCam imaging from JADES. We define clumps as compact off-centre surface-brightness peaks with characteristic sizes of a few hundred parsecs, and develop a dendrogram-based clump-finding method with \texttt{astrodendro} to identify them directly in rest-frame UV, blue, and optical images. Clump-detection completeness is quantified through injection--recovery simulations, which we use to define ($M_\star$, $z$)--dependent 50 per cent completeness-limited $f_{\rm clumpy}$, alongside a probability-weighted completeness-corrected statistic for comparison. For the full sample, the blue and optical $f_{\rm clumpy}$ reach $\approx55$ per cent at $z\approx1$--2 and decline to $\lesssim5$ per cent by $z\approx8$--9, while the UV fraction remains higher at $\approx12$ per cent. At $z\approx1$--2, $f_{\rm clumpy}$ increases strongly with stellar mass, from $\approx35$ per cent in the lowest-mass bin to $\approx85$ per cent at $\log_{10}(M_\star/{\rm M_\odot})\geq10$. Most clumpy galaxies host only one or two robustly detected off-centre clumps, while higher-multiplicity systems occur preferentially in massive, strongly star-forming galaxies at low to intermediate redshift. Comparison with major merger rates suggests that mergers could account for a substantial fraction of low-mass ($\log_{10}(M_\star/{\rm M_\odot})<8.5$), high-redshift clumpy galaxies, while the observed fractions in more massive systems ($\log_{10}(M_\star/{\rm M_\odot})>9$) require an additional \textit{in situ} clump-formation channel. Our results are consistent with a transition from merger-dominated clumpy morphologies at early times and low stellar masses to an increasing contribution from \textit{in situ} clump-formation channels, such as disc instabilities, in more massive systems.
\end{abstract}



\section{Introduction}
\label{sec:introduction}

Observations of star-forming galaxies have revealed that they often exhibit irregular morphologies with distinct kiloparsec-scale substructures called \textit{clumps} \citep{Elmergreen_2005, Elmergreen_2007, Forster_Schreiber_2011a, Forster_Schreiber_2011b, Genzel_2011, Wisnioski_2011, Guo_2012, Wuyts_2012, Ikeda_2026, Tan_2025}. The majority of these clumps are bright in the rest-frame ultraviolet (UV) and are found to be sites of intense star formation \citep{Guo_2018, Huertas-Company_2020, Vanzella_2021, Sattari_2023, Kalita_2025a, Claeyssens_2025}, with the fraction of clumpy galaxies peaking at redshift $z \approx 2$ \citep[e.g.,][]{Guo_2015, Shibuya_2016, Mawatari_2026, Sok_2026a, de_la_Vega_2026a}, coinciding with the peak of the cosmic star formation rate density \citep{Madau_2014}. The stellar mass growth history of galaxies over cosmic time is governed by the combination of star formation and the accretion of satellites through galaxy mergers \citep[e.g.,][]{Oser_2010, Duncan_2019, Duan_2025, Puskas_2025}. Consequently, the observed clump population may comprise a mixture of physically distinct structures, including disc fragments, compact star clusters, filamentary star-forming regions, minor satellites, and merger-induced star-forming features \citep[e.g.,][]{Mandelker_2017, Mandelker_2018, Nakazato_2024}. With recent \textit{JWST} observations \citep[e.g.,][]{Harikane_2025, Sok_2026a, de_la_Vega_2026a, Zhu_2026} revealing a large number of clumpy systems at $z > 3$, a natural question arises regarding the \textit{in situ} or \textit{ex situ} origin of these substructures, particularly as a function of redshift, stellar mass, and rest-frame wavelength. 

The main formation mechanisms proposed by observational studies and simulations include: \textit{(i)} fragmentation (e.g., disc instabilities), and \textit{(ii)} mergers. A leading model for \textit{in situ} clump formation is gravitational collapse in gas-rich turbulent systems, including disc fragmentation \citep{Agertz_2009, Dekel_2009, Ceverino_2010, Genel_2012, Tamburello_2015, Inoue_2016, Mayer_2016, Renaud_2021, van_Donkelaar_2022, Mayer_2025}, but also fragmentation within dense inflowing filaments (taking place in the galaxy) or circumgalactic medium (CGM) overdensities which could contribute to this early clump or cluster population \citep[see, e.g.,][]{Mandelker_2017, Mandelker_2018, van_Donkelaar_2026}. Here, \textit{in situ} refers to the clump forming locally within or around the galaxy, irrespective of whether the material from which it formed was supplied by the CGM or inflowing filaments, rather than to the accretion of a pre-existing stellar system. Violent disc instability (VDI) is commonly used to describe fragmentation in continuously gas-fed, marginally stable discs that do not reach equilibrium due to ongoing cold accretion \citep[e.g.,][]{Dekel_2009, Bournaud_2014, Mandelker_2017, Krumholz_2018}. At high redshift, these discs form by cosmological accretion of cold gas from the intergalactic medium \citep{Birnboim_2003, Keres_2005, Bournaud_2009, Nelson_2013}, and due to their high gas fractions and disc masses \citep[e.g.,][]{Tamburello_2015, Renaud_2021, van_Donkelaar_2022, Mayer_2025}, as well as low or marginal Toomre \textit{Q} stability parameters (\citealt{Toomre_1964}; $Q \lesssim 1$ across most of the disc), the gas collapses under gravitational instabilities and creates clumps. Recent analytical work extends this framework beyond the characteristic fastest-growing Toomre mode, predicting an intrinsic spectrum of clump masses and sizes from the full range of unstable modes \citep{Ferrara_2026}. This fragmentation picture is consistent with the high gas fractions observed or inferred in high-$z$ galaxies \citep{Tacconi_2010, Tacconi_2020, Wang_2022, Parlanti_2023, Danhaive_2025}, as well as the prevalence of disc-like morphologies \citep{Ferreira_2023, Kartaltepe_2023} and rotationally supported gas-rich systems revealed through spectroscopy \citep{Forster_Schreiber_2011a, Forster_Schreiber_2011b, Forster_Schreiber_2018, Wisnioski_2011, Wisnioski_2019, Ubler_2019, Rizzo_2024, Rowland_2024, Danhaive_2025, Cescon_2026}. These observations do not directly demonstrate fragmentation, but show that many high-redshift galaxies host extended gas-rich discs where gravitational instabilities can develop, potentially aided by turbulence or external perturbations. Resolved studies at cosmic noon further reveal extended molecular gas discs and clumpy spiral or ring structures with ordered kinematics, in some cases accompanied by evidence for radial gas inflows \citep[e.g.,][]{Nestor_Shachar_2025, Parlanti_2026, Chen_2026}. \textit{Ex situ} origins of clumps include minor or major mergers and the accretion of satellites such as dwarf galaxies \citep{Di_Matteo_2008, Puech_2010, Calabro_2019, Estrada-Carpenter_2024, Nakazato_2024}. Galaxy interactions can also trigger clump formation within the gas of the interacting systems through tidal forces, gas compression, and induced turbulence \citep[e.g.,][]{Teyssier_2010, Renaud_2014, Renaud_2022, Puskas_2026}. Recent \textit{JWST} observations suggest that even very minor interactions (mass ratio $\lesssim 1/10$) may trigger clump formation, extending this channel below the range typically considered in merger studies \citep{de_la_Vega_2026b}. High-resolution simulations further show that mergers can drive gas fragmentation and the formation of compact star-forming regions and star clusters \citep[e.g.,][]{Renaud_2015, Lahen_2022}. The merger-related clump formation channel is supported by the increasing frequency of galaxy mergers with redshift \citep{Mundy_2017, Duncan_2019, Dalmasso_2024,  Duan_2025, Puskas_2025}, particularly at lower stellar masses. Accreted satellites can leave remnants within the host galaxy, which was found both in simulations \citep[e.g.,][]{Mandelker_2017} and indicated by observations \citep{Ribeiro_2017, Zanella_2019}. Metallicity variations relative to the gas within the disc of the host galaxy can provide a further diagnostic, although negative offsets can arise from either satellite accretion or metal-poor gas accretion and the subsequent triggering of \textit{in situ} clump formation \citep[e.g.,][]{Estrada-Carpenter_2025, Sok_2025, Sok_2026b}. Previous studies argued that fragmentation-driven clump formation is more efficient at cosmic noon than at higher redshift due to larger discs, however, recent observations and simulations (see, e.g., \citealt{Mayer_2025} at $z \approx 7.6$) pushed its relevance further back in redshift. Mergers may become increasingly important compared to internal fragmentation at higher redshifts and lower masses. However, the competing effects of these clump formation mechanisms at $z>3$ and at different stellar masses are still unclear and require a large statistical sample to address this properly.

The presence of star-forming clumps within galaxies that retain an underlying disc component can affect the evolution of their hosts and ultimately alter their morphologies. Clumps that are massive enough can survive galactic feedback and, through dynamical friction and gravitational torques (due to interactions), they can migrate to the galactic centre, eventually causing the formation of a proto-bulge or contributing to bulge growth \citep{Immeli_2004, Dekel_2009, Krumholz_2010, Ceverino_2012, Inoue_2012, Mandelker_2014, Mandelker_2017, Dekel_2022, Zhu_2026}. Lower mass clumps might be dissipated on their own or destroyed by strong feedback and outflows \citep[$\approx\!100$~Myr, see, e.g.,][]{Murray_2010, Genel_2012, Hopkins_2012, Buck_2017, Oklopcic_2017, Fensch_2021}, although the densest high-redshift clumps may resist supernova feedback \citep{Ferrara_2026}. Powerful, mass-loaded outflows have also been observed directly from individual giant star-forming clumps at $z\approx2$ \citep[e.g.,][]{Newman_2012}. The survival, abundance, and young stellar content of clumps may therefore provide a sensitive probe of the strength of stellar feedback \citep{Shin_2023}, while repeated clump formation and disruption may contribute to the growth of galactic thick discs \citep{Noguchi_1998, Bournaud_2007, Inoue_2014}. In gas-rich discs, which sustain clump formation through fragmentation, the clumps themselves can also affect the dynamical state of the discs by feedback and their gravitational potential, causing higher turbulence. Turbulence has been observed in high-redshift galaxies by recent studies \citep[e.g.,][]{Ubler_2019, Parlanti_2023, Danhaive_2025}.

One observational quantity that could reveal which formation mechanism predominates at different redshifts and stellar masses is the \textit{fraction of clumpy galaxies}, or clumpy fraction, denoted by $f_{\rm clumpy}$ \citep[e.g.,][]{Guo_2015, Shibuya_2016}. Studies using \textit{Hubble Space Telescope} (\textit{HST}) data have well established the evolution of $f_{\rm clumpy}$ for UV-selected clumps with redshift up to $z \approx 3$, and found that it increases up to cosmic noon ($z \approx 2$) where it peaks at $\approx 60$ per cent and starts to decline at increasing redshift \citep{Elmergreen_2007, Puech_2010, Guo_2012, Guo_2015, Shibuya_2016, Ribeiro_2017, Huertas-Company_2020, Sok_2022, Martin_2023, Sattari_2023}. However, beyond $z>3$, different studies found discrepant results regarding the magnitude and redshift evolution of $f_{\rm clumpy}$. A number of studies found a decreasing trend in $f_{\rm clumpy}$ with redshift \citep{Conselice_2009, Oesch_2010, Shibuya_2016}; whereas others found a flat evolution \citep{Curtis-Lake_2016, Bowler_2017, Ribeiro_2017}. Many of these studies focused on narrow redshift ranges to measure $f_{\rm clumpy}$, applying different selection methods and criteria, making meaningful quantitative comparison difficult.

The launch of the \textit{James Webb Space Telescope} \citep[\textit{JWST};][]{Gardner_2023} opened a new window on studying the rest-frame UV and optical morphologies of high redshift galaxies beyond $z>5$ \citep[e.g.,][]{Kartaltepe_2023, Morishita_2024, Ormerod_2024, Huertas-Company_2024, Allen_2025, Carreira_2026, Danhaive_2026}. Studies using \textit{JWST} data extended the measurement of the clumpy fraction of galaxies up to $z \approx 12$ using a similar methodology to the one used for \textit{HST} studies. \citet{de_la_Vega_2026a} finds a strong decline of $f_{\rm clumpy}$ with redshift (from $\approx 70$ per cent at $z \simeq 2.75$ to $\approx 10 $ per cent at $z \simeq 7.75$), and higher $f_{\rm clumpy}$ values than \textit{HST} at fixed redshift, which they attribute to the higher sensitivity of \textit{JWST}. \citet{Sok_2026a} studies $f_{\rm clumpy}$ at $0.5 < z < 5$ using data from the CANUCS survey \citep{Willott_2022}, and finds a stellar mass dependent evolution: for massive galaxies with $\log_{10}(M_\star/{\rm M_\odot})\geq10$, $f_{\rm clumpy}$ peaks at $z\approx2$ and decreases towards higher redshift; whereas at $8.5 \leq \log_{10}(M_\star/{\rm M_\odot}) < 10$ the evolution is flat with redshift ($f_{\rm clumpy} \approx 30$ per cent). Large area surveys, such as COSMOS-Web \citep{Casey_2023} allowed the measurement of the physical properties of clumps in the rest-frame NIR out to $z \approx 4$ \citep{Kalita_2025a, Mercier_2026}, finding a decreasing $f_{\rm clumpy}$ with cosmic time ($\approx 40$ per cent at $z \approx 4$ to $\approx 10$ per cent at $z \approx 1$). \citet{Harikane_2025} explores the clumpiness of spectroscopically confirmed galaxies at $z \approx 7-14$, finding $f_{\rm clumpy} \approx 70$ per cent at $z \approx 7$ for the brightest galaxies, possibly originating from merger-induced starburst activity. These results can also be connected to theoretical frameworks for clustered star formation at high redshift, in which high gas surface densities, short dynamical times, and strong turbulence promote rapid gas collapse into compact, bound stellar systems or star-forming clusters, particularly in compact high-redshift discs \citep[e.g.,][]{He_2019, Grudic_2021, Garcia_2025, Horie_2026}. Radiation-hydrodynamic simulations from the \textsc{thesan-zoom} project \citep{Kannan_2025} further show that clumpiness can depend strongly on the observational tracer and star formation phase, with comparatively smooth stellar mass distributions producing highly structured rest-frame UV, optical, and nebular emission \citep{Wang_2026}. Recent studies point to a complex emerging picture, where fragmentation is the dominant mechanism for clump formation in massive disc-like galaxies with high gas fractions up to $z \approx 7$ \citep[e.g., as found by][]{Ferreira_2023, Kartaltepe_2023, Fujimoto_2025, Morishita_2025}, while at lower stellar masses and, in particular, higher redshifts, the merger scenario becomes more prevalent \citep[e.g.,][]{Nakazato_2024, Sok_2026a, de_la_Vega_2026a}.

Moreover, \textit{JWST} has revealed complex, clumpy morphologies at $z>5$ previously undetected by \textit{HST} \citep[e.g.,][]{Chen_2023, Tacchella_2023, Hainline_2024, Tanaka_2024, Baker_2025}. Strong gravitational lensing has uncovered an abundance of high-$z$ clumpy systems \citep{Claeyssens_2023, Claeyssens_2025, Claeyssens_2026, Vanzella_2023, Adamo_2024, Mowla_2024, Abdurrouf_2025, Bradac_2025, Fujimoto_2025, Khullar_2026, Messa_2026, Nakane_2025}. These compact stellar clumps are parsec-scale and likely to be the progenitors of globular clusters \citep[e.g.,][]{Vanzella_2026}, nevertheless, their presence can shed light on the early clumpy star-formation mechanisms.

This work aims to measure the fraction of clumpy galaxies at $z \approx 1-9$ using data from the \textit{JWST} Advanced Deep Extragalactic Survey \citep[JADES;][]{Eisenstein_2026, Rieke_2023} to investigate the predominant clump formation mechanisms across redshift and stellar mass. We develop a self-consistent detection algorithm to identify clumps in rest-frame UV and optical images using JADES DR5 photometry and DR4 spectroscopy. Throughout this paper, we use \textit{clump} as an observational term for a compact, off-centre surface-brightness peak identified in \textit{JWST}/NIRCam imaging. Our detected clumps have a median area-equivalent radius of $\approx \!0.45\,{\rm kpc}$, measured directly from dendrogram regions without correction for the point-spread function (PSF). Given their sizes relative to the imaging resolution, we interpret them as sub-kpc structures or complexes that may contain multiple unresolved star-forming regions or stellar clusters, rather than individual clusters or necessarily gravitationally bound objects. Their origins may include \textit{in situ} fragmentation, merger-induced star formation, or accreted satellites. A companion JADES (DR3) study by \citet{Zhu_2026} investigates clump migration and physical properties using an independent selection method. Here, we focus on the evolution of $f_{\rm clumpy}$ with redshift and host stellar mass, particularly the contribution of major mergers.

The paper is structured as follows. In Section~\ref{sec:data} we describe the dataset used for this study, including the photometric and spectroscopic redshifts, spectral energy distribution (SED) modelling, and sample selection. Section~\ref{sec:Clump finding methodology} presents the clump detection methodology and completeness assessment. In Section~\ref{sec:Fraction of Clumpy Galaxies} we present the clumpy fraction as a function of redshift, stellar mass, and wavelength, and compare our results to the literature. Section~\ref{sec:Contribution of major mergers to clumpy systems} explores the potential contribution of major mergers to clumpy galaxies, while Section~\ref{sec:Clump properties} presents the physical properties of the clumps. In Section~\ref{sec:Discussion} we discuss the inferred clump formation scenarios, comparisons to simulations, and the main selection effects and limitations. Finally, Section~\ref{sec:conclusion} summarises our work and highlights the main findings.

Throughout this paper, we adopt the AB magnitude system \citep{Oke_1983}. We use a standard cosmology with $\Omega_{m} = 0.310$,  $\Omega_{\Lambda} = 0.689$, and $H_0 = 67.66 \mathrm{\ km \ s^{-1} \ Mpc^{-1}}$ \citep{Planck_2020}. Throughout our analysis, we use the \texttt{Astropy} python package \citep{astropy:2022}, and its subpackage \texttt{astropy.cosmology}, where we assume a flat $\Lambda$CDM cosmology with parameters from \citet{Planck_2020}. 

\section{Data}
\label{sec:data}

In this section, we present the dataset used for studying the fraction of clumpy galaxies. We provide a brief overview of the JADES survey and its data products, a compilation of available photometric and spectroscopic redshifts and their comparison, and the SED modelling used to obtain galaxy physical parameters. Finally, we describe the sample selection before we apply the clump detection algorithm, which we present in Section~\ref{sec:Clump finding methodology}.

\subsection{JADES survey}
\label{sec:JADES survey}

For this study we analyse the rich NIRCam imaging dataset from the JADES Data Release 5 \citep[DR5;][]{Alberts_2026, Carreira_2026, Duan_2026, Hainline_2026, Johnson_2026, Robertson_2026, Wu_2026}, which we present briefly in the following. This release consists of \textit{JWST}/NIRCam and \textit{JWST}/MIRI imaging from Cycles 1--4 combined with \textit{HST} photometry covering $\approx\!500 \ {\rm arcmin^2}$ area ($\approx\!320 \ {\rm arcmin^2}$ covered by the six wide-band NIRCam filters) in the GOODS-South and GOODS-North extragalactic fields. 

The DR5 NIRCam imaging, including data reduction, is presented by \citet{Johnson_2026}, while the derived photometric data products are described by \citet{Robertson_2026}, including the source catalogue, source detection, deblending, multi-band photometry, and photometric redshifts. The photometric catalogues are constructed from 35 space-based imaging mosaics including \textit{HST}/ACS, \textit{HST}/WFC3, \textit{JWST}/NIRCam, and \textit{JWST}/MIRI, and contain measurements for approximately 500,000 sources. Sources are identified from signal-to-noise detection images constructed primarily from the long-wavelength NIRCam mosaics, while deblending is aided by higher-resolution short-wavelength NIRCam imaging around $\lambda\!\approx\!2\,\mu{\rm m}$. The segmentation maps are generated by applying custom detection and deblending algorithms to these signal-to-noise images, assigning each detected pixel to a unique source ID and consequently defining the pixels associated with each object. These segmentation maps are then used to define the source catalogue, measure photometry, and mask neighbouring objects. The catalogue provides forced circular-aperture photometry, ellipsoidal Kron photometry \citep{Kron_1980}, and curve-of-growth measurements in each band, together with uncertainty estimates that account for correlated noise in the heterogeneous mosaics. Photometric redshifts are estimated with \texttt{EAZY} \citep{Brammer_2008}, using both small-aperture photometry on the native-resolution mosaics and Kron photometry on common-PSF mosaics. The photometric redshifts are also estimated using \texttt{Prospector} \citep{Johnson_2021}, which we adopt for this study and provide a detailed description in Section~\ref{sec:Galaxy SED modelling}.

For defining photometric centroids and sizes, we use the extensive morphological catalogues by \citet{Carreira_2026}. The morphological catalogue is constructed by fitting each detected source in GOODS-South and GOODS-North independently in all available \textit{JWST}/NIRCam wide-band filters, providing more than three million individual surface-brightness profile fits. The fits are performed with \texttt{pysersic} \citep{Pasha_2023}, which uses Bayesian inference to model each galaxy with a PSF-convolved single component S\'ersic profile \citep{Sersic_1963, Sersic_1968}. For each source and filter, cutouts are generated from the native-resolution DR5 mosaics, using the corresponding segmentation map to mask neighbouring sources, and selecting a model PSF appropriate for the dominant observing programme contributing to the mosaic at that position. The inferred structural parameters include the light-profile centroid, total flux, effective radius, S\'ersic index, axis ratio, and position angle, with both best-fit values and posterior samples provided in the released catalogues \citep{Carreira_2026}. The catalogue also reports quality-of-fit statistics, enabling problematic fits to be identified. We use these morphological measurements to define the global galaxy structure and centroids relevant for our clump analysis, while noting that a single smooth S\'ersic component is not expected to capture non-axisymmetric features such as spiral arms, bars, or clumpy substructure.

\subsection{Photometric and spectroscopic redshifts}
\label{sec:Photometric and spectroscopic redshifts}

\begin{figure}
 \includegraphics[width=\columnwidth]{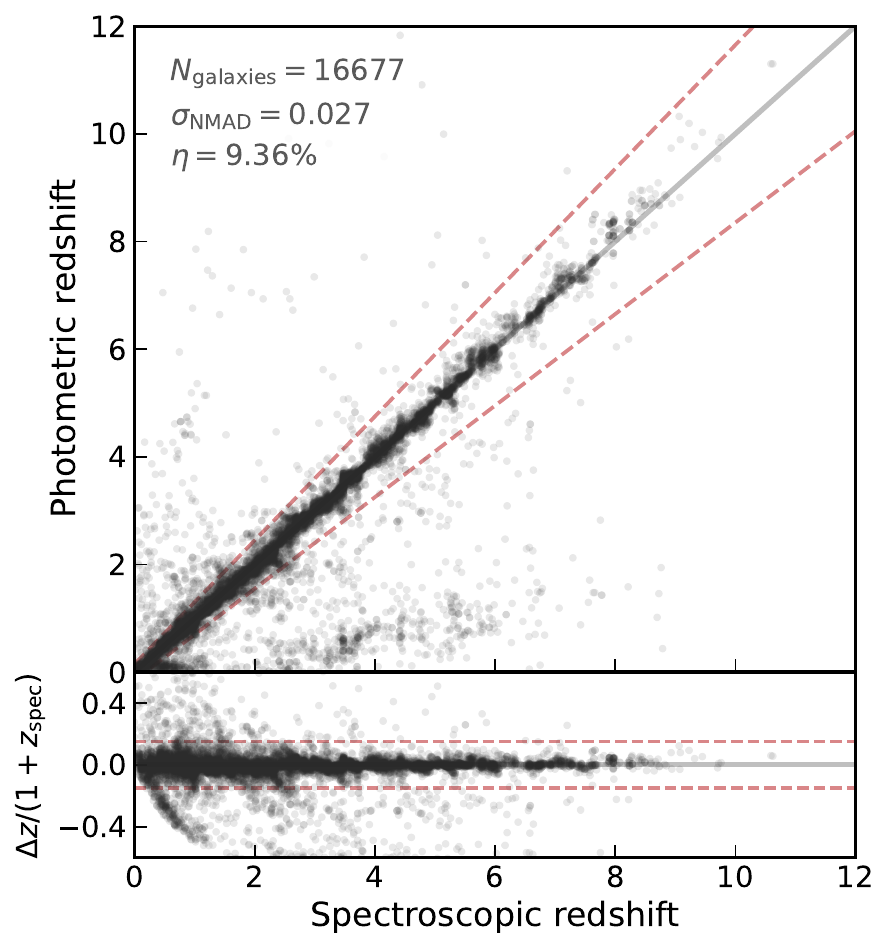}
 \caption{Comparison between the high-quality spectroscopic redshifts compiled in GOODS-South and GOODS-North and the median \texttt{Prospector} redshifts. Only spectroscopic measurements with \texttt{flag\_z\_qual = 1} and \texttt{flag\_match = 1} are shown. The upper panel compares $z_{\rm phot}$ and $z_{\rm spec}$, with the solid grey line showing the one-to-one relation, while the lower panel shows the normalised residuals, $\Delta z/(1+z_{\rm spec})$. The dashed red lines mark the catastrophic outlier threshold, $|\Delta z|/(1+z_{\rm spec}) = 0.15$. The comparison includes 16,677 spectroscopic measurements, with $\sigma_{\rm NMAD} = 0.027$ and an outlier fraction of $\eta = 9.36$ per cent. See Appendix~\ref{app:specz_catalogue} for the complete list of spectroscopic source catalogues and the catalogue construction procedure.}
 \label{fig:specz_vs_photoz}
\end{figure}

In this work, we adopt photometric redshifts estimated by the SED-fitting code \texttt{Prospector} \citep{Johnson_2021} and use spectroscopic redshifts instead where available. In this section, we compare the adopted photometric redshifts with a compilation of available spectroscopic redshifts in the JADES fields. The construction of the spectroscopic-redshift catalogue, including the input catalogues, astrometric matching, quality flag homogenisation, and final catalogue priority ordering, is described in Appendix~\ref{app:specz_catalogue}.

Briefly, the catalogue is constructed by matching each spectroscopic catalogue to the JADES photometric catalogue in the GOODS-South and GOODS-North fields. The different catalogues are homogenised into a common format, and their reliability and coordinate-match quality flags are converted to a unified system, where the highest quality and best coordinate matches are assigned \texttt{flag\_z\_qual = 1} and \texttt{flag\_match = 1}, respectively.

For the comparison below, we use only the best spectroscopic redshift measurements, i.e. satisfying \texttt{flag\_z\_qual = 1} and \texttt{flag\_match = 1} (these flags are defined in Appendix~\ref{app:specz_catalogue}). This selection yields 10,357 spectroscopic redshift measurements in GOODS-South and 6,337 in GOODS-North, giving a total of 16,694 high-quality spectroscopic measurements across the two fields. Since the galaxy physical properties used in this work are taken from the \texttt{Prospector} value-added catalogue (for a brief summary see Section~\ref{sec:Galaxy SED modelling}; \citealt{Duan_2026}), we compare the spectroscopic redshifts directly to the median \texttt{Prospector} photometric redshifts.

After requiring a finite \texttt{Prospector} photo-$z$, the comparison contains 16,677 high-quality spectroscopic redshifts (see Figure~\ref{fig:specz_vs_photoz}). Catastrophic outliers are defined as sources satisfying
\begin{equation}
    \frac{|z_{\mathrm{phot}} - z_{\mathrm{spec}}|}{1 + z_{\mathrm{spec}}} > 0.15 .
\end{equation}
The agreement is good overall, with a catastrophic outlier fraction of $\eta = 9.36$ per cent. We quantify the scatter around the one-to-one relation using the Normalised Median Absolute Deviation (NMAD),
\begin{equation}
 \sigma_{\mathrm{NMAD}} = 1.48 \times \mathrm{median} \left( \left| \frac{\delta z - \mathrm{median}(\delta z)}{1 + z_{\mathrm{spec}}} \right| \right),
\label{eq:sigma_nmad_specz}
\end{equation}
where $\delta z = z_{\mathrm{phot}} - z_{\mathrm{spec}}$. For the combined GOODS-South and GOODS-North sample, we measure $\sigma_{\mathrm{NMAD}} = 0.027$, indicating a tight agreement between the spectroscopic and photometric redshifts estimated by \texttt{Prospector} for the bulk of the sample. See Figure~\ref{fig:specz_vs_photoz} for the one-to-one comparison between the spectroscopic and the photometric redshifts.

Beyond serving as a validation, the spectroscopic redshift compilation is used to construct a catalogue of the best available redshift for each galaxy, as described in Appendix~\ref{app:specz_catalogue}. Where a secure spectroscopic redshift is available, it is preferred over the photometric redshift from \texttt{Prospector}. If multiple spectroscopic redshifts are available for the same object, we first select the measurements with the best combination of redshift quality and coordinate match flags. When several measurements with the same best flags are available and at least two of them agree within $\Delta z_{\mathrm{spec}} \leq 0.1$, we adopt the redshift from the highest-priority catalogue. If only one valid redshift is available, it is automatically assigned as the best spectroscopic redshift. This procedure ensures that the final redshift catalogue uses the most reliable available redshift while avoiding ambiguous spectroscopic identifications, and provides the redshifts used for the sample selection and redshift binning in the clumpy galaxy analysis.

\subsection{Galaxy SED modelling}
\label{sec:Galaxy SED modelling}

Here we briefly describe the SED fitting, from which the JADES DR5 stellar population catalogue is obtained which we use for this work. A full description of the method and the dataset is presented by \citet{Duan_2026}. The catalogue is based on Bayesian SED modelling with \texttt{Prospector} \citep{Johnson_2021}, applied to the JADES DR5 multi-wavelength photometric catalogue in GOODS-South and GOODS-North. The fiducial fits use the \textit{HST}+\textit{JWST}/NIRCam photometry, providing homogeneous stellar population constraints for the full photometric sample. Additional fits are performed by including \textit{JWST}/MIRI photometry or fixing the redshift to a secure spectroscopic redshift where these are available (see Section~\ref{sec:Photometric and spectroscopic redshifts}). For each source, all usable photometric bands are included, and a 5 per cent photometric error floor is applied. The fiducial \textit{HST}+\textit{JWST}/NIRCam configuration combined with fits fixed to the spectroscopic redshift (if available) is adopted for our analysis to ensure uniform stellar masses and the best redshifts across the full galaxy sample.

The SED modelling assumes a \citet{Chabrier_2003} initial mass function (IMF) and uses the FSPS stellar population synthesis framework \citep{Conroy_2009, Conroy_2010}, accelerated with the \texttt{Parrot} emulator \citep{Mathews_2023}. The model includes 18 free parameters describing redshift, stellar mass, stellar- and gas-phase metallicity, a non-parametric star-formation history, dust attenuation and emission, and a possible AGN contribution. Stellar metallicities are assigned a mass-dependent prior based on the local mass--metallicity relation \citep{Gallazzi_2005}, with a mass-dependent mean and scatter that increases towards lower stellar masses, while gas-phase metallicity is allowed to vary independently. Nebular emission is included following \citet{Byler_2017}, with a fixed ionisation parameter, and dust attenuation is modelled using a two-component prescription \citep{Charlot_2000}, accounting separately for birth-cloud and diffuse ISM attenuation. Mid-infrared dust emission \citep{Draine_2007} and AGN torus emission \citep{Nenkova_2008a,Nenkova_2008b,Leja_2018} are also included in the model, although the latter is primarily constrained for sources with MIRI coverage.

\begin{figure}
 \includegraphics[width=\columnwidth]{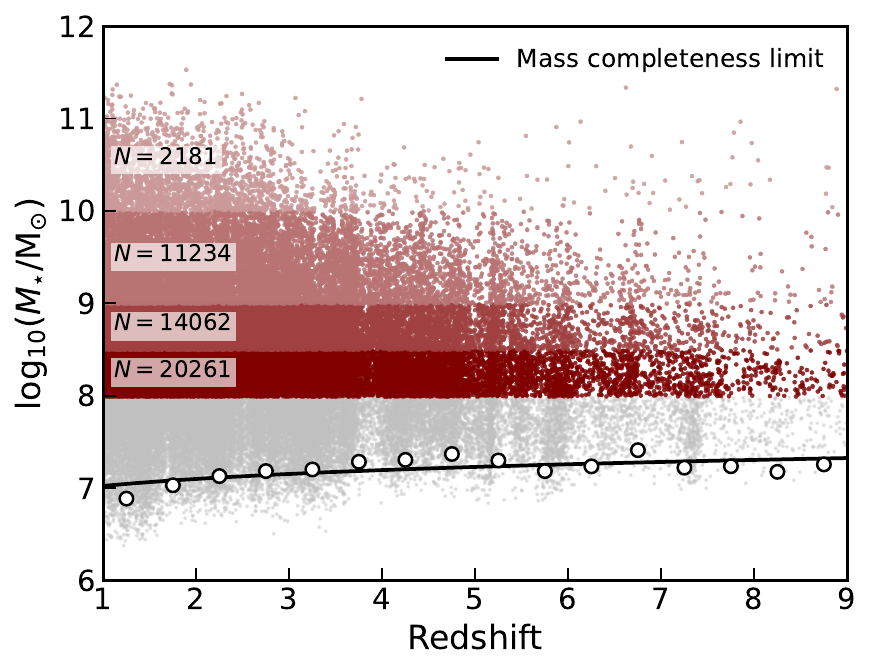}
 \caption{Stellar mass distribution as a function of redshift for the combined GOODS-South and GOODS-North sample. Grey points show the parent catalogue, while coloured points show the initial galaxy sample used for the clumpy fraction analysis, separated by the stellar mass bins, with the number of objects displayed within each bin. The open circles show the binned 90 per cent stellar-mass completeness limits estimated following the method of \citet{Pozzetti_2010} (as described in \citealt{Puskas_2025}), and the black curve shows the fitted redshift-dependent completeness limit. The selected sample remains well above the inferred completeness limit across the full redshift range, corresponding to an approximate 90 per cent completeness threshold of $7 \lesssim \log_{10}(M_\star/{\rm M_\odot}) \lesssim 7.5$.}
 \label{fig:stellar_mass_redshift}
\end{figure}

The star formation histories (SFHs) are described by a non-parametric continuity model with seven age bins \citep{Leja_2017, Leja_2019, Tacchella_2022a}. Instead of assuming a flat expectation for the SFR ratios between neighbouring age bins, the fiducial catalogue adopts a physically motivated star-forming main sequence (SFMS) prior. This assumes that galaxies grow, on average, along the evolving SFMS \citep[e.g.][]{Speagle_2014, Schreiber_2015, Simmonds_2025}, while still allowing substantial deviations, including bursty star formation and rapid quenching, when required by the data. Posteriors are sampled with \texttt{dynesty} \citep{Speagle_2020}, and we adopt the median as the fiducial value for each physical parameter, with uncertainties defined by the $16^{\rm th}$ and $84^{\rm th}$ percentiles.

For the present work, the most relevant quantities are the photometric or spectroscopic redshifts and stellar masses used to define our redshift and stellar-mass bins. The catalogue validation presented by \citet{Duan_2026} shows that the inferred stellar masses are well constrained by the broad \textit{HST}+\textit{JWST}/NIRCam wavelength coverage, while comparisons with spectroscopic redshifts demonstrate robust photometric redshift performance. We therefore use these uniformly derived stellar population properties throughout our clumpy fraction analysis. To visualise the stellar mass distribution of our sample, we plot the stellar masses of the combined GOODS-South and GOODS-North sample as a function of redshift in Figure~\ref{fig:stellar_mass_redshift}. 

\begin{figure*}
 \includegraphics[width=\linewidth]{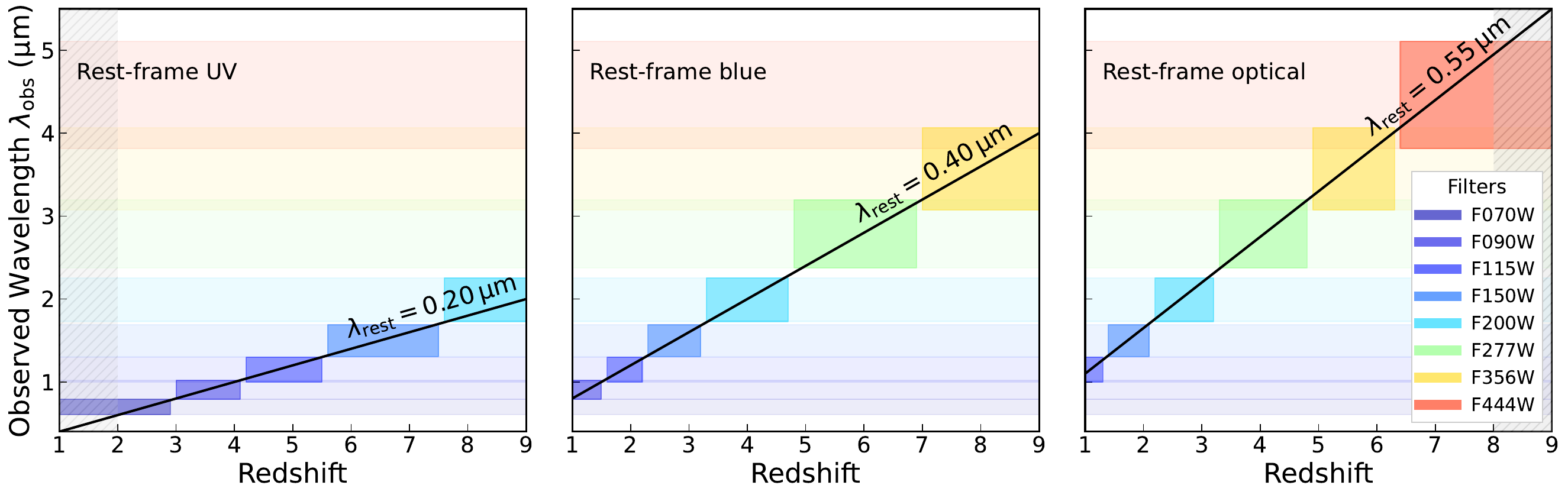}
 \caption{Selection of the NIRCam wide-band filters used to probe the three target rest-frame wavelengths as a function of redshift. The three panels correspond to the rest-frame UV, blue, and optical wavelengths at $\lambda_{\rm rest}=0.2$, $0.4$, and $0.55~\mu{\rm m}$, respectively. In each panel, the black line shows the target observed-frame wavelength, while the shaded regions show the wavelength coverage of the available JADES NIRCam filters. For each galaxy, we select the filter closest to the target observed-frame wavelength. The filter set provides near-continuous coverage over most of $1 \leq z \leq 9$, except for the rest-frame UV at $z \lesssim 2$ and the rest-frame optical at $z \gtrsim 8$ (indicated by the hatched regions). In these regimes we still use F070W and F444W, respectively, as the closest available filters.}
 \label{fig:filter_selection}
\end{figure*}

We estimate the stellar-mass completeness following the method of \citet{Pozzetti_2010}, as described in detail in \citet{Puskas_2025}. Briefly, this is done by computing in redshift bins the limiting stellar mass each galaxy would have if its magnitude in the F444W band were equal to the local survey depth ($5\sigma$--flux limit). This uses the observed range of galaxy mass-to-light ratios to estimate the stellar mass at which galaxies of different intrinsic properties would remain detectable. We then use the limiting-mass distribution of the faintest galaxies to define the 90 per cent stellar-mass completeness limit in each bin and fit a logarithmic curve to obtain a continuous limit. Our sample has a high stellar mass completeness, with a 90 per cent limit between $7 \lesssim \log_{10}(M_\star/{\rm M_\odot}) \lesssim 7.5$ across the full redshift range.

\subsection{Sample selection}
\label{sec: Sample selection}

Finally, we present the sample pre-selection on which we later run the clump finder algorithm, as described in Section~\ref{sec:Clump finding methodology}. Our parent sample is constructed by combining the photometric catalogue, size measurements, centroids from the morphological catalogue, and the derived physical properties from the \texttt{Prospector} catalogue. 

We require galaxies to have reliable photometric coverage, redshift estimates, stellar masses, and morphological information for the clump analysis. The fiducial photometric-redshift \texttt{Prospector} catalogue is first combined with the spectroscopic-redshift \texttt{Prospector} catalogue, giving priority to the spec-$z$ catalogue where available. We then apply a set of quality cuts for the \texttt{Prospector} parent catalogue, requiring ${\rm SNR_{F444W}} \geq 5$ ($\texttt{flag\_snr\_5}=1$) and coverage by at least six NIRCam bands ($\texttt{Nbands\_nircam}\geq6$). Furthermore, we require the redshift to be well constrained ($\texttt{flag\_good\_redshift\_30}=1$), which is defined as $z_{84}-z_{16} \leq 0.30 \times z_{50}$, where the lower indices denote percentiles. Finally, we require a well-constrained stellar mass ($\texttt{flag\_good\_mass\_dex1}=1$), which is defined as the stellar mass posterior width ($84^{\rm th} - 16^{\rm th}$) being less than 1 dex. The final sample is restricted to galaxies in the redshift range $1 \leq z \leq 9$, using the median redshift estimate $z_{\rm red,50}$, and to stellar masses $\log_{10}(M_\star/{\rm M_\odot}) \geq 8$, using the median stellar mass estimate. We note that this choice of stellar mass limit is well above the previously estimated 90 per cent completeness limit. These cuts ensure that the galaxies have sufficient wavelength coverage for SED fitting with reliable derived physical properties, and are in the conservative mass completeness range.

For each galaxy in the final sample, we analyse the rest-frame UV, blue, and optical morphologies at three fixed rest-frame wavelengths: $\lambda_{\rm rest}=0.2$, $0.4$, and $0.55\,\mu{\rm m}$. For each galaxy and rest-frame wavelength, the corresponding observed-frame wavelength is
\begin{equation}
    \lambda_{\rm obs} = \lambda_{\rm rest}(1+z),
\end{equation}
where $z$ is the median photometric redshift or the spectroscopic redshift, if available. Rather than interpolating between images, we select the available NIRCam filter whose central wavelength is closest to $\lambda_{\rm obs}$, using the set of available NIRCam filters: F070W, F090W, F115W, F150W, F200W, F277W, F356W, and F444W. The filter choice is therefore redshift-dependent and is made independently for each galaxy and rest-frame wavelength (see Figure~\ref{fig:filter_selection}). The JADES NIRCam wide-band filter set provides an almost complete coverage for the three target wavelengths for the full redshift range of $1 \leq z \leq 9$, except for the rest-frame UV at $z \lesssim 2$ and the rest-frame optical at $z \gtrsim 8$. Therefore, we decide to drop the rest-frame UV measurement at $z=1-2$ as it is not covered by F070W, while retaining the rest-frame optical measurement at $z=8-9$ which is still partially covered by F444W. We do acknowledge that our results in the latter case may be less reliable due to the partial filter coverage.

We then create cutouts for each galaxy in the final sample in the filters probing the target rest-frame wavelengths, centred on the sky positions from the photometric catalogue. The cutout size is defined by the larger dimension of the source-extraction bounding box, with an additional padding of 10 pixels around the source (using the $0.03''$/pix images). The same square cutout size is used to extract images at all rest-frame wavelengths from the corresponding JADES NIRCam mosaics using their WCS. Matching segmentation-map cutouts are extracted using the same dimensions and centres, with pixels outside the main segment filled with zeroes.

This procedure allows us to probe approximately the same rest-frame emission across the full redshift range using JADES NIRCam imaging. The three wavelength choices sample progressively older stellar populations: the rest-frame UV is most sensitive to recent, unobscured star formation, while the 0.4 and 0.55 µm bands trace the rest-frame optical light, which is less dominated by the youngest stars and more closely reflects the underlying stellar structure. This enables a consistent comparison of clumpy structure as a function of redshift, stellar mass, and wavelength.

\section{Clump finding methodology}
\label{sec:Clump finding methodology}

In this section we present the clump detection methodology which we apply to identify clumpy galaxies and to measure the clump properties. We develop and apply a dendrogram-based clump detection approach tailored to high-redshift \textit{JWST} imaging, as presented in Section~\ref{sec:Clump detection algorithm}. Then we perform a completeness assessment by running simulations with randomised artificial clump injection to our real galaxy images and try to recover them with the same clump detection algorithm, as described in Section~\ref{sec:Completeness assessment}.

\subsection{Clump detection algorithm}
\label{sec:Clump detection algorithm}

Several methods exist in the literature to identify bright clumps in imaging data. We list some of the main methods briefly for an overview, before motivating our choice of clump detection algorithm.

\textit{Contrast-image method:} clumps are identified by smoothing the galaxy image, subtracting this smooth component to create a high-contrast residual image, masking pixels with low signal-to-noise, running \texttt{SExtractor} \citep{Bertin_1996} within a dilated segmentation map, and selecting off-centre clumps above a fractional luminosity threshold, typically $L_{\rm clump}/L_{\rm gal} \geq 0.08$ \citep[e.g.,][]{Guo_2015, Shibuya_2016}. 

\textit{Fourier high-pass filtering method:} clumps are detected by Fourier-transforming the rest-frame NUV image, filtering out the low-frequency host galaxy light, and applying source detection to the remaining high-frequency structure, with additional cuts on contiguity, signal-to-noise, excluding the central clump, lying within the galaxy segmentation map, and completeness-limited fractional luminosity \citep[e.g.,][]{Sattari_2023, de_la_Vega_2026a}. 

\textit{Normalised-profile (wavelet/starlet) method:} Clumps are identified by subtracting large-scale galaxy light using a wavelet/starlet decomposition \citep{Starck_2007, Melchior_2018} and then detecting compact peaks in the resulting high-contrast residual image. In some implementations this is applied after an initial normalised-profile selection of clumpy galaxy candidates, while in others it is applied directly to rest-frame UV or NIR images \citep{Wuyts_2012, Kalita_2025b, Sok_2022, Sok_2025, Sok_2026a}. 

\textit{Sérsic-residual or watershed methods:} clumps are defined as statistically significant residual substructures after subtracting a smooth two-dimensional Sérsic model, with Gaussian-smoothed residuals with a threshold of $3\sigma$, local maxima used as watershed markers \citep{Meyer_1994}, and spurious regions removed using flux-fraction and connected-pixel cuts \citep[e.g.,][]{Zhu_2026}.

All these methods involve a smoothing and subtracting step to some degree with different assumptions regarding the smoothing kernel and other threshold choices in various parameter spaces. Our goal was to implement a clump finding algorithm which skips the smoothing and subtracting process and can be directly applied on the NIRCam imaging cutouts with a minimum number of parameter choices. Therefore, we chose the \texttt{Python} package \texttt{astrodendro} \citep{Robitaille_2019} to extract clumps from the images directly. \texttt{astrodendro} identifies hierarchical structure in a 2-dimensional image by following how contiguous regions of emission grow and merge as the intensity threshold is lowered. In this representation, the smallest structures without underlying substructures are defined as \textit{leaves}, larger connected structures are \textit{branches}, and the full connected region forms the \textit{trunk} of the \textit{dendrogram}. The dendrogram construction process is in practice an isophotal-like analysis as the structure is built from the brightest pixel, and lower-flux pixels are subsequently assigned to the leaves and branches \citep{Goodman_2009}. The package has previously been used to identify mock H$\alpha$ clumps and H{\sc ii} regions in simulations \citep{Tacchella_2022b, McClymont_2024}, as well as star-forming clumps in low-redshift ram-pressure-stripped galaxies \citep{Giunchi_2023} and interacting dwarf galaxies \citep{Lassen_2025}. It has also been applied to galaxies at $z\approx0.5$--3 \citep[e.g.,][]{Meng_2020, Sharma_2021} and in other simulation-based studies \citep[e.g.,][]{Oklopcic_2017}. In this work we extend the usage of \texttt{astrodendro} up to $z \approx 9$.

In the following, we describe the steps of the clump detection algorithm and the parameter choices that we made.

\begin{enumerate}[1.]
    \item \textit{Object cutouts.} We identify clumps independently in each of the three rest-frame wavelength cutouts (see Section~\ref{sec: Sample selection}), at $\lambda_{\rm rest}=0.2$, $0.4$, and $0.55\,\mu{\rm m}$ (rest-frame UV, blue, and optical). We also create cutouts from the segmentation map (see Section~\ref{sec:JADES survey}) for each object, with the same size as the imaging cutouts.
    
    \item \textit{Validity checks.} Before running the clump finding algorithm, we apply a set of quality checks to ensure that the science cutouts and segmentation maps are suitable for analysis. For each galaxy, we load the corresponding segmentation map cutout and define the galaxy segmentation region as the central segment, i.e. the segmentation label at the centre of the cutout. Galaxies are excluded if the central segmentation label is zero. Finally, we exclude galaxies if any of their three rest-frame wavelength cutouts are missing. This latter criterion is needed since we would like to analyse how the clumpy fraction varies as a function of wavelength for the same sample of galaxies simultaneously. It is a strict requirement, which almost halves our sample. Furthermore, we exclude galaxies for which any pixel within the central segmentation region has a non-finite or zero value in the science image, ensuring that the galaxy region used for the clump analysis is fully covered by valid imaging data. 

    \par\hspace*{1em}After applying the segmentation map validity checks, the sample contains 47,738 galaxies. Requiring valid cutouts in all three rest-frame wavelength bands and fully finite, non-zero science pixels within the central segmentation region reduces the final sample on which we run the clump detection code to 24,064 galaxies. The final sample has similar redshift, stellar mass, and field distributions to the parent sample.

\begin{figure*}
 \includegraphics[width=\linewidth]{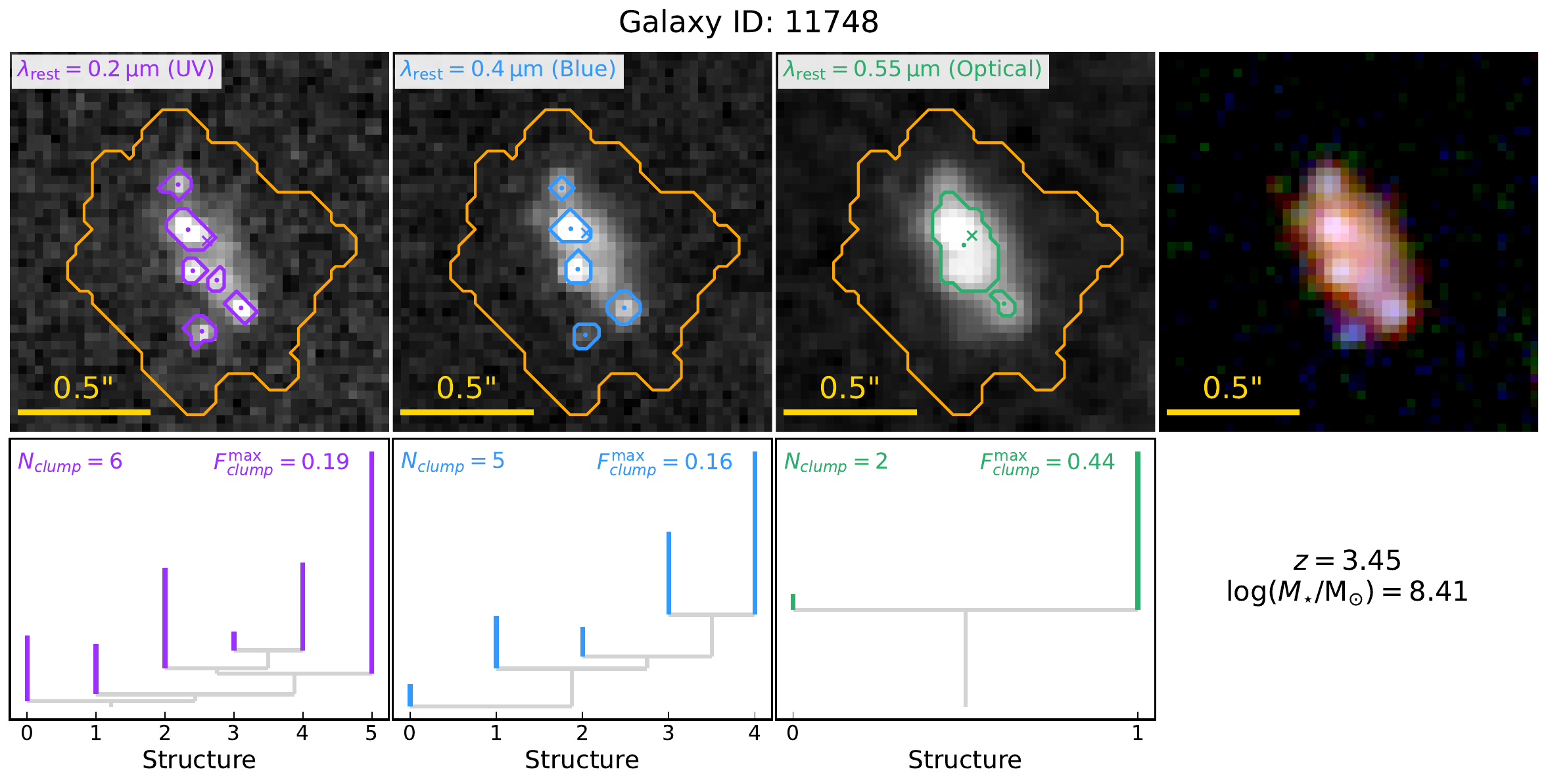}
 \caption{Example cutouts of a clumpy galaxy at $z_{\rm phot}=3.45$ with $\log_{10}(M_\star / {\rm M_\odot}) = 8.41$ (and ${\rm SFR_{10}}=1.9 \ {\rm M_\odot/yr}$) at the three rest-frame wavelengths used in this work: UV, blue, and optical (at $\lambda_{\rm rest}=0.2$, $0.4$, and $0.55\,\mu{\rm m}$, respectively), together with the combined RGB image constructed from these cutouts. The coloured contours for each band mark the clumps identified by the detection algorithm using the \texttt{astrodendro} package, shown in purple, blue, and green for increasing rest-frame wavelength. The orange contour indicates the segmentation region assigned to the central galaxy in the photometric catalogue, which defines the area used to measure the total galaxy flux and to associate detected structures with the target galaxy. The lower panels show the corresponding dendrogram tree structure for each wavelength, with the selected clump leaves highlighted in the same colour as in the image cutouts. The scale bar in each image corresponds to 0.5 arcsec (3.8 pkpc). The small crosses indicate the morphological centroids of the galaxies, while the dots indicate the flux-weighted centres of the individual clumps.}
 \label{fig:clump_dendrogram}
\end{figure*}

    \item \textit{Dendrogram-based clump detection.} For each galaxy in our sample, we run \texttt{astrodendro} on the science cutouts corresponding to the three rest-frame wavelengths to identify compact substructures. We interpret the dendrogram leaves as clump candidates, since they correspond to local surface-brightness peaks that remain distinct from neighbouring emission, down to a specified contrast level. The dendrogram construction is controlled primarily by three parameters: \textit{(i)} the minimum pixel value \texttt{min\_value} required for inclusion in the dendrogram, \textit{(ii)} the minimum significance or contrast \texttt{min\_delta} required for a structure to be considered independent, and \textit{(iii)} the minimum number of connected pixels \texttt{min\_npix}. 
    
    \par\hspace*{1em}The background noise level is estimated directly from the cutouts using sigma-clipped statistics (with a choice of 3$\sigma$). Our parameter choices for computing the dendrograms are: a minimum intensity threshold of $3\sigma$, a minimum contrast of $1\sigma$, and a minimum area of 4 connected pixels. We consider only dendrogram leaves, corresponding to terminal structures in the dendrogram hierarchy, as clump candidates. To ensure that the detected structures on the entire cutout are associated with the target galaxy, we retain only leaves for which at least 25 per cent of their pixels overlap with the central segmentation mask. This method provides a direct, segmentation-like identification of substructures without first constructing a smoothed galaxy model or residual image.

    \item \textit{Flux fraction selection.} At this stage, we have a set of clump candidates associated with the central galaxy. Although the dendrogram parameters were chosen to select significant compact structures, some detected candidates may still be too faint relative to their host galaxy to be physically meaningful. Therefore, we make an additional clump flux cut, motivated by the total flux of the galaxy. The fluxes of the leaves are directly available from the dendrogram structure ($F_{\rm clump}$), and it is simply given by the summed flux of the pixels belonging to each dendrogram leaf. For consistency, we refrain from using the circular or Kron fluxes from the photometric catalogue (Section~\ref{sec:JADES survey}), and instead we measure the total galaxy flux as the sum of the pixel fluxes within the central segment ($F_{\rm galaxy}$). We only retain clump candidates which satisfy the criterion $F_{\rm clump}/F_{\rm galaxy} > 0.001$ (same as $L_{\rm clump}/L_{\rm galaxy}$ due to ratio, which we use hereafter). See more about different flux fraction threshold choices later in this section.

    \item \textit{Clump photometry and properties.} For each retained clump, we measure its flux, fractional flux, pixel area, physical area, pixel position, sky position, and distance from the galaxy centre. Clump centroids are computed as flux-weighted centroids using the positive pixel fluxes within each dendrogram leaf as weights. Physical areas and projected distances are computed using the angular scale at the galaxy redshift. The projected distance of each clump from the galaxy centre or galactocentric distance ($r_{\rm clump}$) is measured relative to the morphological centroid in the corresponding filter probing that rest-frame wavelength. We also normalise $r_{\rm clump}$ by the maximum radial extent of the central segmentation region ($r_{\rm seg,max}$) measured from the same morphological centroid.

    \item \textit{Central and non-central clumps.} We flag central clumps using the morphology-derived centroid. A clump is classified as central if its dendrogram leaf contains the morphology centroid. If no clump directly contains the centroid, the closest clump is instead flagged as central, provided that its centroid lies within 3 pixels ($0.09''$) of the morphology centroid. If neither condition is satisfied, all detected clumps in that galaxy and wavelength are treated as off-centre. This central flag is used only to identify and exclude central structures from the subsequent off-centre clump analysis.
    
\end{enumerate}

The final measurements are saved in an HDF5 catalogue, with one group per galaxy and separate subgroups for the three rest-frame wavelengths. For each wavelength, we store the galaxy-level quantities, including the physical pixel scale, morphology centroid coordinates, and the maximum segmentation radius, together with the measurements for each clump. Diagnostic plots are also generated, showing the cutouts, segmentation contours, clump contours, dendrogram trees, and RGB images (see Figure~\ref{fig:clump_dendrogram}).

Since \texttt{astrodendro} identifies local surface-brightness peaks rather than physically distinct star forming regions, some selected leaves may trace compact regions within spiral arms, asymmetric discs, tidal features, or residuals associated with source segmentation and deblending. The injection--recovery simulations (see Section~\ref{sec:Completeness assessment}) quantify the detectability of PSF-like clumps embedded in real galaxy backgrounds, but do not directly measure the purity of the detected catalogue or distinguish between these different morphological origins. We therefore interpret the detections as compact off-centre surface-brightness substructures, rather than assigning every individual detection a unique physical origin.

We note that the flux (or luminosity) fraction threshold choice in the clump selection step is a crucial part of this work, as the clumpy fraction measurement will directly depend on this parameter choice, i.e. if a clump is above or below this threshold the galaxy will be determined as clumpy or not. Therefore, we treat this choice with special care and look at different thresholds used in the literature. \citet{Guo_2015} define clumps as structures with $L_{\rm clump}/L_{\rm galaxy} > 0.08$, i.e. define clumps as whose UV luminosity is brighter than 8 per cent of the total UV luminosity of the galaxy. This choice was based on redshifting M101 to observing conditions of their CANDELS sample and comparing the resulting local H\,{\sc ii} region clumps to real high-redshift clumps. \citet{Guo_2015} found that clumps above $L_{\rm clump}/L_{\rm galaxy}\simeq0.08$ could no longer be explained by redshifted local star-forming regions alone, and therefore adopted this as their conservative clump threshold. Other studies using \textit{HST} data adopted the same threshold choice of 8 per cent, e.g., \citet{Shibuya_2016}, \citet{Ribeiro_2017}, \citet{Guo_2018}, \citet{Sok_2022}, and \citet{Sattari_2023}. Other studies use a more aggressive cut at 5 per cent for the fractional contribution of the clump luminosity to the total galaxy UV luminosity \citep[see, e.g.,][]{Wuyts_2012, Sok_2026a}. These choices are somewhat arbitrary and have an effect on the number of clumpy galaxies and the clump multiplicity. The improved sensitivity and spatial resolution of \textit{JWST}/NIRCam allow us to probe lower-contrast substructures than was typically possible with earlier imaging. At this current stage we set a very low (aggressive) threshold of $L_{\rm clump}/L_{\rm galaxy} > 0.001$ to include even the very faint clump candidates. We revisit this choice in Section~\ref{sec:Fraction of Clumpy Galaxies} when we calculate the clumpy fractions after the completeness assessment.

\subsection{Completeness assessment}
\label{sec:Completeness assessment}

\begin{figure*}
 \includegraphics[width=\linewidth]{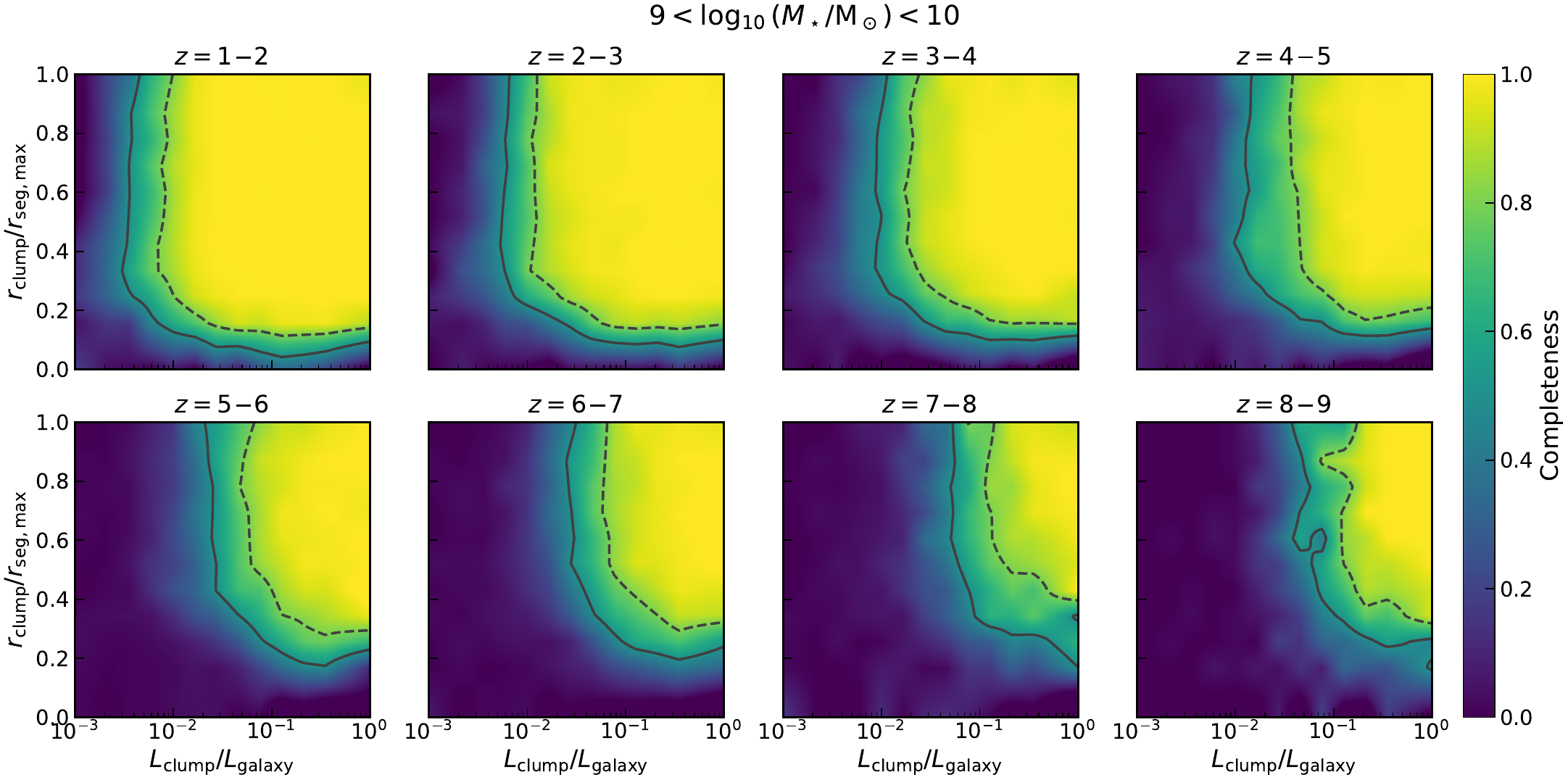}
 \caption{Two-dimensional clump detection completeness maps in the plane of clump luminosity fraction, $L_{\rm clump}/L_{\rm galaxy}$, and normalised galactocentric radius, $r_{\rm clump}/r_{\rm seg,max}$, shown for galaxies in the stellar mass bin $9.0 \leq \log_{10}(M_\star/{\rm M_\odot}) < 10.0$ at the blue rest-frame wavelength, $\lambda_{\rm rest}=0.4 \ \mu{\rm m}$. Each panel corresponds to a different redshift bin. The maps are derived from the Monte Carlo injection--recovery simulations, in which artificial PSF-like clumps with varying luminosity ratios and galactocentric radii are injected into real galaxy cutouts and recovered using the same dendrogram-based clump finding algorithm as applied to the science sample. The colour scale indicates the recovery fraction, $C=N_{\rm rec}/N_{\rm inj}$, while the solid and dashed contours mark the 50 per cent and 80 per cent completeness limits, respectively. Completeness is generally highest for bright clumps at larger galactocentric radii, and declines towards fainter clumps, smaller radii, and higher redshift. The corresponding one-dimensional completeness curves (Figures~\ref{fig:flux_completeness} and \ref{fig:radial_completeness}) further illustrate the stronger dependence on clump luminosity fraction and the weaker radial dependence, with reduced recovery near the galaxy centre.
}
 \label{fig:2d_completeness}
\end{figure*}

\begin{table*}
    \centering
    \caption{Minimum clump to galaxy fractional luminosity, $(L_{\rm clump}/L_{\rm galaxy})_{50}$, expressed as a percentage, required to recover at least 50 per cent of injected clumps as a function of redshift, stellar mass, and rest-frame wavelength. We use these values for each respective bin to select a 50 per cent complete clump sample for our fiducial clumpy galaxy fraction measurement. Ellipses indicate bins excluded from subsequent analysis (see Section~\ref{sec:Clumpy fraction}).}
    \label{tab:lclump_lgal_50_limits}
    \resizebox{\linewidth}{!}{%
    \begin{tabular}{ccccc c ccccc}
        \cline{1-5}\cline{7-11}
        \noalign{\vskip 2pt}
        Redshift bin & $\log_{10}(M_\star/{\rm M_\odot})$ 
        & UV 
        & Blue 
        & Optical
        &&
        Redshift bin & $\log_{10}(M_\star/{\rm M_\odot})$ 
        & UV 
        & Blue 
        & Optical \\[2pt]
        \cline{1-5}\cline{7-11}
        \noalign{\vskip 1pt}
        \cline{1-5}\cline{7-11}
        \noalign{\vskip 2pt}

        \multirow{4}{*}{$1$--$2$} 
        & $8$--$8.5$  & $\cdots$  & $\geq 4.1\%$ & $\geq 2.8\%$
        && \multirow{4}{*}{$5$--$6$} 
        & $8$--$8.5$  & $\geq 12\%$  & $\geq 13\%$  & $\geq 18\%$ \\
        & $8.5$--$9$  & $\cdots$ & $\geq 1.6\%$ & $\geq 1.2\%$
        &&
        & $8.5$--$9$  & $\geq 5.9\%$ & $\geq 6.3\%$ & $\geq 9.7\%$ \\
        & $9$--$10$  & $\cdots$ & $\geq 0.4\%$ & $\geq 0.3\%$
        &&
        & $9$--$10$  & $\geq 3.8\%$ & $\geq 3.9\%$ & $\geq 4.8\%$ \\
        & $>10$      & $\cdots$ & $\geq 0.2\%$ & $\geq 0.2\%$
        &&
        & $>10$      & $\cdots$     & $\cdots$     & $\cdots$     \\
        \cline{1-5}\cline{7-11}

        \multirow{4}{*}{$2$--$3$} 
        & $8$--$8.5$  & $\geq 13\%$  & $\geq 4.9\%$ & $\geq 4.0\%$
        && \multirow{4}{*}{$6$--$7$} 
        & $8$--$8.5$  & $\geq 13\%$  & $\geq 16\%$  & $\geq 29\%$ \\
        & $8.5$--$9$  & $\geq 7.3\%$ & $\geq 2.7\%$ & $\geq 2.2\%$
        &&
        & $8.5$--$9$  & $\geq 7.3\%$ & $\geq 8.4\%$ & $\geq 15\%$ \\
        & $9$--$10$  & $\geq 2.3\%$ & $\geq 0.7\%$ & $\geq 0.6\%$
        &&
        & $9$--$10$  & $\geq 3.8\%$ & $\geq 4.5\%$ & $\geq 7.4\%$ \\
        & $>10$      & $\geq 2.4\%$ & $\geq 0.3\%$ & $\geq 0.2\%$
        &&
        & $>10$      & $\cdots$     & $\cdots$     & $\cdots$     \\
        \cline{1-5}\cline{7-11}

        \multirow{4}{*}{$3$--$4$} 
        & $8$--$8.5$  & $\geq 9.7\%$ & $\geq 6.4\%$ & $\geq 6.8\%$
        && \multirow{4}{*}{$7$--$8$} 
        & $8$--$8.5$  & $\geq 12\%$  & $\geq 21\%$  & $\geq 29\%$ \\
        & $8.5$--$9$  & $\geq 4.4\%$ & $\geq 2.9\%$ & $\geq 3.1\%$
        &&
        & $8.5$--$9$  & $\geq 8.0\%$ & $\geq 13\%$  & $\geq 18\%$ \\
        & $9$--$10$  & $\geq 2.4\%$ & $\geq 1.1\%$ & $\geq 1.2\%$
        &&
        & $9$--$10$  & $\geq 8.1\%$ & $\geq 8.9\%$ & $\geq 10\%$ \\
        & $>10$      & $\geq 2.0\%$ & $\geq 0.4\%$ & $\geq 0.4\%$
        &&
        & $>10$      & $\cdots$     & $\cdots$     & $\cdots$     \\
        \cline{1-5}\cline{7-11}

        \multirow{4}{*}{$4$--$5$} 
        & $8$--$8.5$  & $\geq 9.3\%$ & $\geq 7.6\%$ & $\geq 10\%$
        && \multirow{4}{*}{$8$--$9$} 
        & $8$--$8.5$  & $\geq 12\%$  & $\geq 25\%$  & $\geq 35\%$ \\
        & $8.5$--$9$  & $\geq 5.0\%$ & $\geq 4.1\%$ & $\geq 5.1\%$
        &&
        & $8.5$--$9$  & $\geq 6.0\%$ & $\geq 12\%$  & $\geq 16\%$ \\
        & $9$--$10$  & $\geq 2.7\%$ & $\geq 1.8\%$ & $\geq 2.3\%$
        &&
        & $9$--$10$  & $\geq 4.6\%$ & $\geq 8.4\%$ & $\geq 10\%$ \\
        & $>10$      & $\geq 2.3\%$ & $\geq 1.0\%$ & $\geq 1.0\%$
        &&
        & $>10$      & $\cdots$     & $\cdots$     & $\cdots$     \\
        \cline{1-5}\cline{7-11}
    \end{tabular}%
    }
\end{table*}

In this section we describe how we assess the completeness of our clump detection algorithm. The detectability of clumps depends on their luminosity, their position within the host galaxy, the observed wavelength, and the properties of the host galaxy population. We therefore quantify the clump-detection completeness by injecting fake clumps onto a set of randomly selected real galaxy images, regardless if they are clumpy or not to begin with, and run the clump detection algorithm to recover them, similar to the approaches used in the literature \citep[e.g.,][]{Guo_2012, Guo_2015, Guo_2018, Shibuya_2016, Soto_2017, Sattari_2023, Sok_2026a, de_la_Vega_2026a}. These simulations quantify the recovery completeness of injected clumps but do not directly measure the purity of the detected catalogue. An explicit purity estimate would require a separate false-positive analysis using, for example, noise-only or smooth mock galaxy images.

We perform these Monte Carlo injection--recovery simulations independently in the same redshift and stellar mass bins used for the clumpy fraction measurements, i.e. $1 \leq z < 9$ in bins of $\Delta z=1$, and stellar mass bins of $8.0 \leq \log_{10}(M_\star/{\rm M_\odot}) < 8.5$, $8.5 \leq \log_{10}(M_\star/{\rm M_\odot}) < 9.0$, $9.0 \leq \log_{10}(M_\star/{\rm M_\odot}) < 10.0$, and $\log_{10}(M_\star/{\rm M_\odot}) \geq 10.0$. The simulations are run separately at the three rest-frame wavelengths $\lambda_{\rm rest}=0.2$, $0.4$, and $0.55\,\mu{\rm m}$. We exclude the highest-mass bin at $z\geq5$, where the number of galaxies is too small for a robust estimate (see Figure~\ref{fig:stellar_mass_redshift}). We randomly select 100 galaxies per redshift--stellar mass bin, or all galaxies if fewer than 100 are available (affecting only four high-$z$ bins), resulting in a sample of 2,546 galaxies. For each galaxy, we inject artificial clumps into the science cutouts and attempt to recover them using the same dendrogram-based clump finding algorithm applied to the real data. We repeat this process 250 times per galaxy with varying clump properties to sample the parameter space. An injected clump is considered recovered if a detected off-centre clump lies within 2 pixels of its injected position. Central detections are excluded, consistent with our science clump selection.

Artificial clumps are modelled as point sources using the model PSF (mPSF) corresponding to the nearest observed NIRCam filter, following previous works from the literature \citep[e.g.,][]{Guo_2015, Guo_2018, Shibuya_2016, Sattari_2023, de_la_Vega_2026a}. The NIRCam mPSFs were generated from simulated star images constructed to match the real observations, rather than measured directly from stars in the science mosaics (for a detailed description, we refer the reader to \citealt{Robertson_2026}). Artificial clumps are injected only within the central galaxy segmentation region. Their galactocentric radii are sampled uniformly in radius over the range $2~{\rm pix} \leq r_{\rm inj} \leq r_{\rm seg,max}$, where $r_{\rm seg,max}$ is the maximum radial extent of the central galaxy segmentation region measured from the morphological centroid. The clump luminosity is assigned by sampling the luminosity ratio log-uniformly over the range $10^{-3} \leq L_{\rm clump}/L_{\rm gal} \leq 0.75$, from which the clump luminosity is calculated by inputting the galaxy luminosity.

For each injection trial we save the injected position and radius (normalised by $r_{\rm seg,max}$), the injected luminosity ratio, the recovery flag, the distance to the closest detected clump, and the measured properties of the matched detection. The completeness is given by
\begin{equation}
    C = \frac{N_{\rm rec}}{N_{\rm inj}},
\end{equation}
where $N_{\rm inj}$ is the number of successful injection trials in a given bin and $N_{\rm rec}$ is the number of recovered clumps. We then compute completeness curves both as a function of clump luminosity, $L_{\rm clump}/L_{\rm galaxy}$, and as a function of normalised galactocentric radius, $r_{\rm clump}/r_{\rm seg,max}$. Uncertainties on the completeness values are computed from the 68 per cent credible interval of the beta posterior distribution, assuming a non-informative Jeffreys prior for the binomial recovery fraction \citep{Cameron_2011}. We fit the completeness as a function of clump luminosity, marginalised over galactocentric radius, with a sigmoid model \citep{Sok_2026a},
\begin{equation}
    C(x) =
    \frac{1}{1+\exp[-k(\log_{10}x - x_0)]},
    \qquad
    x = \frac{L_{\rm clump}}{L_{\rm gal}},
\end{equation}
where $x_0$ gives the logarithmic luminosity ratio at 50 per cent completeness and $k$ controls the steepness of the transition. From these fits we derive the 50 per cent and 80 per cent completeness limits, denoted by $L_{50}$ and $L_{80}$, where
\begin{equation}
    L_C = 10^{x_0 + k^{-1}\ln[C/(1-C)]}.
\end{equation}
These limits provide an empirical estimate of the clump luminosity above which our detections are complete at a given level, and are later used to define completeness-limited clumpy fractions.

The corresponding one-dimensional completeness curves are shown in Appendix~\ref{app:completeness_assessment}. Figure~\ref{fig:flux_completeness} shows the completeness as a function of clump luminosity fraction for the rest-frame blue wavelength. As expected, the recovery fraction increases strongly with $L_{\rm clump}/L_{\rm galaxy}$, while the limiting luminosity ratio becomes larger at higher redshift and in lower-mass galaxies. We list the 50 per cent completeness limits for the clump luminosity fractions in Table~\ref{tab:lclump_lgal_50_limits}. Figure~\ref{fig:radial_completeness} shows the corresponding radial completeness curves. The completeness shows a much weaker dependence on the galactocentric radius than on the luminosity ratio. Although the completeness is reduced at small radii where injected clumps are more likely to be blended with the galaxy centre or bulge (mostly at $r_{\rm clump}/r_{\rm seg,max} \lesssim 0.25$), it stays relatively flat at larger radii for all redshifts and stellar masses. In addition to the one-dimensional completeness curves, we construct two-dimensional completeness maps in the plane of clump luminosity ratio and galactocentric radius. These maps quantify the joint dependence of the recovery fraction on both parameters and are used to derive luminosity ratio limits at fixed radius. Figure~\ref{fig:2d_completeness} shows the two-dimensional completeness surfaces in the $L_{\rm clump}/L_{\rm galaxy}$ and $r_{\rm clump}/r_{\rm seg,max}$ plane. These maps confirm that clump recovery is primarily driven by clump luminosity, with a secondary dependence on radius, and that the completeness decreases towards higher redshift.

\section{Fraction of Clumpy Galaxies}
\label{sec:Fraction of Clumpy Galaxies}

In this section we present our results on the fraction of clumpy galaxies over the studied redshift range, which is the main result of this paper. We start by defining the clumpy fraction and presenting the fiducial results in Section~\ref{sec:Clumpy fraction}. We then use the completeness estimates from Section~\ref{sec:Completeness assessment} to construct a probability-weighted statistic for comparison in Section~\ref{sec:Completeness correction}. Furthermore, we discuss the redshift and stellar mass dependence of the clumpy fractions in Section~\ref{sec:Redshift evolution and stellar mass dependence}, and compare them to results from the literature in Section~\ref{sec:Comparison to literature}.

\subsection{Clumpy fraction}
\label{sec:Clumpy fraction}

We now study the fraction of clumpy galaxies ($f_{\rm clumpy}$) as a function of redshift at UV, blue, and optical rest-frame wavelengths. The fraction of clumpy galaxies is defined as the ratio between the number of galaxies hosting off-centre clumps and the total number of galaxies in a given redshift, stellar mass, and rest-frame wavelength bin. 

Clumps are identified using the \texttt{astrodendro} package as presented in Section~\ref{sec:Clump detection algorithm}. Clumpy galaxies are defined as galaxies hosting at least one off-centre clump identified by the algorithm, that satisfies a certain luminosity fraction threshold compared to the host galaxy. As discussed above, these threshold choices vary between studies, and heavily affect the resulting clumpy fraction. For example, galaxies are defined as clumpy if their clump contains at least 8 per cent \citep[e.g.,][]{Guo_2015} or 5 per cent \citep[e.g.,][]{Wuyts_2012} of the rest-frame UV luminosity of the host galaxy. Our threshold choice of 0.1 per cent is very permissive and we need to make additional cuts in order to define a physically meaningful clumpy fraction.

For our fiducial clumpy fraction measurement, we follow the general approach of \citet{de_la_Vega_2026a}. Rather than choosing a single global value for the fractional luminosity threshold, we instead define a bin-dependent threshold corresponding to the 50 per cent completeness limit derived from the clump injection--recovery simulations described in Section~\ref{sec:Completeness assessment}. This is done independently in each redshift, stellar mass, and rest-frame wavelength bin. We denote this threshold as $L_{50}(z, M_\star, \lambda_{\rm rest})$, and Table~\ref{tab:lclump_lgal_50_limits} contains the values for each respective bin. A galaxy is then classified as clumpy in the completeness-limited sample if it hosts at least one off-centre clump satisfying the luminosity fraction criterion that
\begin{equation}
    \frac{L_{\rm clump}}{L_{\rm galaxy}} > L_{50}(z, M_\star, \lambda_{\rm rest}).
\end{equation}
The corresponding clumpy fraction is then defined as
\begin{equation}
    f_{\rm clumpy}^{50} = \frac{N_{\rm clumpy}^{50}}{N_{\rm gal}},
\end{equation}
where $N_{\rm clumpy}^{50}$ is the number of galaxies hosting at least one off-centre clump above the 50 per cent completeness limit, and $N_{\rm gal}$ is the total number of galaxies in the same bin.

No completeness correction is applied in this fiducial measurement. Instead, the purpose of this approach is to restrict the measurement to a regime where the injection--recovery simulations indicate that clump recovery is at least 50 per cent complete in luminosity. We therefore interpret $f_{\rm clumpy}^{50}$ as a conservative, completeness-limited observed clumpy fraction. It gives an empirically motivated reliable measurement, which might underestimate the underlying intrinsic clumpy fractions. The statistical uncertainties are estimated from the 68 per cent CI of the binomial distribution, following the same Bayesian approach used for the completeness estimates \citep{Cameron_2011}.

\subsection{Completeness correction}
\label{sec:Completeness correction}

Although our fiducial clumpy fraction measurement is based on a 50 per cent completeness-limited clump sample, we also compute a probability-weighted completeness-corrected statistic based on detected clumps, as a complementary comparison to our fiducial results. 

Several previous studies applied completeness corrections to their reported clumpy fractions. For example, \citet{Guo_2015} applies a correction globally to each bin, rather than at the galaxy level, using
\begin{equation}
    f_{\rm clumpy}^{\rm corr}=f_{\rm clumpy}^{\rm obs}+\frac{1}{n_c}\left(\frac{1}{X}-1\right) f_{\rm clumpy}^{\rm obs}-\frac{1}{n_c}\left(\frac{1}{X}-1\right)\left(f_{\rm clumpy}^{\rm obs}\right)^2,
    \label{eq:Guo_2015_correction}
\end{equation}
where $X$ is the completeness and $n_c$ is the average number of clumps per galaxy in the bin. The second term accounts for additional clumpy galaxies expected from undetected clumps, while the third prevents double-counting missed clumps in galaxies already identified as clumpy. Previous studies \citep[e.g.][]{Shibuya_2016, Guo_2018} often assume an intrinsic average of $n_c = 2$. However, this approach requires a global luminosity fraction threshold, which can substantially affect the inferred clumpy fraction, as discussed above.

We therefore adopt an alternative correction that does not require a global luminosity fraction threshold and instead applies inverse-completeness weights at the clump level before constructing a galaxy-level statistic. This correction uses the two-dimensional completeness maps derived in Section~\ref{sec:Completeness assessment}, which quantify the recovery probability as a function of clump luminosity fraction and normalised galactocentric radius,
\begin{equation}
    C = C\left(L_{\rm clump}/L_{\rm galaxy}, r_{\rm clump}/r_{\rm seg,max}\right).
\end{equation}
The correction is applied separately in each redshift, stellar mass, and rest-frame wavelength bin, using the corresponding completeness map. For each detected off-centre clump $j$ in galaxy $i$, we assign a weight that is inversely proportional to the completeness
\begin{equation}
    w_{ij} = \frac{1}{C(\ell_{ij}, r_{ij})},
\end{equation}
with the luminosity ratio and galactocentric radius of clump \textit{j} being
\begin{equation}
    \ell_{ij} = \frac{L_{\rm clump}^{ij}}{L_{\rm galaxy}^{i}},
    \qquad
    r_{ij} = \frac{r_{\rm clump}^{ij}}{r_{\rm seg,max}^{i}}.
\end{equation}
Only off-centre clumps are included in the correction. Unlike the 50 per cent completeness-limited analysis described above, this probability-weighted correction includes clumps down to $C=0.2$, provided that their completeness estimates are sufficiently reliable. We therefore only use clumps falling in completeness-map cells with $C \geq 0.2$ and with at least five injected clumps contributing to the cell. Clumps outside these valid completeness regions are excluded.

We first compute the completeness-corrected \textit{clump incidence} for each galaxy \textit{i}, defined as
\begin{equation}
    \lambda_i = \sum_{j} \frac{1}{C(\ell_{ij}, r_{ij})},
    \label{eq:lambda_route2}
\end{equation}
where the sum is over detected off-centre clumps \textit{j} in galaxy $i$ that pass the completeness quality cuts. Galaxies with no detected valid clumps have $\lambda_i=0$. This quantity represents the inverse-completeness-weighted clump incidence from valid detections, rather than a complete reconstruction of the intrinsic clump population. However, because the clumpy fraction is binary at the galaxy level, a direct average of $\lambda_i$ would estimate the mean number of clumps per galaxy, rather than the fraction hosting at least one clump.

Since a clumpy galaxy is defined as hosting at least one off-centre clump, we convert the weighted clump incidence into a galaxy-level statistic by assuming Poisson counting statistics. Treating $\lambda_i$ as an effective Poisson mean, we define the corresponding probability weight for galaxy $i$ as
\begin{equation}
    p_i = 1 - e^{-\lambda_i}.
    \label{eq:route2_probability}
\end{equation}
The completeness-corrected clumpy fraction is then defined as the average of these probability weights over all galaxies in the bin,
\begin{equation}
    f_{\rm clumpy}^{\rm corr} = \frac{1}{N_{\rm gal}} \sum_{i=1}^{N_{\rm gal}} p_i
    = \frac{1}{N_{\rm gal}} \sum_{i=1}^{N_{\rm gal}} \left[1-e^{-\lambda_i}\right],
    \label{eq:route2_fraction}
\end{equation}
where all galaxies in the bin contribute to the denominator. Thus, galaxies with no detected valid clumps contribute $p_i=0$, rather than being excluded from the calculation. Uncertainties on $f_{\rm clumpy}^{\rm corr}$ are estimated by bootstrapping the galaxies in each bin and recomputing the mean of the $p_i$ distribution. We use the scatter of the bootstrap realisations as the uncertainty on the corrected fraction. As a consistency check, we also compute a simple binomial-like uncertainty, although the bootstrap uncertainty is preferred because it operates directly on the probability-weighted galaxy sample.

This estimator has several useful properties, but also some limitations. First, it is applied at the galaxy level rather than the global bin level and is bounded between 0 and 1 by construction, unlike a completeness-corrected mean clump abundance. Second, it incorporates the luminosity- and radius-dependent completeness of each detected clump. Third, the method avoids imposing a single hard luminosity-fraction threshold on the detected clumps. However, this measurement should not be interpreted as a direct completeness-corrected version of a binary threshold-based clumpy fraction. The inverse-completeness weights increase the contribution of detected clumps in regions where the recovery probability is low, but only for clumps that fall within the reliable completeness regime. Crucially, the method cannot recover galaxies whose clumps are all missed, since these contribute \(\lambda_i=0\) and hence \(p_i=0\), even if they intrinsically host clumps. Moreover, the conversion from weighted clump incidence to a probability weight assumes Poisson counting statistics, so a galaxy with a single detected clump contributes \(p_i=1-e^{-1/C}\) rather than being assigned a binary value of unity. We therefore interpret this quantity as a probability-weighted completeness-corrected statistic based on detected clumps, rather than an unbiased reconstruction of the intrinsic clumpy fraction. Its purpose is to provide a complementary comparison of the observed evolutionary trends under a different treatment of detected clumps, rather than an independent validation of the fiducial results.

\begin{figure}
 \includegraphics[width=\columnwidth]{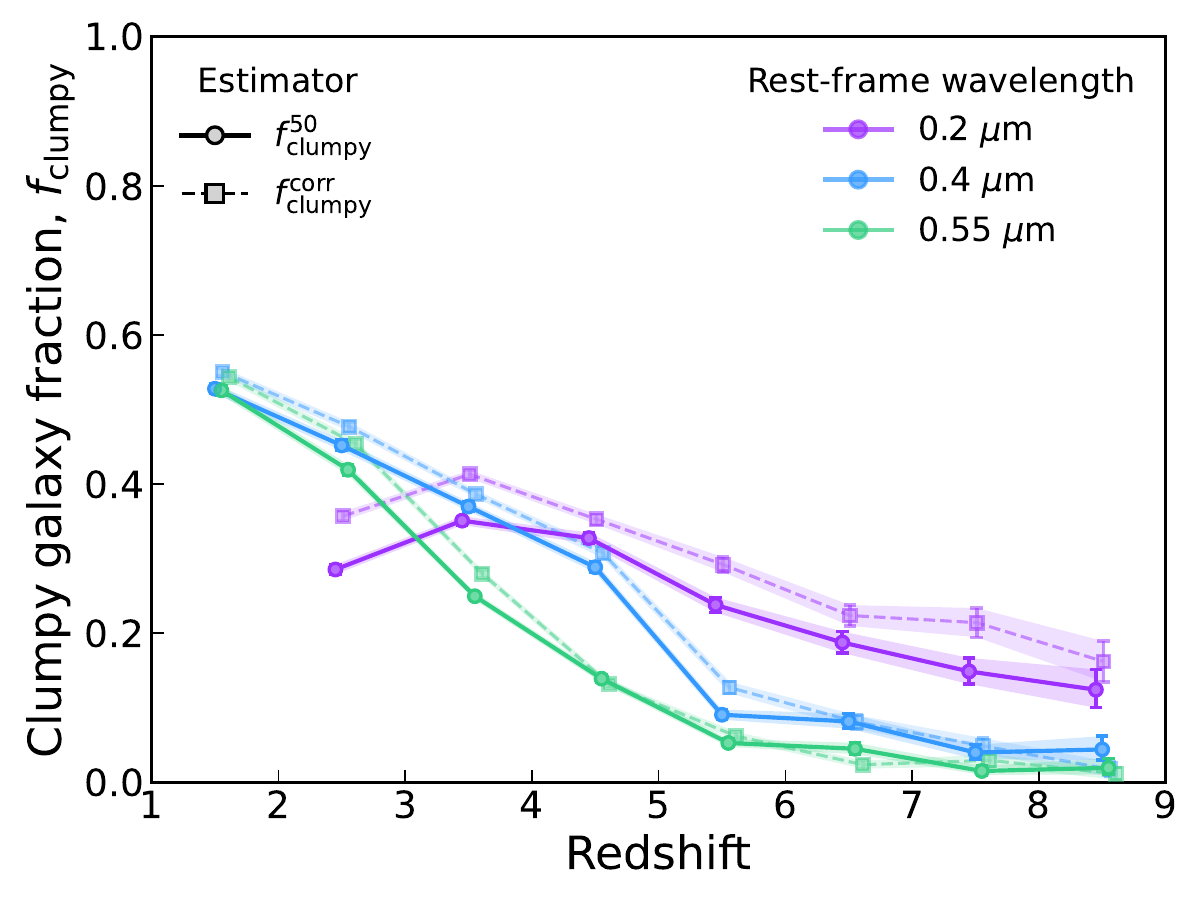}
 \caption{The fraction of clumpy galaxies for the full sample at different rest-frame wavelengths as a function of redshift. Results from the two methods are displayed, where circle markers and solid lines show fractions calculated using the fiducial 50 per cent complete sample ($f_{\rm clumpy}^{50}$), while square markers and dashed lines show results for the probability-weighted completeness-corrected statistic ($f_{\rm clumpy}^{\rm corr}$). Overall the two methods yield similar results and trends qualitatively, with the largest quantitative difference being at the rest-frame UV clumpy fractions, which are most affected by incompleteness. The $z<2$ UV points are excluded due to insufficient rest-frame coverage.}
 \label{fig:clumpy_fraction_comp50_weights_combined}
\end{figure}

Figure~\ref{fig:clumpy_fraction_comp50_weights_combined} compares the fiducial 50 per cent completeness-limited clumpy fractions with the probability-weighted completeness-corrected values. Overall, the two approaches give similar qualitative trends with redshift and rest-frame wavelength. This qualitative agreement suggests that the main evolutionary trends persist under an alternative weighting of detected clumps, within the limitations of the correction discussed above. The largest quantitative differences are seen at rest-frame UV wavelengths, where clump detectability is most sensitive to the luminosity fraction threshold and inverse-completeness weighting has the strongest effect. The rest-frame UV data points at $z=1-2$ are excluded from the figure and subsequent analysis due to insufficient rest-frame UV wavelength coverage (see Section~\ref{sec: Sample selection}), as discussed in the following section. The values of the completeness-limited and completeness-corrected clumpy fractions, together with their uncertainties, are tabulated in Appendix~\ref{app:clumpy_fraction_values}.

\subsection{Redshift evolution and stellar mass dependence}
\label{sec:Redshift evolution and stellar mass dependence}

\begin{figure*}
 \includegraphics[width=\linewidth]{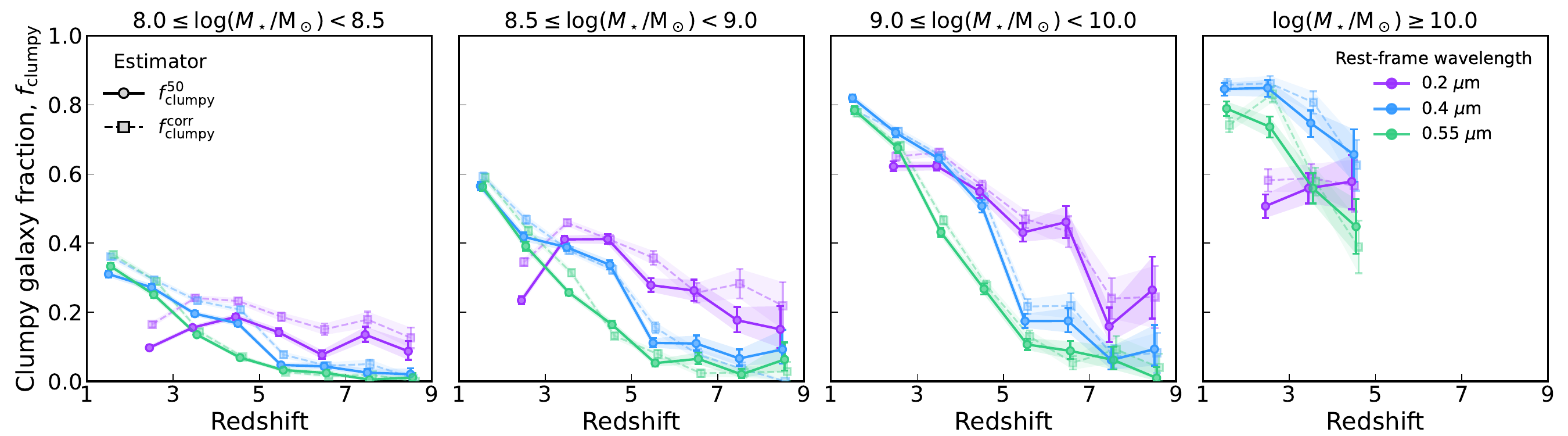}
 \caption{Redshift evolution of clumpy fractions in four stellar mass bins. Circle markers and solid lines denote the fiducial completeness-limited fractions, while square markers and dashed lines show the completeness-corrected results. In general, we find decreasing clumpy fractions with increasing redshift at all stellar masses. Furthermore, we find higher clumpy fractions for higher stellar masses at fixed redshift. In the highest stellar mass bin at $\log_{10}(M_\star/{\rm M_\odot})\geq10$, we do not display results at $z>5$ due to the small number of galaxies. The rest-frame UV measurements at $z=1-2$ are also excluded due to insufficient F070W wavelength coverage (Section~\ref{sec: Sample selection}). The two clumpy fraction estimators ($f_{\rm clumpy}^{50}$ and $f_{\rm clumpy}^{\rm corr}$) yield broadly similar results and consistent evolutionary trends.}
 \label{fig:clumpy_fraction_comp50_weights}
\end{figure*}

Here we discuss the redshift evolution and stellar mass dependence of our derived clumpy fractions. Figure~\ref{fig:clumpy_fraction_comp50_weights_combined} shows the redshift evolution and comparison between the fiducial 50 per cent complete and the completeness corrected clumpy fraction measurements. These fractions are combined for all stellar mass bins, but are kept separately for the three rest-frame wavelengths. 

The fractions generally decline from $z\approx1$ to $z\approx9$, with a clear dependence on rest-frame wavelength. In the case of the fiducial results, the fractions measured at blue and optical rest-frame wavelengths are the highest at $\approx \! 55$ per cent in the $z=1-2$ redshift bin and decline to $\approx \! 5$ per cent at $z=8-9$. In the case of the rest-frame UV clumpy fractions, the evolution is not monotonic, with a turnover at $z\approx3-4$, where the fraction reaches $\approx \! 35$ per cent. At lower redshifts, the UV clumpy fraction is below the blue and optical clumpy fractions. However, this scaling reverses beyond $z > 3$, and the fractions decrease with increasing rest-frame wavelength at a given redshift bin. The rest-frame UV fraction stays relatively high at $\approx \! 12$ per cent even in the highest redshift bin at $z=8-9$, above the blue and optical fractions. This suggests that compact, UV-bright star-forming structures remain relatively common even in the earliest galaxies probed by our sample, despite the overall decline in clumpy fractions towards higher redshifts.

The interpretation of the rest-frame UV clumpy fractions at lower redshifts is complicated by the limited wavelength coverage of the available NIRCam filters. As shown in Figure~\ref{fig:filter_selection}, at $z<3$ the rest-frame UV is probed only by F070W, which covers a smaller fraction of the JADES footprint. Moreover, at $z<2$, F070W no longer reaches the target wavelength of $\lambda_{\rm rest}=0.2\,\mu{\rm m}$, and we therefore exclude the rest-frame UV measurements in the $z=1-2$ redshift bin from our analysis. The requirement for each galaxy to have coverage at all three target wavelengths (see Section~\ref{sec:Clump detection algorithm}) may introduce additional selection effects, potentially contributing to the different behaviour of the UV clumpy fractions at lower redshifts.

The wavelength dependence at higher redshifts is expected, as the imaging resolution is higher in the rest-frame UV than at optical wavelengths, allowing more structure to be resolved and identified. Moreover, clumps associated with young star-forming regions are expected to appear more prominently in the rest-frame UV than in the optical, which more closely traces the underlying stellar population \citep{Wang_2026}. At lower redshifts, the relatively lower UV clumpy fractions may partly result from the stronger attenuation of far-UV emission by dust, which can obscure young star-forming clumps \citep{Calzetti_2000, Salim_2020}. The higher optical clumpy fractions may also reflect a contribution from older stellar populations, including more evolved structures or accreted stellar components associated with mergers \citep{Wuyts_2012, Cibinel_2015, Kalita_2025b}.

The completeness-corrected clumpy fractions broadly agree with the 50 per cent completeness-limited measurements, but are systematically higher in several bins. The largest differences occur at rest-frame UV wavelengths, where the corrected fractions are higher by approximately 5 percentage points on average. This is expected, since the values in Table~\ref{tab:lclump_lgal_50_limits} indicate that the UV clump identification is the least complete, especially at lower stellar masses of $\log_{10}(M_\star / {\rm M_\odot})< 9$, where we expect a much larger number of galaxies (see the stellar mass distribution in Figure~\ref{fig:stellar_mass_redshift}). Therefore, the rest-frame UV clumpy fractions get the highest corrections on average according to the completeness correction method presented in Section~\ref{sec:Completeness correction}. Nevertheless, this comparison shows that the broad evolutionary trends persist when applying an alternative completeness weighting to the detected clumps.

\begin{figure}
 \includegraphics[width=\columnwidth]{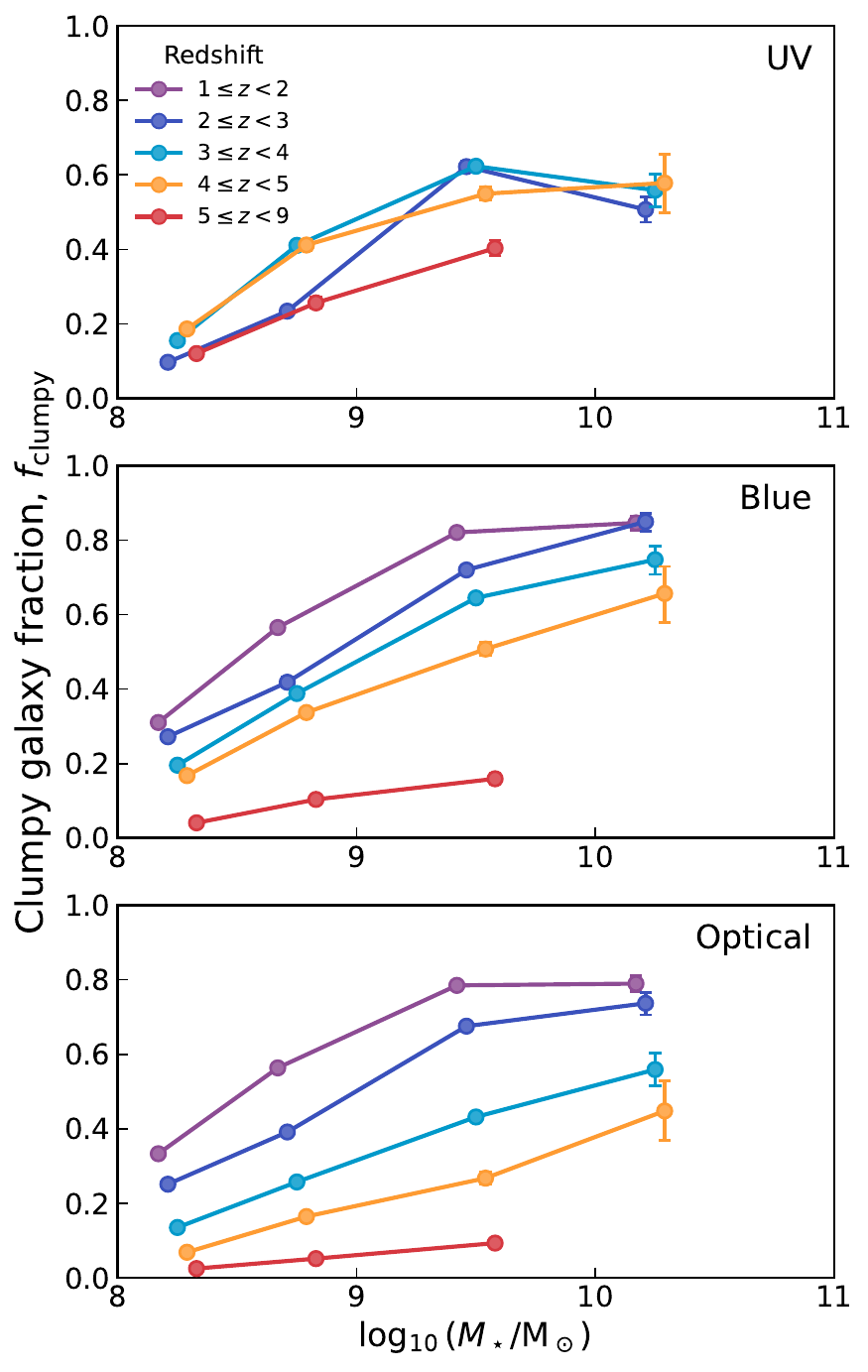}
 \caption{Completeness-limited clumpy fractions as a function of stellar mass at three rest-frame wavelengths, colour-coded by redshift. Fractions generally increase with stellar mass and decline with redshift, particularly in the blue and optical, while UV fractions flatten at the highest masses. The highest-redshift bin combines $z=5-9$. UV measurements at $z=1-2$ and highest-mass measurements at $z\geq5$ are excluded, as discussed in the text.}
 \label{fig:f_clumpy_stellar_mass}
\end{figure}

\begin{figure*}
 \includegraphics[width=0.8\linewidth]{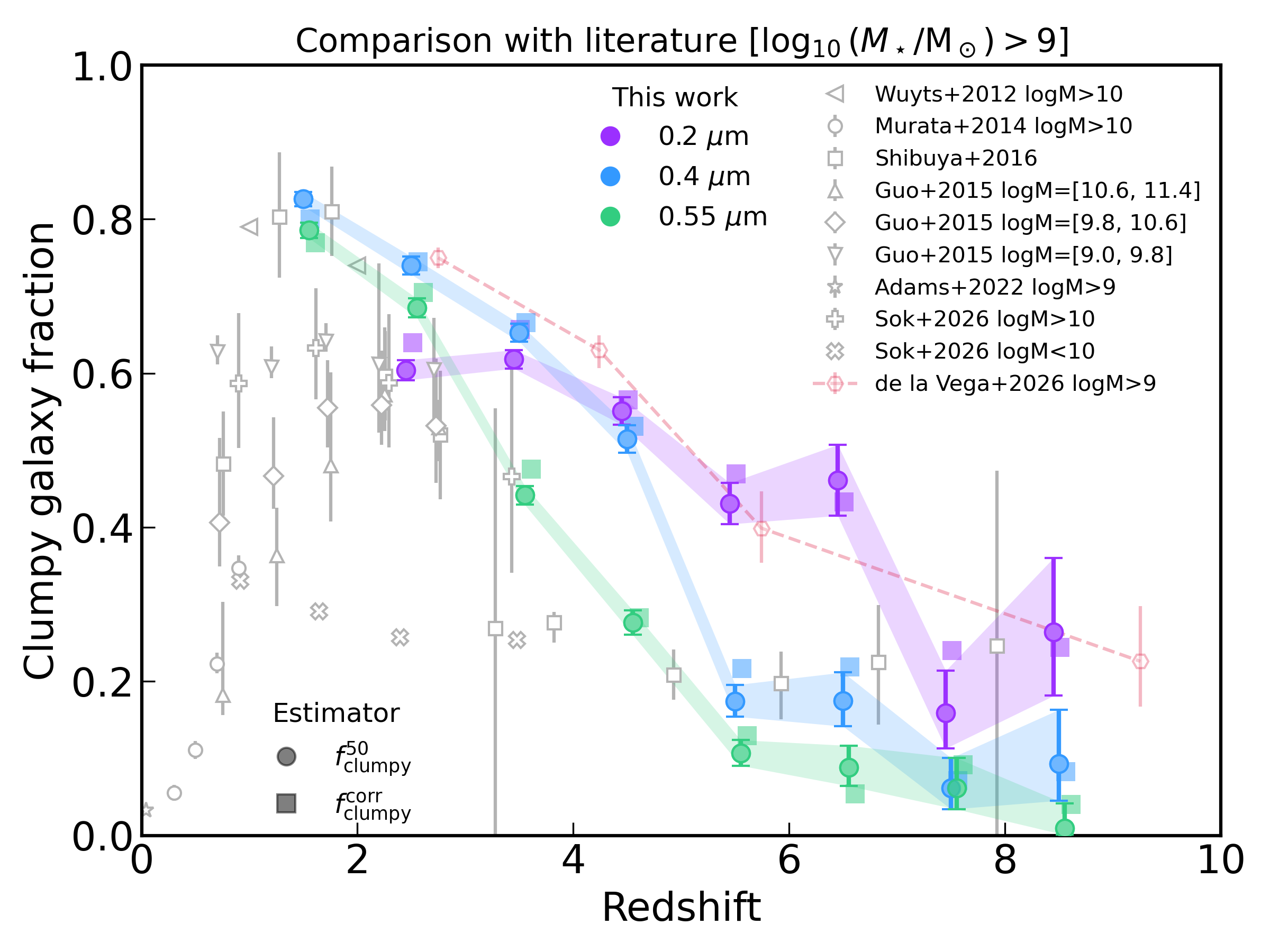}
 \caption{Fraction of clumpy galaxies as a function of redshift displaying results from this study and other works from the literature. We only include galaxies with $\log_{10}(M_\star/{\rm M_\odot})>9$ for our clumpy fraction calculations to make it easier to compare with other literature results, in particular with the fractions of \citet{de_la_Vega_2026a}, who employ a similar methodology for \textit{JWST} data at the same stellar mass threshold for their sample. We denote our fractions for the 50 per cent complete sample with filled circles and the completeness corrected fractions by the filled squares. We find that our results are in good overall agreement with the literature values, in particular our rest-frame UV ($\lambda_{\rm rest}=0.2~{\rm \mu m}$) fractions and the clumpy fractions from \citet{de_la_Vega_2026a} at $\lambda_{\rm rest}=0.23~{\rm \mu m}$.}
 \label{fig:clumpy_fraction_comp50_weights_literature}
\end{figure*}

If we now split the clumpy fractions by stellar mass as well, we obtain the trends displayed in the four panels of Figure~\ref{fig:clumpy_fraction_comp50_weights}. As before, we plot both the fiducial 50 per cent completeness-limited and the completeness-corrected clumpy fractions for comparison. In the case of the highest stellar mass bin at $\log_{10}(M_\star / {\rm M_\odot})\geq 10$, we only measure the clumpy fractions for $z\approx1-5$, as above this redshift range the number of galaxies in our catalogue declines steeply, preventing reliable number statistics. There is a clear and strong scaling of clumpy fractions with stellar mass, with the highest fractions generally measured for the most massive galaxies. In the $z=1-2$ bin, the blue and optical $f_{\rm clumpy}$ decrease from almost $\approx \! 85$ per cent at $\log_{10}(M_\star / {\rm M_\odot})\geq 10$, to $\approx \! 80$ per cent at $9\leq\log_{10}(M_\star / {\rm M_\odot})<10$, $\approx \! 60$ per cent at $8.5\leq\log_{10}(M_\star / {\rm M_\odot})<9$, and $\approx \! 35$ per cent at $8\leq\log_{10}(M_\star / {\rm M_\odot})<8.5$. The rest-frame wavelength dependence remains broadly similar to the combined results, with the UV fractions generally being the highest, followed by the blue and optical fractions, at $z\gtrsim4$ for $\log_{10}(M_\star / {\rm M_\odot})<10$. In the highest stellar mass bin, however, the rest-frame blue fractions remain the highest across the redshift range probed, while the UV and optical fractions are comparable at $z\approx3-4$. Interestingly, the turnover in the rest-frame UV clumpy fractions around $z\approx3-4$ also shows some stellar mass dependence, being more pronounced at lower and intermediate masses.

To further illustrate the stellar mass dependence, Figure~\ref{fig:f_clumpy_stellar_mass} shows the fiducial completeness-limited clumpy fractions as a function of stellar mass in five redshift bins, combining the measurements at $z=5-9$ to improve the number statistics. The blue and optical fractions show a clear increase with stellar mass at all redshifts, while the rest-frame UV fractions exhibit a flattening or slight turnover at the highest stellar masses, particularly at $z\approx2-4$. At $z=5-9$, the UV clumpy fractions remain substantially higher than the blue and optical fractions across all three stellar mass bins, indicating that the observed wavelength dependence persists when accounting for stellar mass. The completeness-corrected results in Figure~\ref{fig:clumpy_fraction_comp50_weights} follow similar trends and broadly agree with the fiducial results, as in the case of the combined fractions. The magnitude of the corrections varies across stellar mass and redshift but is generally largest in the rest-frame UV, as expected from the completeness assessment.

\subsection{Comparison to literature}
\label{sec:Comparison to literature}

We compare our $f_{\rm clumpy}$ measurements to results from several other works in the literature. A perfect quantitative comparison is not possible, as different studies measure clumpy fractions using different methods, rest-frame wavelengths, clump definitions, completeness treatments, and stellar mass selections. Therefore, we limit our clumpy fraction measurements to $\log_{10}(M_\star / {\rm M_\odot})\geq 9$ to make them more comparable with the literature results. The comparison is therefore primarily qualitative, focusing on the overall trends with redshift.

We present our $f_{\rm clumpy}$ measurements and literature comparison in Figure~\ref{fig:clumpy_fraction_comp50_weights_literature}. Overall, our measurements are broadly consistent with the range of values reported in previous studies, although they are generally higher than many earlier \textit{HST}-based measurements. At $z\approx1-3$, our clumpy fractions are high, with values of $f_{\rm clumpy}\approx0.6-0.8$ depending on rest-frame wavelength and estimator (50 per cent complete or completeness corrected). 

These values exceed or overlap with the upper range of previous \textit{HST} measurements from \citet{Wuyts_2012}, \citet{Guo_2015}, \citet{Shibuya_2016}, and more recent \textit{JWST} measurements by \citep{Sok_2026a}, although the scatter between literature values is substantial. Part of the scatter is expected from the different stellar mass limits and clump selection criteria used in these studies. As shown in Figure~\ref{fig:f_clumpy_stellar_mass}, the clumpy fraction depends strongly on stellar mass, and therefore measurements using different mass selections should not be expected to agree exactly. Interestingly, some previous \textit{HST} studies find decreasing UV clumpy fractions with increasing stellar mass \citep[e.g.,][]{Guo_2015, Sattari_2023}, in contrast to the predominantly positive mass dependence found here. This further highlights the difficulty of directly comparing results obtained using different observational samples and detection methods.

The generally higher fractions measured with \textit{JWST} may partly result from its improved sensitivity, allowing fainter and lower-contrast structures to be identified that were previously undetected in \textit{HST} imaging. This interpretation is supported by \citet{de_la_Vega_2026a}, who directly compare clump detections in \textit{HST} and \textit{JWST} images and find that the increased sensitivity to fainter clumps has a stronger effect on the recovered clumpy fractions than differences in spatial resolution alone. Nevertheless, differences in the adopted relative clump luminosity thresholds and galaxy selections may also contribute to the observed discrepancies.

At lower redshifts, the measurements from \citet{Murata_2014} and \citet{Adams_2022} lie below the values found at $z\gtrsim1$, following the general picture that $f_{\rm clumpy}$ increases from the local Universe towards cosmic noon. Our lowest-redshift bin is therefore consistent with an extension of this rising trend towards $z\approx1-2$, where clumpy morphologies become substantially more common. We note, however, that the comparison at these redshifts is especially sensitive to sample selection and methodology. For example, \citet{Murata_2014} require at least three distinct components to classify a galaxy as clumpy, compared to our criterion of at least one off-centre clump.

At $z>3$, we find a qualitatively similar overall decline towards higher redshifts as \citet{Shibuya_2016}, although our measured $f_{\rm clumpy}$ are generally higher, particularly in the rest-frame UV. The most direct comparison at high redshift is with \citet{de_la_Vega_2026a}, who measure clumpy fractions using publicly available JADES DR3 data \citep{Rieke_2023, Bunker_2024, DEugenio_2025, Eisenstein_2025}, using a similar stellar mass limit of $\log_{10}(M_\star / {\rm M_\odot})>9$ and a rest-frame UV wavelength close to our $\lambda_{\rm rest}=0.2\,\mu{\rm m}$ measurement. We find good agreement between our rest-frame UV fractions and their measurements over the overlapping redshift range. Both studies find high clumpy fractions of order $\approx \! 60$--$75$ per cent at $z\approx3$, followed by a gradual decline towards higher redshift. Some quantitative differences remain, potentially reflecting differences in the sample selection and clump detection methods. Nevertheless, the consistent evolution is encouraging, suggesting that the high-redshift UV clumpy fraction trends are recovered by independent detection methods applied to JADES imaging.

The comparison also highlights the importance of rest-frame wavelength. While our rest-frame UV fractions remain relatively high out to the highest redshifts, the blue and optical fractions decline more rapidly beyond $z\gtrsim4$. This wavelength dependence is not directly captured in most literature measurements, which often use a single rest-frame band or a different wavelength selection. The agreement between the literature UV measurements and our $\lambda_{\rm rest}=0.2\,\mu{\rm m}$ fractions is consistent with the interpretation that rest-frame UV imaging preferentially traces young, star-forming clumps, while longer wavelengths are more sensitive to the underlying older stellar populations.

Finally, the agreement remains good between our completeness-limited and completeness-corrected $f_{\rm clumpy}$ in the literature comparison sample. The corrected values are generally higher, as expected, but they follow the same evolution and remain within the broad range spanned by previous work. This provides a complementary indication that the evolutionary trends are similar under different treatments of the detected clump population. Taken together, the comparison indicates that our measurements are consistent with previous results, while revealing generally higher $f_{\rm clumpy}$ than many \textit{HST}-based studies and extending the rest-frame wavelength-dependent view of clumpy galaxy evolution out to $z\approx9$ with \textit{JWST} observations.


\section{Contribution of major mergers to clumpy systems}
\label{sec:Contribution of major mergers to clumpy systems}

\begin{figure*}
 \includegraphics[width=\linewidth]{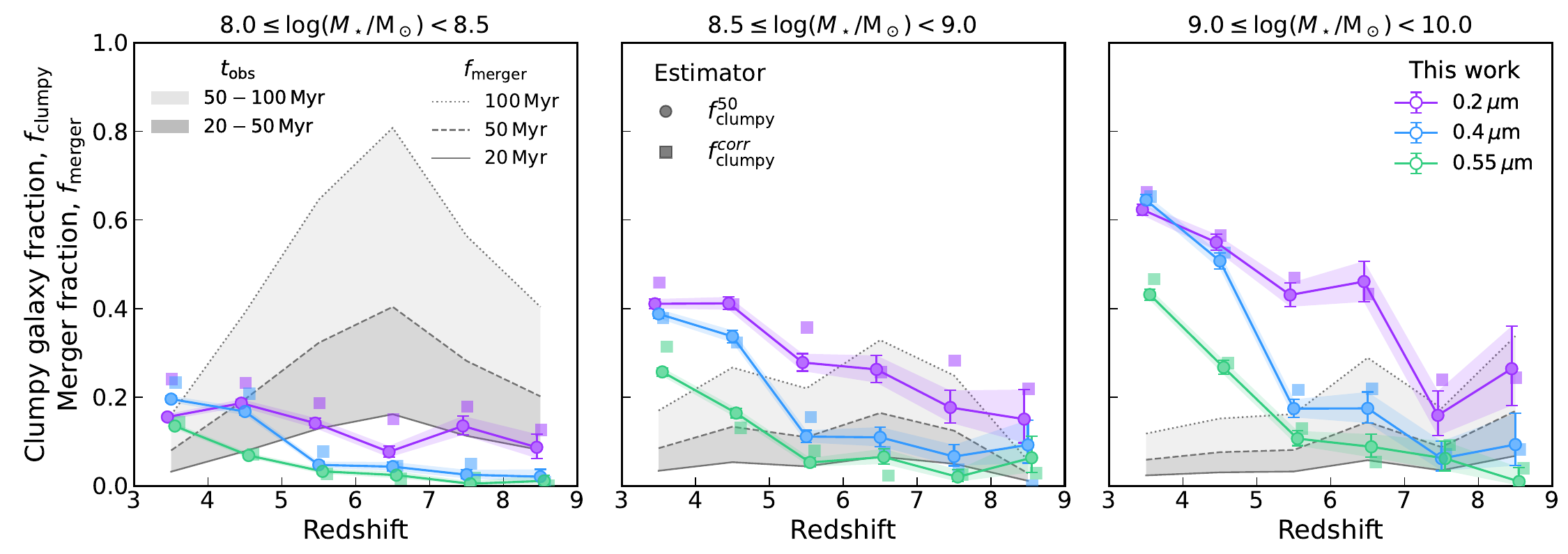}
 \caption{Comparison between the observed clumpy fractions and the potential contribution expected from major mergers in three stellar-mass bins. Circles and squares show the completeness-limited and completeness-corrected clumpy fractions, respectively, at the three rest-frame wavelengths considered. The major merger fractions are estimated from the merger rates of \citet{Puskas_2025}, assuming merger-induced clump observability timescales of $t_{\rm obs}=20$, 50, and 100 Myr. The darker and lighter grey bands indicate the corresponding 20--50 and 50--100 Myr ranges, respectively. These estimates represent model-dependent upper limits for the adopted timescales under the assumption that every major merger produces a detectable clumpy phase. The comparison suggests that major mergers could account for a substantial fraction of clumpy systems at low stellar masses and high redshift, whereas additional formation channels, including \textit{in situ} gravitational instabilities, are required where the observed clumpy fractions exceed the merger-driven expectation.}
 \label{fig:clumpy_fraction_merger_fraction}
\end{figure*}

Having measured the redshift evolution of the clumpy galaxy fraction, we now investigate whether the observed clumpy systems could plausibly be produced by major mergers. In this scenario, a galaxy may appear clumpy either because an accreted companion is identified as an off-centre clump, or because the interaction triggers disturbed, spatially concentrated star formation within the system. Major mergers are therefore a natural external mechanism for producing clumpy morphologies, especially at high redshift where gas fractions and merger rates are expected to be high \citep[e.g.,][]{Calabro_2019, Tacconi_2020}. Minor interactions may also trigger \textit{in situ} clump formation \citep{de_la_Vega_2026b}, and are not captured by the major merger rates considered here. However, clumps can also form internally through gravitational instabilities in gas-rich discs or feedback-regulated star formation. The comparison below is therefore not intended to uniquely identify the origin of individual clumps, but rather to estimate whether the observed abundance of clumpy galaxies is consistent with a major merger-driven origin.

We use the major merger rates measured by \citet{Puskas_2025}, which were derived for the same JADES fields over a similar redshift and stellar mass range. These merger rates are based on close-pair fractions for major companions with stellar mass ratios $\mu\geq1/4$ and projected separations of 5--30 pkpc. The close-pair fractions were converted into merger rates by assuming a merger observability timescale, such that the merger rate per galaxy is
\begin{equation}
    \mathcal{R}_{\rm M}(z,M_\star) = \frac{f_{\rm pair}(z,M_\star)}{\tau_{\rm pair}(z)},
\end{equation}
where $f_{\rm pair}$ is the close-pair fraction and $\tau_{\rm pair}$ is the observability timescale over which a merger is identified as a close pair \citep[see][]{Snyder_2017, Duncan_2019, Conselice_2022, Puskas_2025}. The resulting quantity, $\mathcal{R}_{\rm M}$, gives the average number of major mergers per galaxy per unit time. The best-fitting redshift evolution of the merger rate derived by \citet{Puskas_2025} is given by
\begin{equation}
    \mathcal{R}_{\rm M}(z) = 0.0249 \times (1+z)^{2.85} e^{-0.16(1+z)}\,{\rm Gyr}^{-1},
\end{equation}
for the full stellar mass range of $\log_{10}(M_\star/{\rm M}_\odot)=[8,10]$.

To estimate the fraction of galaxies that could be observed as clumpy due to major mergers, we multiply the merger rate by the timescale over which the merger-induced clumpy phase remains observable,
\begin{equation}
    f_{\rm merger}(z,M_\star,t_{\rm obs}) = \mathcal{R}_{\rm M}(z,M_\star) \times t_{\rm obs}.
    \label{eq:fmerge}
\end{equation}
Here $t_{\rm obs}$ is not the close-pair observability timescale used to derive the merger rate, but the duration for which the resulting clumpy morphology or merger-induced clump remains detectable in our images. This timescale is uncertain and depends on the physical nature of the clumps, the gas content of the galaxy, the merger stage, and the wavelength at which the system is observed, since rest-frame UV and optical/near-IR selections can trace different clump populations \citep[e.g.,][]{Kalita_2025b}. 

Simulations generally suggest that bright star-forming clumps can span a broad range of lifetimes. Low-mass or low-density clumps may be rapidly disrupted by stellar feedback on timescales of a few free-fall times, corresponding to only a few tens of Myr in some cases \citep[e.g.,][]{Mandelker_2017, Oklopcic_2017}. Similarly, merger-induced off-centre clumps can be transient, merging within a dynamical time of several tens of Myr \citep[e.g.,][]{Nakazato_2024}. Other simulations find dispersal timescales closer to $\approx 100$ Myr for UV-bright clumpy star-forming regions \citep[e.g.,][]{Buck_2017}, while more massive or dense clumps may survive for several hundred Myr and migrate within the disc \citep[e.g.,][]{Tamburello_2015, Mandelker_2017}. We therefore consider three representative values at $t_{\rm obs}=20$, 50, and 100 Myr. The uncertainty on the merger-driven fraction is propagated directly from the uncertainty on the merger rate, while treating $t_{\rm obs}$ as a fixed parameter.

This calculation should be interpreted as a model-dependent upper-limit estimate of the possible contribution of major mergers to $f_{\rm clumpy}$ for the adopted values of $t_{\rm obs}$. It assumes that every major merger produces an observable clumpy phase which lasts for exactly $t_{\rm obs}$. In reality, not all mergers will necessarily produce clumps that satisfy our detection criteria. Furthermore, the close-pair-selected merger rates trace galaxies before coalescence, whereas the clumpy phase may occur before, during, or after the final coalescence. Therefore, Equation~\ref{eq:fmerge} does not provide a one-to-one prediction for $f_{\rm clumpy}$, but rather an estimate of whether the merger rate is high enough to account for the observed abundance of clumpy galaxies.

\begin{figure}
 \includegraphics[width=\columnwidth]{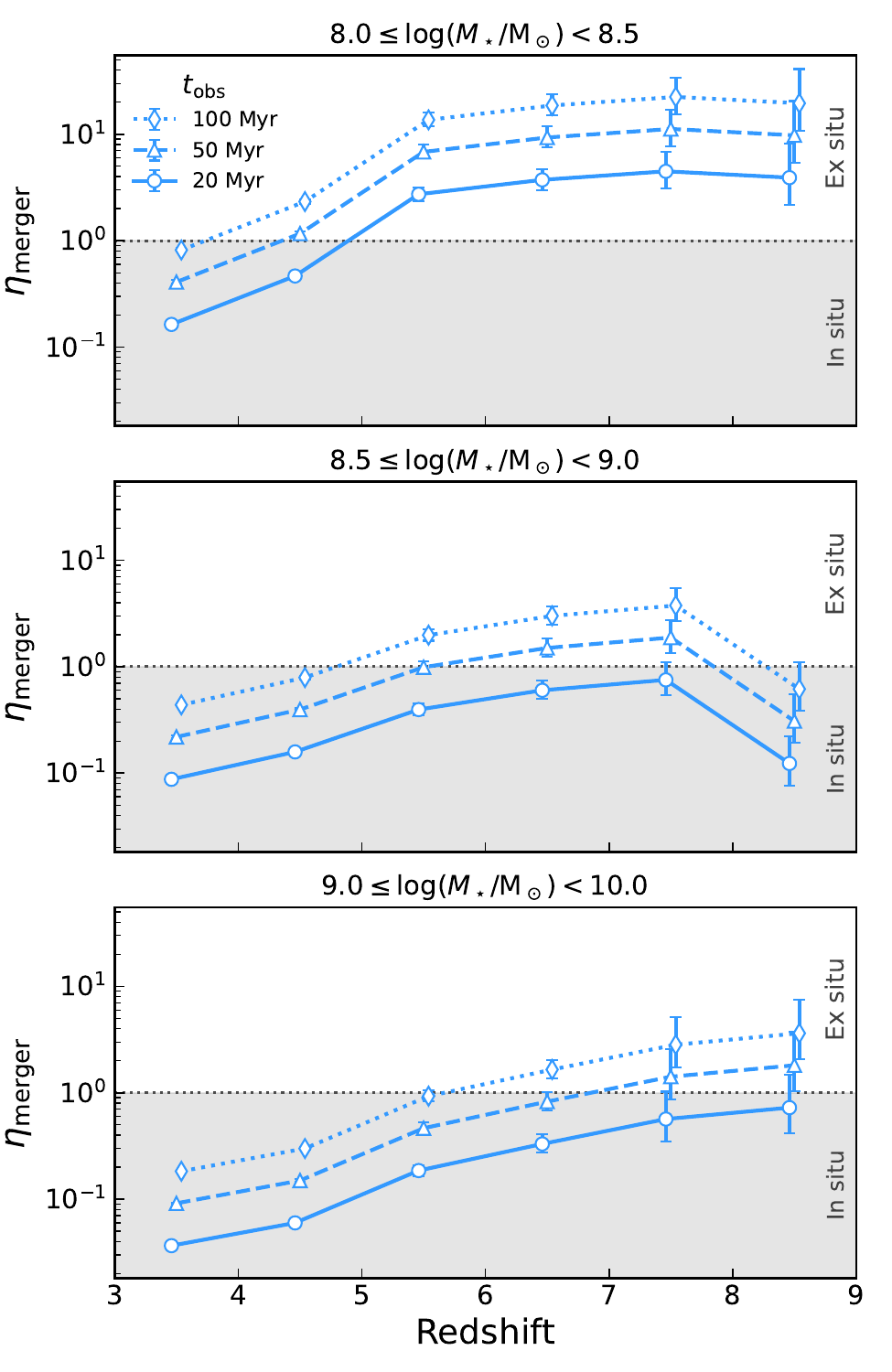}
 \caption{Redshift evolution of $\eta_{\rm merger}=f_{\rm merger}/f_{\rm clumpy}$ in three stellar mass bins, using the fiducial clumpy fractions at $\lambda_{\rm rest}=0.4\,\mu{\rm m}$. The inferred major merger fractions assume clump observability timescales of 20, 50, and 100 Myr, shown with different markers and line styles. The dotted line marks $\eta_{\rm merger}=1$, below which major mergers alone are insufficient to explain the observed clumpy fractions. The \textit{in situ} and \textit{ex situ} labels refer only to this major merger comparison. Error bars reflect uncertainties on $f_{\rm clumpy}$.}
 \label{fig:eta_merger}
\end{figure}

Figure~\ref{fig:clumpy_fraction_merger_fraction} compares the observed clumpy fractions with the merger-driven fractions inferred from Equation~\ref{eq:fmerge}, using the stellar-mass-dependent merger rates of \citet{Puskas_2025}. We restrict the comparison to $z\geq3$, where these merger rates were measured, and consider three stellar-mass bins: $8 \leq \log_{10}(M_\star/{\rm M_\odot}) < 8.5$, $8.5 \leq \log_{10}(M_\star/{\rm M_\odot}) < 9$, and $9 \leq \log_{10}(M_\star/{\rm M_\odot}) < 10$. We exclude the highest-mass bin, $\log_{10}(M_\star/{\rm M_\odot})\geq10$, because both the merger-rate and clumpy-fraction measurements become increasingly limited by small-number statistics at high redshift.

In the lowest stellar mass bin, $8 \leq \log_{10}(M_\star/{\rm M_\odot}) < 8.5$, the inferred merger-driven fraction rises rapidly with redshift, becoming comparable to or exceeding the observed clumpy fractions at $z\gtrsim5$--6, particularly for $t_{\rm obs}=50$--100 Myr. This is consistent with simulations showing that gravitational interactions can strongly disturb the morphologies of low-mass galaxies \citep[e.g.,][]{Martin_2021}. However, because $f_{\rm merger}$ is an upper-limit estimate, this agreement does not necessarily imply that the observed clumps are of merger origin, but rather that major mergers could plausibly be a dominant formation channel.

In the intermediate stellar mass bin, $8.5 \leq \log_{10}(M_\star/{\rm M_\odot}) < 9.0$, the inferred merger-driven fraction is generally lower than in the lowest-mass bin, reaching $\approx 0.3$ at $z\approx6$--7 for $t_{\rm obs}=100$ Myr. This is comparable to the UV $f_{\rm clumpy}$ and exceeds the blue and optical fractions, although shorter $t_{\rm obs}$ suggest that major mergers alone cannot explain the observed abundance of clumpy systems.

In the highest stellar mass bin considered here, $9 \leq \log_{10}(M_\star/{\rm M_\odot}) < 10$, the observed $f_{\rm clumpy}\approx0.3$--0.65 at $z\approx3-5$ exceed the merger-driven expectation of $\lesssim0.2$, even for $t_{\rm obs}=100$ Myr. This suggests that additional mechanisms, particularly \textit{in situ} clump formation through gravitational instabilities in gas-rich discs, play an important role in more massive galaxies. This interpretation is consistent with the higher incidence of disc-like morphologies and ordered rotation among more massive high-redshift galaxies \citep[e.g.,][]{Forster_Schreiber_2018, Wisnioski_2019, Ferreira_2023, Kartaltepe_2023, Danhaive_2025}, which provide favourable conditions for disc fragmentation. Major mergers may still contribute, particularly at high-$z$, and may indirectly enhance clump formation by driving gas inflows and triggering star formation.

To quantify the potential relative contribution of major mergers to the observed clumpy fractions, we define the ratio
\begin{equation}
    \eta_{\rm merger} = \frac{f_{\rm merger}}{f_{\rm clumpy}},
    \label{eq:eta_merger}
\end{equation}
where $f_{\rm clumpy}$ is the measured clumpy fraction in the same redshift and stellar mass bin. If $\eta_{\rm merger}\ll1$, the major merger rate is insufficient to explain the observed clumpy fraction, even under the optimistic assumption that every merger produces an observable clumpy phase. If $\eta_{\rm merger}\approx1$, major mergers could account for a substantial fraction of clumpy systems. If $\eta_{\rm merger}>1$, the major merger rate is high enough that all observed clumpy systems could in principle be merger related, but this does not imply that all mergers produce clumpy galaxies or that all clumpy galaxies are merger-driven.

Figure~\ref{fig:eta_merger} shows the redshift evolution of $\eta_{\rm merger}$ for the three stellar mass bins, using the fiducial clumpy fractions at $\lambda_{\rm rest}=0.4\,\mu{\rm m}$. Major mergers could account for a substantial fraction of low-mass clumpy systems at high redshift, particularly for longer $t_{\rm obs}$, but are generally insufficient at higher stellar masses. The \textit{in situ} and \textit{ex situ} labels in the figure refer only to whether major mergers could account for the observed clumpy fractions. Even where $\eta_{\rm merger}<1$, minor mergers, flybys, or interaction-induced star formation may contribute, but these are not accounted for in the merger rates of \citet{Puskas_2025} and remain beyond the scope of this comparison.

Overall, the comparison suggests a mass-dependent contribution of major mergers to the observed clumpy population. At low stellar masses and high redshifts, the major merger rate is sufficiently high that mergers could account for a large fraction of clumpy systems, although their actual contribution remains uncertain. At higher stellar masses, the observed clumpy fractions exceed the maximum merger-driven contribution for plausible clump observability timescales, suggesting that \textit{in situ} processes play an increasingly important role. This supports a picture in which clumpy morphologies at early times arise from a combination of external interactions and internal disc instabilities, with major mergers becoming less capable of explaining the observed clumpy population towards higher stellar mass.


\section{Clump properties}
\label{sec:Clump properties}

In this section, we investigate the physical properties and distributions of the clumps themselves. First, we focus on the clump luminosity function in bins of redshift, stellar mass, and rest-frame wavelength in Section~\ref{sec:Clump luminosity functions}. We then discuss the clump multiplicity of galaxies as a function of stellar mass and SFR in Section~\ref{sec:Clump multiplicity}. Finally, we present the size distribution of the detected clumps in Section~\ref{sec:Clump sizes}.

\subsection{Clump luminosity functions}
\label{sec:Clump luminosity functions}

\begin{figure*}
 \includegraphics[width=0.8\linewidth]{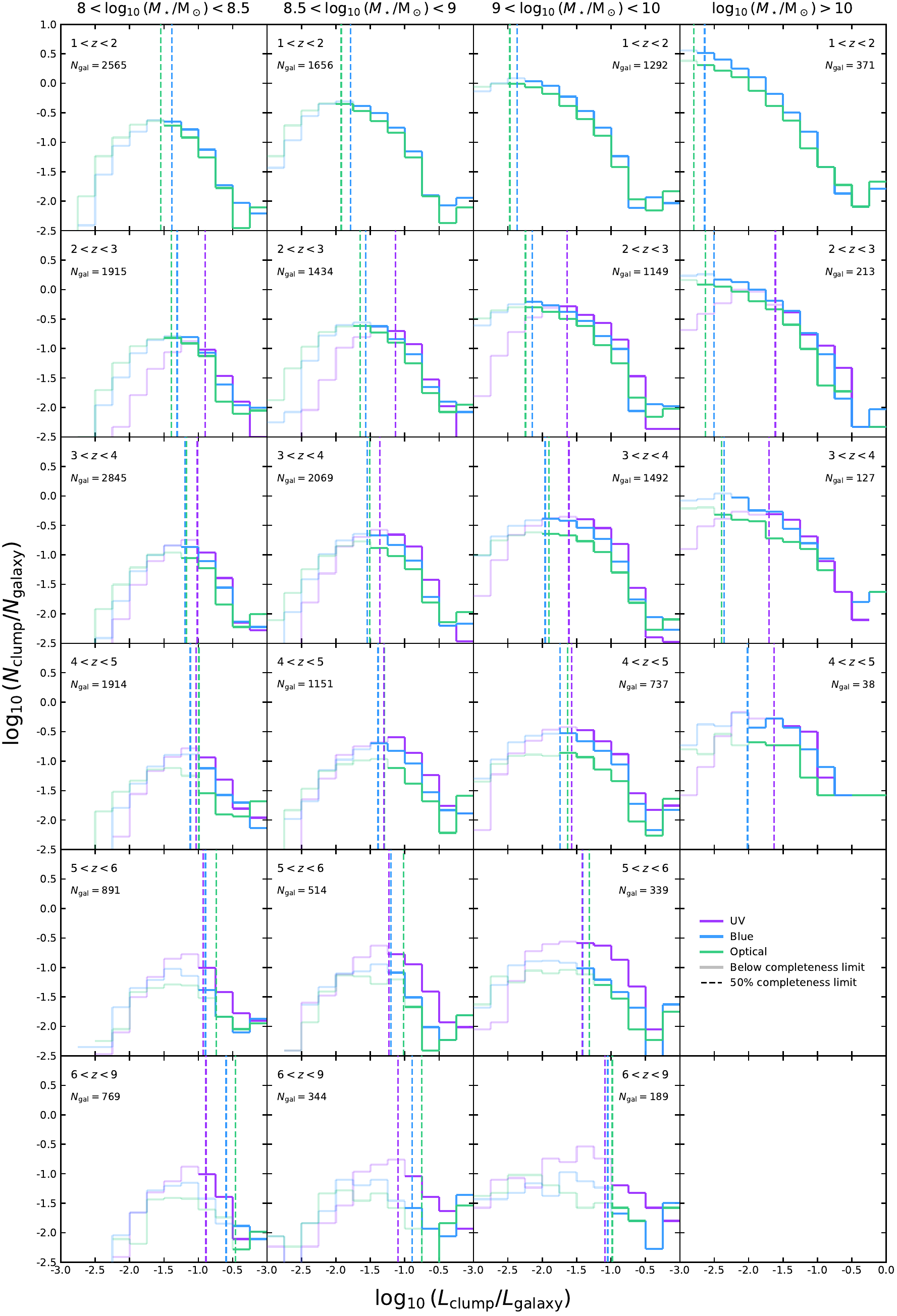}
\caption{Clump luminosity functions as a function of redshift, stellar mass, and rest-frame wavelength. Each panel shows the number of non-central clumps per galaxy as a function of fractional luminosity, $\log_{10}(L_{\rm clump}/L_{\rm galaxy})$, normalised by the number of galaxies in the corresponding redshift--mass bin. Columns correspond to stellar-mass bins and rows to redshift bins, with the redshift range and number of galaxies indicated in each panel. The UV, blue, and optical measurements are shown in purple, blue, and green, respectively. Darker histogram segments show clumps above the corresponding 50 per cent completeness limit, while lighter segments indicate the observed distribution below this limit. Vertical dashed lines mark the 50 per cent completeness thresholds derived from the injection--recovery simulations. The final row combines the $6<z<7$, $7<z<8$, and $8<z<9$ bins to improve statistics. The highest-mass bin is omitted at $z \geq 5$, as it is excluded from the completeness analysis. The UV measurements are not shown at $z<2$ due to missing NIRCam wavelength coverage.}
 \label{fig:clump_luminosity_function}
\end{figure*}

\begin{figure*}
 \includegraphics[width=\linewidth]{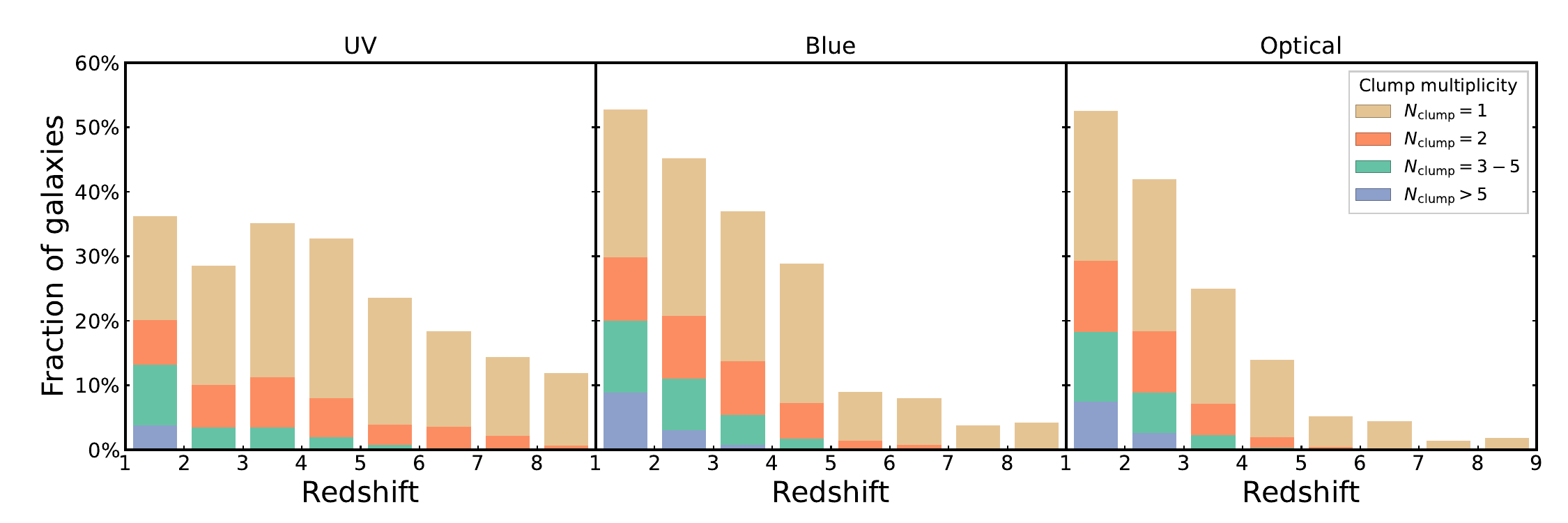}
 \caption{Clump multiplicity as a function of redshift and rest-frame wavelength. Each bar shows the fraction of galaxies in each redshift bin, with coloured components summing to the clumpy fraction. Columns correspond to the UV, blue, and optical measurements. Galaxies are grouped by the number of detected non-central clumps above the bin-dependent 50 per cent completeness limit, with categories $N_{\rm clump}=1$, $N_{\rm clump}=2$, $N_{\rm clump}=3$--$5$, and $N_{\rm clump}>5$. The total bar height represents the clumpy fraction, while the coloured components show increasing clump multiplicity. The same galaxy can occupy different multiplicity categories in different wavelength panels because clumps are detected independently in each rest-frame band.}
 \label{fig:clump_multiplicity_redshift}
\end{figure*}

First, we examine the distribution of clump luminosities relative to their host galaxies. Following previous works, we use the fractional luminosity function, defined as the number of clumps per galaxy as a function of $L_{\rm clump}/L_{\rm galaxy}$ \citep[e.g.][]{Guo_2015, Kalita_2025b}. This quantity is useful because it is independent of the absolute luminosity scale of the host population and directly connects to the clump-selection thresholds commonly used in the literature. In contrast to the clumpy fractions presented in Section~\ref{sec:Clumpy fraction}, here we retain the full observed clump population and show explicitly which part of the distribution lies above the 50 per cent completeness limit in each redshift--mass--wavelength bin.

Figure~\ref{fig:clump_luminosity_function} shows the fractional luminosity functions of the detected clumps in bins of redshift, stellar mass, and rest-frame wavelength. The distributions span a broad range in fractional luminosity, from $L_{\rm clump}/L_{\rm galaxy}\approx10^{-3}$ to values approaching $L_{\rm clump}/L_{\rm galaxy}\approx0.3$--$0.5$, with most of the observed clumps lying at $L_{\rm clump}/L_{\rm galaxy}\approx0.01$--$0.1$. The luminosity functions are generally continuous over this range, rather than showing a clear natural division between ordinary substructure and the brightest clumps. This supports the view that any single luminosity-fraction threshold used to define a clump is a necessary practical observational choice rather than a sharp physical boundary. The 50 per cent completeness limits therefore provide a more transparent and observationally motivated way of defining the robust part of the observed distribution.

The completeness limits indicated by the vertical dashed lines in Figure~\ref{fig:clump_luminosity_function} vary substantially with redshift, stellar mass, and wavelength. In most bins, the faint end of the luminosity function falls below the 50 per cent completeness threshold, especially at high redshift and in the lowest-mass galaxies. At $1<z<2$, the optical completeness limits decrease from $L_{\rm clump}/L_{\rm galaxy}\approx0.03$--$0.04$ in the lowest-mass bin to $\approx0.002$ for $\log_{10}(M_\star/{\rm M_\odot})\geq 10$. By $6<z<9$, these limits rise to $\approx 0.3-0.35$ in the lowest-mass galaxies. This reflects the increasing difficulty of detecting low-contrast clumps in smaller and fainter galaxies, and explains why fixed luminosity-fraction cuts can select different intrinsic clump populations in different regions of parameter space. At low and intermediate redshifts ($1<z<5$), the completeness limits in the higher-mass bins generally reach fractional luminosities of a few per cent or less, allowing the luminosity functions to be traced over a wide dynamic range. At higher redshifts ($z>5$), the limits typically rise to $L_{\rm clump}/L_{\rm galaxy}\gtrsim0.05$ and can exceed 0.1 in the lowest-mass galaxies, such that only the brighter portion of the clump population remains accessible. The observed distributions should therefore be interpreted as lower limits on the abundance of faint clumps.

The wavelength dependence of the luminosity functions provides additional information on the nature of the clumps. In the UV, the fractional luminosity is most sensitive to unobscured recent star formation, while the optical wavelengths trace somewhat older stellar populations and are less dominated by the youngest stars. The differences between the UV, blue, and optical distributions therefore suggest variations in the stellar populations and star formation activity of the detected clumps. At low and intermediate redshifts ($2<z<5$), the rest-frame optical distributions are broadly comparable to the UV at lower fractional luminosities, indicating that clumpy substructure is also prominent at longer wavelengths. However, at larger fractional luminosities ($L_{\rm clump}/L_{\rm galaxy}\gtrsim0.05$--$0.1$), the UV distributions generally become more prominent, particularly at $z>3$, consistent with clumps associated with enhanced recent star formation. These differences may also partly reflect wavelength-dependent completeness and variations in the underlying host-galaxy light.

The luminosity functions also illustrate the stellar-mass dependence seen in the clumpy fractions. Higher-mass galaxies tend to host a more prominent bright tail and a larger number of detectable clumps per galaxy, particularly at intermediate redshifts. Similar trends have been found in recent \textit{JWST} studies, with more massive galaxies hosting larger numbers of clumps \citep[e.g.,][]{Zhu_2026} and preferentially containing more massive clumps \citep[e.g.,][]{Claeyssens_2025, Kalita_2025b}, while \citet{Sok_2026a} also report flatter clump stellar mass-function slopes in more massive hosts. In lower-mass galaxies, the luminosity functions are generally lower in normalisation and more strongly affected by the completeness limit, implying that only the brightest clumps can be robustly identified at the highest redshifts.

We do not fit analytic functions to the luminosity functions in this work. Previous studies have fitted \citet{Schechter_1976} functions or power laws to luminosity or clump mass distributions, but the inferred faint-end slope is highly sensitive to incompleteness and to the details of the clump-finding algorithm \citep{Guo_2015}. Since our primary goal is to compare the observed clump population across redshift, stellar mass, and wavelength, we instead use the luminosity functions as a diagnostic of the detected clump population and of the completeness-limited regime used in our clumpy fraction measurements. The main conclusion from Figure~\ref{fig:clump_luminosity_function} is that clumpy galaxies host a continuous range of substructures with varying fractional luminosities, rather than a distinct population of exceptionally bright clumps. The greater prominence of UV-bright clumps at $z>3$ and of brighter clumps in more massive galaxies suggests that the contribution of clumpy substructure to the stellar light is linked to the evolving star formation activity and physical conditions within galaxies.

\subsection{Clump multiplicity}
\label{sec:Clump multiplicity}

\begin{figure*}
 \includegraphics[width=\linewidth]{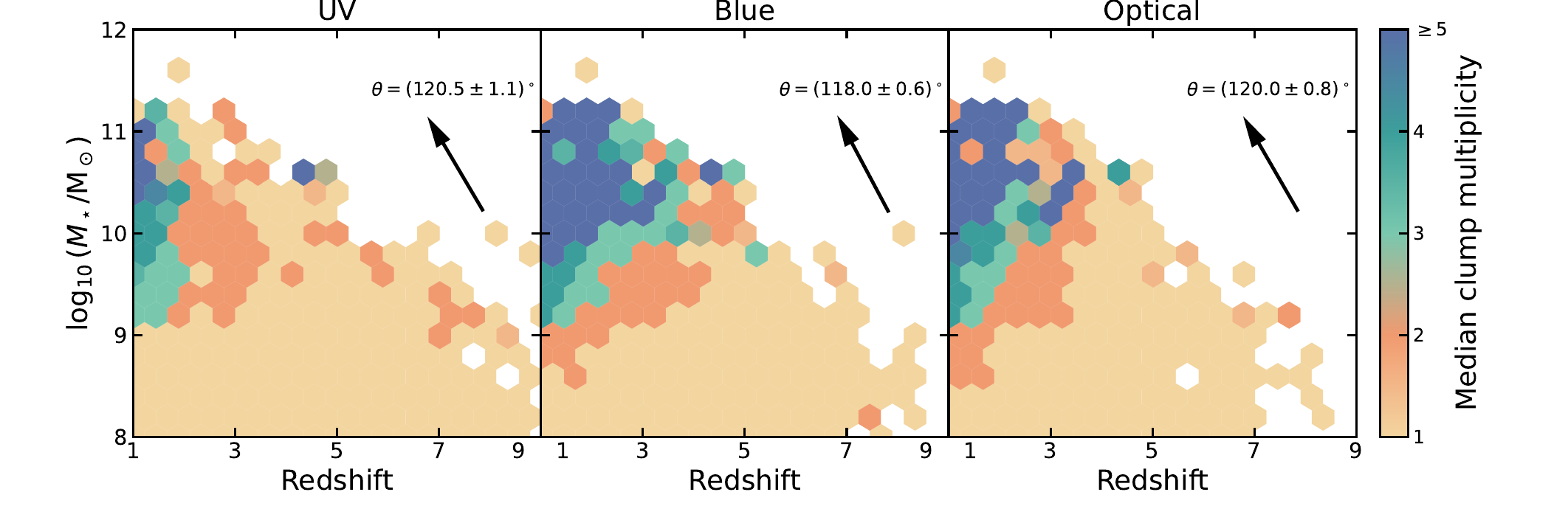}
 \caption{Clump multiplicity in the stellar mass--redshift plane for galaxies hosting at least one off-centre clump, shown at the UV, blue, and optical rest-frame wavelengths. Hexbins are coloured by median multiplicity, with $N_{\rm clump}\geq5$ assigned to blue and $N_{\rm clump}=1$ to yellow. Arrows indicate the directions defined by the partial Pearson correlations with stellar mass and redshift, holding the other variable fixed. The arrow angles $\theta$, measured anticlockwise from the positive-redshift direction, are $(120.5\pm1.1)^\circ$, $(118.0\pm0.6)^\circ$, and $(120.0\pm0.8)^\circ$, respectively, with uncertainties from paired bootstrap resampling. These indicate that multiplicity increases with stellar mass and decreases with redshift, with a stronger dependence on stellar mass.}
 \label{fig:clump_multiplicity_mass}
\end{figure*}

\begin{figure*}
 \includegraphics[width=\linewidth]{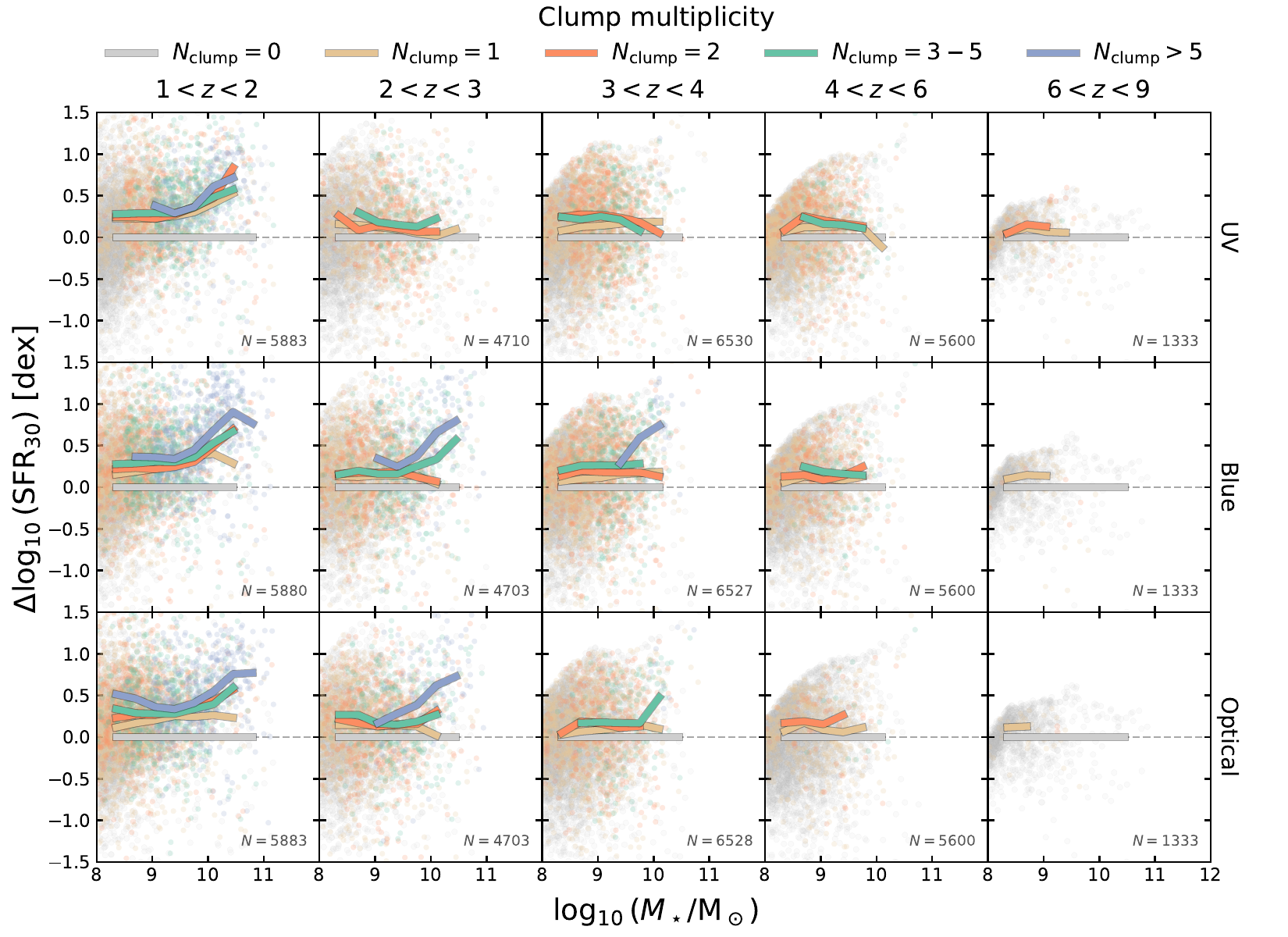}
 \caption{Relative SFR enhancement as a function of stellar mass, clump multiplicity, redshift, and rest-frame wavelength. Each panel shows the offset in $\log_{10}({\rm SFR}_{30})$, where ${\rm SFR}_{30}$ denotes the SFR averaged over the past 30 Myr, relative to the running median of non-clumpy galaxies at similar stellar masses, with redshift bins in columns and UV, blue, and optical measurements in rows. Individual galaxies are shown as faint points, coloured by the number of non-central clumps above the bin-dependent 50 per cent completeness limit, using the categories in Figure~\ref{fig:clump_multiplicity_redshift}. Solid curves show the running median offsets for each multiplicity category, and the dashed line indicates zero enhancement. Panel annotations indicate the number of galaxies with valid SFR offsets.}
 \label{fig:clump_multiplicity_sfr}
\end{figure*}

Next, we investigate whether clumpy galaxies typically host only a single or multiple clumps. We use the completeness-limited clump catalogue to count the non-central clumps in each galaxy at each rest-frame wavelength. Figure~\ref{fig:clump_multiplicity_redshift} shows the multiplicity distribution as a function of redshift, while Figure~\ref{fig:clump_multiplicity_mass} explores its dependence on stellar mass and redshift, and Figure~\ref{fig:clump_multiplicity_sfr} examines the relationship between clump multiplicity and SFR at fixed stellar mass.

Most clumpy galaxies in our sample contain only one or two detected clumps above the 50 per cent completeness limit. As shown in Figure~\ref{fig:clump_multiplicity_redshift}, galaxies with higher multiplicities, especially those with $N_{\rm clump}>5$, are rare and preferentially found at lower redshifts ($z<3$). The multiplicity distribution also depends on rest-frame wavelength. Since clumps are detected independently in the UV, blue, and optical images, a galaxy can move between multiplicity classes depending on wavelength. We retain the UV measurements at $1<z<2$ for the multiplicity analysis, where F070W probes somewhat redder rest-frame wavelengths ($\approx0.23-0.35\,\mu{\rm m}$), while excluding them from the $f_{\rm clumpy}$ analysis to preserve consistent wavelength coverage. The UV panels are most sensitive to recent, unobscured star formation, while the optical panels trace somewhat older stellar populations. Differences in multiplicity between wavelength panels may reflect variations in the underlying stellar populations, as well as dust attenuation and wavelength-dependent detection sensitivity.

Figure~\ref{fig:clump_multiplicity_mass} illustrates the dependence of clump multiplicity on stellar mass and redshift. Among galaxies hosting at least one detected clump, multiplicity generally increases with stellar mass at fixed redshift and decreases with redshift at fixed stellar mass. The partial Pearson correlations yield similar arrow angles of $\theta\approx120^\circ$ across all three rest-frame wavelengths, indicating a stronger dependence on stellar mass than redshift. This may partly reflect the larger sizes and luminosities of massive galaxies, which can host extended star-forming structures containing several detectable clumps. This trend is also consistent with \citet{Zhu_2026}, who independently find that more massive JADES galaxies tend to host more clumps.

Figure~\ref{fig:clump_multiplicity_sfr} shows the SFR enhancement of galaxies with different clump multiplicities relative to non-clumpy galaxies at similar stellar masses. Galaxies hosting more clumps generally exhibit higher SFRs, particularly at lower redshifts ($z<3$) and in the blue and optical measurements, with the strongest enhancements at higher stellar masses. This association suggests enhanced star-forming activity in high-multiplicity systems, although the trend becomes less clear at higher redshifts due to limited statistics. As shown in Appendix~\ref{app:clump_multiplicity_sfms}, the enhancement persists for SFRs averaged over 10, 30, and 100 Myr, reaching $\approx0.2-0.3$ dex for $N_{\rm clump}>5$ at $1<z<2$ in the blue and optical, but weakens slightly with increasing averaging timescale.

The dependence of clump multiplicity on projected galaxy axis ratio, shown in Appendix~\ref{app:multiplicity_axis_ratio}, provides insight into possible formation mechanisms. In rotating discs, projection effects would favour detection of multiple clumps in more face-on systems (higher $B/A$). Instead, we find that multiplicity tends to decrease with increasing $B/A$, particularly in the blue and optical measurements at lower redshifts. This trend may indicate that some elongated galaxies are morphologically disturbed, consistent with a contribution from interactions or mergers, although observational biases cannot be excluded.

Together, these results suggest that higher clump multiplicities are associated with higher stellar masses and stronger star formation activity, potentially reflecting both gravitational fragmentation in gas-rich discs and merger-related processes. At lower stellar masses and higher redshifts, galaxies typically host only one detectable clump, although the more restrictive completeness limits may conceal additional faint substructures. Thus, multiplicity complements $f_{\rm clumpy}$ by revealing the number of substructures detected within galaxies.

\subsection{Clump sizes}
\label{sec:Clump sizes}

Finally, we examine the projected sizes of the detected clumps. For each non-central clump in the 50 per cent completeness-limited catalogue, we compute an area-equivalent circular radius,
\begin{equation}
R_{\rm clump,circ} = \sqrt{\frac{A_{\rm clump}}{\pi}},
\label{eq:clump_circular_radius}
\end{equation}
where $A_{\rm clump}$ is the projected physical area assigned to the clump in ${\rm kpc}^2$. The area is obtained from the number of pixels belonging to each dendrogram leaf and converted to physical units using the angular pixel scale and the angular-diameter distance at the galaxy redshift. We emphasise that this quantity is not a Sérsic effective radius and is not corrected for the PSF. It should instead be interpreted as the characteristic projected size of the detected clump region, providing a useful measure for comparing the apparent spatial extent of clumps across wavelengths.

\begin{figure}
 \includegraphics[width=\columnwidth]{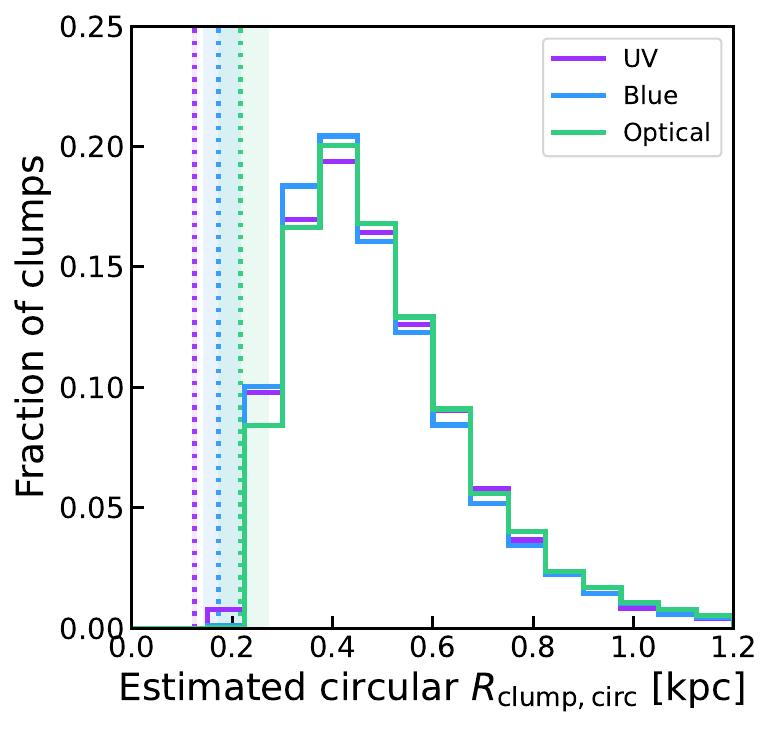}
 \caption{Distribution of the estimated circular radii of non-central clumps in the 50 per cent completeness-limited catalogue, defined as $R_{\rm clump,circ}=\sqrt{A_{\rm clump}/\pi}$. UV, blue, and optical clumps are shown in purple, blue, and green, respectively. Dotted lines indicate the median physical scale corresponding to $0.5\times{\rm FWHM}_{\rm PSF}$, with shaded regions showing the 16--84 percentile range. The radii are derived from the dendrogram segmentation areas and are not corrected for PSF broadening. Median radii are 0.464, 0.458, and 0.475 kpc for the UV, blue, and optical samples, respectively.}
 \label{fig:clump_circular_reff}
\end{figure}

Figure~\ref{fig:clump_circular_reff} shows that the clump size distribution is strongly concentrated at sub-kpc scales, with relatively few clumps extending to larger radii. The three rest-frame wavelength distributions are broadly similar, with median radii of $R_{\rm clump,circ} \approx 0.46$--$0.48\,{\rm kpc}$, indicating that the characteristic projected sizes do not change dramatically between the UV and optical. These characteristic sizes are broadly consistent with the independent JADES clump measurements of \citet{Zhu_2026}, although their residual-based selection, size definitions, and PSF corrections differ from those adopted here. The compact nature of the detected structures is consistent with both \textit{in situ} fragmentation and merger-induced clump formation, and does not by itself distinguish between these mechanisms.

The distributions divided by stellar mass and redshift are shown in Figure~\ref{fig:clump_circular_radius_mass_redshift} in the Appendix~\ref{app:clump_radii}. At $1 \leq z < 3$, the three wavelength distributions are generally similar, whereas at $5 \leq z < 9$, the optical clumps exhibit larger median radii, particularly in low- and intermediate-mass galaxies. In the two lowest stellar-mass bins, the median optical radii reach $\approx 0.6$--$0.8\,{\rm kpc}$, compared with $\approx 0.3\,{\rm kpc}$ in the UV. Such relatively extended structures are qualitatively compatible with a larger contribution from accreted substructures or merger-induced star formation \citep{Calabro_2019, de_la_Vega_2026b}, consistent with the merger-related interpretation discussed in Section~\ref{sec:discussion_origin}. However, the measured sizes are sensitive to the wavelength-dependent PSF, surface-brightness sensitivity, clump-to-host contrast, and small-number statistics, particularly at high redshift. These differences should therefore be interpreted cautiously and cannot independently establish a merger origin.

The observed sizes are also affected by resolution and blending. Individual detected clumps may correspond to smaller star-forming regions or clusters that are not resolved into separate components. Indeed, recent lensed observations suggest that some kpc-scale ``giant'' clumps can comprise multiple $\approx\!100$--300~pc substructures \citep{Kalita_2026a, Kalita_2026b}. Furthermore, observations of strongly lensed galaxies have revealed extremely compact (pc-scale), intensely star-forming stellar clusters at high redshift \citep[e.g.,][]{Vanzella_2023, Adamo_2024, Claeyssens_2025, Messa_2026}, which could potentially represent the underlying substructure of the unlensed clumps presented here. The measured radii should therefore not necessarily be interpreted as the intrinsic sizes of individual star-forming units. Higher-resolution observations, or strongly lensed samples, are required to determine whether these clumps are single bound structures or unresolved associations of smaller stellar clusters \citep[e.g.,][]{Khullar_2026, Vanzella_2026}.

Despite these caveats, the size distributions indicate that the detected clump population is dominated by compact, high-contrast regions with fractional luminosities of a few per cent and typical projected radii below a kiloparsec. Together with the luminosity functions and multiplicity trends, these measurements provide a complementary view of the clump population, while highlighting the limitations of inferring their physical origins from unresolved sizes alone.

\section{Discussion}
\label{sec:Discussion}

We have developed a dendrogram-based framework for identifying clumps in galaxies and applied it to $>24,000$ galaxies in JADES over the redshift range $1\leq z\leq9$. The redshift and stellar-mass dependence of $f_{\rm clumpy}$ provides insight into the physical mechanisms responsible for these subgalactic structures. Below, we discuss the possible origins of clumps in the context of previous observations and simulations \citep{Guo_2015, Shibuya_2016, Ribeiro_2017, Mandelker_2017, Nakazato_2024, de_la_Vega_2026a, Sok_2026a}.

\subsection{Comparison with observations and simulations}
\label{sec:discussion_comparison}

\begin{figure*}
 \includegraphics[width=\linewidth]{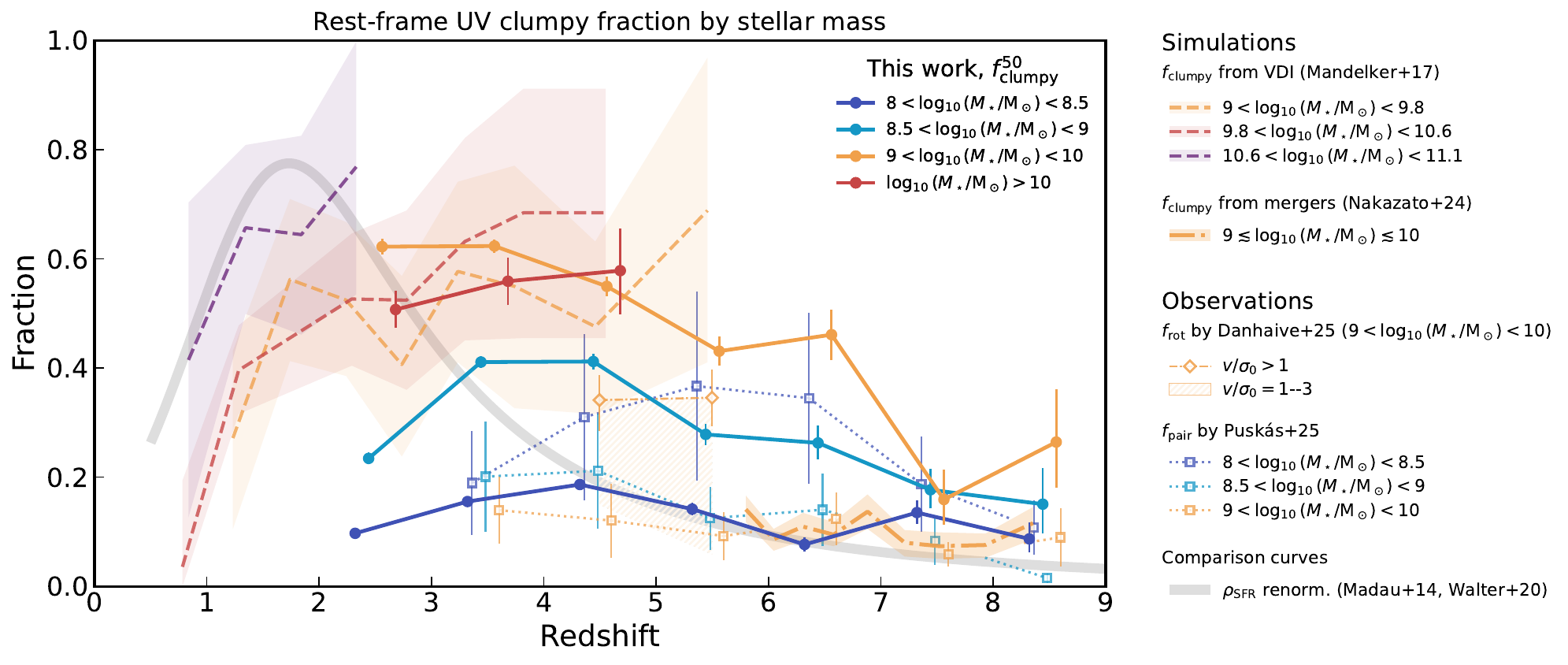}
 \caption{Evolution of the rest-frame UV clumpy fraction as a function of redshift and stellar mass. Coloured circles connected by solid lines show the 50 per cent completeness-limited measurements from this work, $f_{\rm clumpy}^{50}$, in four stellar-mass bins, with dark blue, cyan, orange, and red indicating increasing stellar mass. Where possible, similar colour shades indicate comparable stellar-mass ranges across the literature datasets. Error bars show the 68 per cent binomial confidence intervals, with small horizontal offsets applied for clarity. Dashed curves and shaded regions show simulated clumpy fractions from the VDI-dominated discs of \citet{Mandelker_2017} and the merger-driven, [O~\textsc{iii}]-bright systems of \citet{Nakazato_2024}. Orange diamonds and a hatched region indicate the fractions of rotationally supported galaxies from \citet{Danhaive_2025} for different $v/\sigma_0$ thresholds. Open squares connected by dotted lines show the major merger close-pair fractions from \citet{Puskas_2025}, tracing major interactions rather than clumpiness. The broad grey curve shows the cosmic SFRD evolution using the \citet{Madau_2014} functional form with parameters from \citet{Walter_2020}, renormalised for visual comparison.}
 \label{fig:simulation_comparison}
\end{figure*}

In Figure~\ref{fig:simulation_comparison}, we place our rest-frame UV clumpy fraction measurements in the context of simulations, observed disc and galaxy close-pair fractions, and the cosmic star formation rate density (SFRD). This comparison focuses on the rest-frame UV, $\lambda_{\rm rest}=0.2~{\rm \mu m}$, because this wavelength most directly traces young star-forming clumps and gives the closest comparison to the UV clump selections used in much of the literature \citep{Guo_2012, Guo_2015, Shibuya_2016, de_la_Vega_2026a}. The coloured circles connected by solid lines show our fiducial 50 per cent completeness-limited clumpy fractions, $f_{\rm clumpy}^{50}$, in four stellar-mass bins, with dark blue, cyan, orange, and red indicating increasing stellar mass. As discussed in Section~\ref{sec:Redshift evolution and stellar mass dependence}, we find a strong stellar-mass dependence \citep{Guo_2015, Sok_2026a, de_la_Vega_2026a}. At fixed redshift, more massive galaxies are more likely to be clumpy, while the lowest-mass galaxies show a much lower and comparatively flatter clumpy fraction. In the highest-mass bin, $\log_{10}(M_\star/{\rm M_\odot})\geq10$, we only show measurements at $z<5$, where the sample remains sufficiently populated for a reliable completeness-limited measurement.

The data for comparison in Figure~\ref{fig:simulation_comparison} are taken from both observations and simulations, and therefore should be interpreted qualitatively rather than as a one-to-one comparison. The orange diamonds and shaded region show the fraction of rotationally supported galaxies from \citet{Danhaive_2025} for $9<\log_{10}(M_\star/{\rm M_\odot})<10$, adopting $v/\sigma_0>1$ and thresholds spanning $v/\sigma_0>1$--3, respectively. They find a rotationally supported fraction of approximately 34 per cent at $z\sim4.5$--5.5, indicating that rotationally supported systems already constitute a substantial population at high redshift. Although rotational support does not necessarily imply a settled disc, these measurements provide useful observational context for the interpretation of \textit{in situ} clump formation. We also show the major merger close-pair fractions from \citet{Puskas_2025}, indicated by open squares connected by dotted lines, which were measured in the same stellar-mass bins (except for $\log_{10}(M_\star / {\rm M_\odot})\geq10$) as in this work using JADES DR3 data. These are not clumpy fractions, but rather an empirical tracer of the incidence of major interactions (stellar mass ratio $\mu \geq 1/4$; see also \citealt{Mundy_2017, Duncan_2019, Duan_2025}). Their implications for merger-driven clump formation were further explored in Section~\ref{sec:Contribution of major mergers to clumpy systems}. Finally, the broad grey curve shows a renormalised cosmic SFRD evolution based on the functional form of \citet{Madau_2014}. Following the approach adopted by \citet{Sok_2026a}, we use a Madau--Dickinson-like functional form,
\begin{equation}
    f_{\rm SFRD,renorm}(z) = a\, \frac{(1+z)^b}{1+\left[(1+z)/c\right]^d},
    \label{eq:renorm_sfrd}
\end{equation}
where the parameters are fixed to the values reported by \citet{Walter_2020}, $b=2.88$, $c=2.75$, and $d=5.88$, while the amplitude $a$ is refit only to renormalise the curve to the scale of the observed clumpy fractions. This curve is therefore not used as a physical measurement of the SFRD in absolute units, but as a comparison to the redshift dependence expected if clump formation broadly tracks the global availability of cold gas and star formation \citep{Madau_2014, Walter_2020, Sok_2026a}.

The two simulation comparisons in Figure~\ref{fig:simulation_comparison} represent different physical regimes. The fragmentation-driven comparison is based on the cosmological zoom-in simulations of \citet{Mandelker_2017}, which study massive, gas-rich, high-redshift galaxies over approximately $1\lesssim z\lesssim6$. These build on the VDI framework developed by \citet{Dekel_2009} and demonstrated in cosmological disc simulations by \citet{Ceverino_2010}. Following \citet{Mandelker_2014}, clumps are identified as three-dimensional overdensities in the gas distribution, and their origins are classified using their dark-matter content: \textit{in situ} clumps are interpreted as products of disc fragmentation, while \textit{ex situ} clumps correspond to accreted satellites or minor mergers. \citet{Mandelker_2017} find that most off-centre clumps form through VDI, although \textit{ex situ} clumps contribute a non-negligible fraction of the clump mass and star formation. The simulated curves in Figure~\ref{fig:simulation_comparison} are therefore best interpreted as the expected clumpy fractions for galaxies in a regime where massive, gas-rich, unstable discs have already formed, rather than an exclusively \textit{in situ} contribution. This is most relevant to our intermediate- and higher-mass bins, where our measured $f_{\rm clumpy}$ values are highest, broadly consistent with the mass dependence reported by \citet{de_la_Vega_2026a}.

The merger-driven comparison is based on \citet{Nakazato_2024}, who use the FirstLight cosmological zoom-in simulations \citep{Ceverino_2017} to study [O~\textsc{iii}]-bright clumps at $5.5<z<9.5$ in galaxies selected to have $M_\star=(0.5-6)\times10^{10}~{\rm M_\odot}$ at $z=5$. Their clumps are identified through high surface densities of stars younger than 10 Myr, motivated by the detectability of rest-frame optical [O~\textsc{iii}] emission with \textit{JWST}. In contrast to the lower-$z$ VDI-dominated picture of \citet{Mandelker_2017}, \citet{Nakazato_2024} find that most detectable high-$z$ clumps form during major mergers, where tidal debris and gas compression produce young, off-centre star-forming structures. Such merger-induced gas compression and enhanced star formation have also been investigated in numerical simulations \citep{Di_Matteo_2008, Renaud_2015}, while \citet{Calabro_2019} find observational evidence for enhanced clumpiness in merging galaxies. The resulting clumpy systems are short-lived, typically merging within a few tens of Myr. Importantly, \citet{Nakazato_2024} do not find clumpy systems formed through cold-accretion-driven VDI at $z\gtrsim6$, arguing that fragmentation requires a sufficiently massive disc with a large disc mass fraction. They suggest that disc fragmentation may become important in their most massive systems, for example at $M_\star\approx10^{10}~{\rm M_\odot}$ by $z\approx5.5$. Their results therefore provide a useful comparison to our high-$z$, intermediate-mass measurements, where merger-driven clump formation may play an important role, as also suggested by \citet{de_la_Vega_2026a}. However, differences in stellar-mass selection and clump identification limit a direct quantitative comparison, particularly with our lowest-mass galaxies.

\subsection{A transition in clump formation mechanisms}
\label{sec:discussion_origin}

Taken together, our measurements and the simulations of \citet{Mandelker_2017} and \citet{Nakazato_2024} suggest a transition in the dominant clump formation pathway with both cosmic time and stellar mass, broadly consistent with recent observational studies \citep{de_la_Vega_2026a, Sok_2026a}. At lower redshifts and higher stellar masses, especially around cosmic noon, the high values of $f_{\rm clumpy}$ are difficult to explain through major mergers alone, although interactions have long been considered a possible origin of clumpy morphologies \citep{Puech_2010, Ribeiro_2017}. In Section~\ref{sec:Contribution of major mergers to clumpy systems}, we showed that the merger-driven fraction inferred from the major merger rates of \citet{Puskas_2025} generally falls below the observed $f_{\rm clumpy}$ in the higher-mass bins, even when assuming longer clump observability timescales. This implies that additional formation channels beyond major mergers are required. Minor mergers may also contribute through accreted satellites or interaction-induced star formation, as discussed by \citet{Mandelker_2014} and \citet{Guo_2015}, and more recently by \citet{de_la_Vega_2026b}. Nevertheless, the most natural interpretation for the high $f_{\rm clumpy}$ in massive galaxies is \textit{in situ} gravitational fragmentation in gas-rich discs, as predicted by the VDI framework of \citet{Dekel_2009} and demonstrated in numerical simulations \citep{Agertz_2009, Ceverino_2010, Bournaud_2014, Inoue_2016, Mandelker_2017}. Recent analytical models by \citet{Ferrara_2026} similarly predict a broad intrinsic clump mass spectrum and mass--size relations from gravitational instability, with subsequent evolution influenced by dynamical friction and feedback. This interpretation is also supported by the companion JADES study of \citet{Zhu_2026}, who find clump mass functions and radial trends consistent with formation in turbulent discs, followed by migration, disruption, or structural evolution, as explored in earlier theoretical studies \citep{Ceverino_2012, Mandelker_2014, Dekel_2022}.

At the low-mass and high-redshift end, the interpretation changes. In our lowest stellar-mass bin, $8\leq\log_{10}(M_\star/{\rm M_\odot})<8.5$, the clumpy fraction remains comparatively low and nearly flat with redshift. However, as shown in Section~\ref{sec:Contribution of major mergers to clumpy systems}, the merger-driven fraction inferred from the major merger rates of \citet{Puskas_2025} can become comparable to the observed clumpy fraction at high redshift, suggesting that major mergers could account for a substantial fraction of these systems. This remains sensitive to the assumed pair observability timescales \citep{Lotz_2010, Snyder_2017, Duncan_2019}, as well as the duration of the detectable clumpy phase and the assumption that every merger produces detectable clumps. This interpretation is qualitatively consistent with \citet{Nakazato_2024}, who find predominantly merger-induced clumps at $z\approx6-9$, although their simulations target galaxies selected to reach higher stellar masses. Minor mergers may also contribute by triggering off-centre star formation \citep{Renaud_2015, Calabro_2019}, as recently illustrated by \citet{de_la_Vega_2026b}.

The intermediate-mass regime, $9\lesssim\log_{10}(M_\star/{\rm M_\odot})\lesssim10$, is likely where the transition becomes most apparent. In this mass range, our clumpy fractions are substantially higher than in the lowest-mass bin, while the comparison with the major merger rates of \citet{Puskas_2025} indicates that mergers alone are increasingly insufficient towards lower redshifts. At higher redshifts, however, merger-induced clump formation could still contribute substantially, consistent with the FirstLight simulations of \citet{Nakazato_2024}. We therefore interpret this as a mixed regime, in which both merger-related processes and \textit{in situ} gravitational fragmentation contribute, with the latter becoming increasingly important as galaxies assemble more massive, gas-rich discs, as suggested by \citet{Mandelker_2017}. This picture is broadly consistent with \citet{de_la_Vega_2026a}, who argue that disc fragmentation dominates at lower redshifts and $\log_{10}(M_\star/{\rm M_\odot})\gtrsim9$, while mergers become more important at higher redshifts and intermediate stellar masses.

\subsection{Gravitational instabilities and the role of gas}
\label{sec:discussion_physics}

It is important, however, not to interpret disc fragmentation and VDI too narrowly. Classical VDI is often framed in terms of fragmentation in a marginally stable disc with Toomre parameter $Q \! \approx \! 1$ \citep{Toomre_1964, Dekel_2009}, but high-redshift galaxies are not necessarily isolated, axisymmetric systems. Using cosmological simulations, \citet{Inoue_2016} showed that clumps can form in highly perturbed, non-linear discs even when the smoothed pre-collapse regions have $Q$ values of order unity to a few. They discuss several possible mechanisms, including non-axisymmetric instabilities, rapidly decaying turbulence, and externally stimulated instabilities associated with minor mergers or counter-rotating gas streams. Stellar feedback can further regulate clump formation and survival, with simulations demonstrating a strong dependence on feedback prescriptions \citep{Mayer_2016, Oklopcic_2017, Mandelker_2017}. Therefore, when we refer to a disc fragmentation or VDI-like origin, we mean \textit{in situ} gravitational fragmentation in gas-rich, turbulent systems, including externally stimulated instabilities, rather than only the idealised linear Toomre instability of an isolated smooth disc.

This broader \textit{in situ} interpretation is physically motivated by the conditions of high-redshift galaxies. Observations show that molecular gas fractions increase strongly with redshift \citep{Tacconi_2010, Tacconi_2020, Wang_2022}, while \textit{JWST} imaging reveals substantial disc-like populations even at early epochs \citep{Ferreira_2023, Kartaltepe_2023}. Kinematic studies further indicate that rotationally supported, often turbulent systems are already present at high redshift \citep{Forster_Schreiber_2018, Wisnioski_2019, Ubler_2019, Parlanti_2023, Danhaive_2025, Lee_2025}. In sufficiently massive discs, higher gas surface densities can reduce the Toomre stability parameter and increase the characteristic fragmentation scale, promoting the formation of massive star-forming clumps \citep{Dekel_2009, Ceverino_2010, Mandelker_2017, Krumholz_2018}. At the same time, simulations by \citet{Di_Matteo_2008} and \citet{Renaud_2015} show how gas-rich mergers can trigger enhanced star formation through tidal compression, while \citet{Nakazato_2024} demonstrate the formation of merger-induced clumps at high redshift. Therefore, high gas content does not uniquely favour one channel: in massive discs it promotes \textit{in situ} fragmentation, while in low-mass disturbed systems it can amplify the clump-forming effect of galaxy interactions.

Our results therefore support a gradual, mass- and redshift-dependent transition in clump origin, rather than a simple division between merger-driven clumps at high redshift and fragmentation-driven clumps at later times. At fixed redshift, interactions and accreted companions may contribute more strongly towards lower stellar masses \citep{Ribeiro_2017, Zanella_2019, Nakazato_2024}, whereas sufficiently massive, gas-rich discs may undergo \textit{in situ} fragmentation even at early epochs, as demonstrated in the simulations of \citet{Mayer_2025}. This is also compatible with the model of \citet{Krumholz_2018}, in which turbulence in high-redshift discs is driven predominantly by gravitational instabilities and radial transport, with stellar feedback becoming increasingly important at later times. The contrasting evolutionary trends in our sample, from nearly constant clumpy fractions at low stellar masses to a pronounced increase towards cosmic noon in more massive galaxies, are also consistent with this picture. Our comparison with the major merger rates of \citet{Puskas_2025} is central to this interpretation: where the inferred merger-driven fraction is comparable to $f_{\rm clumpy}$, a merger origin is plausible, whereas a substantial excess of $f_{\rm clumpy}$ over the merger expectation requires additional formation channels, most naturally \textit{in situ} fragmentation. 
Figure~\ref{fig:simulation_comparison} therefore supports an increasingly important contribution from gravitational fragmentation at higher stellar masses and later cosmic times, broadly consistent with \citet{de_la_Vega_2026a, Sok_2026a}. However, minor mergers remain unconstrained by our comparison, the highest-mass regime at $z\gtrsim5$ is poorly sampled, and the merger-rate measurements do not extend below $z=3$.

\section{Summary \& Conclusions}
\label{sec:conclusion}

In this work, we investigated the evolution of the clumpy galaxy fraction, $f_{\rm clumpy}$, using deep \textit{JWST}/NIRCam imaging from JADES. Our sample comprises 24,064 galaxies at $z \approx 1$--9 with stellar masses of $\log_{10}(M_\star/{\rm M_\odot}) \geq 8$. We identified clumps as compact, off-centre surface-brightness peaks using a dendrogram-based detection method with \texttt{astrodendro} at three rest-frame wavelengths: UV ($0.2\,\mu m$), blue ($0.4\,\mu m$), and optical ($0.55\,\mu m$). We quantified clump-detection completeness through injection--recovery tests and adopted bin-dependent luminosity thresholds corresponding to 50 per cent completeness. We also derived a probability-weighted completeness-corrected statistic from detected clumps for comparison with the fiducial results. Our main findings are summarised below.

\begin{enumerate}[(i)]
    \item The clumpy fraction generally declines with increasing redshift, with a strong dependence on rest-frame wavelength. In the full sample, the blue and optical $f_{\rm clumpy}$ decrease from $\approx55$ per cent at $z\approx1$--2 to $\approx5$ per cent at $z\approx8$--9, while the UV $f_{\rm clumpy}$ remains higher at the earliest epochs, reaching $\approx12$ per cent at $z\approx8$--9. The probability-weighted completeness-corrected measurements show similar evolution, although they do not account for galaxies whose clumps remain entirely undetected.

    \item The evolution of $f_{\rm clumpy}$ depends strongly on stellar mass. At $z\approx1$--2, $f_{\rm clumpy}$ increases from $\approx35$ per cent in the lowest-mass bin to nearly $\approx85$ per cent at $\log_{10}(M_\star/{\rm M_\odot}) \geq 10$. The lowest-mass galaxies show relatively weak evolution in the rest-frame UV, whereas more massive systems exhibit a pronounced increase in clumpiness towards cosmic noon, followed by a decline at higher redshifts.

    \item Most clumpy galaxies host only one or two detectable off-centre clumps, with higher multiplicities preferentially found in more massive and actively star-forming systems. These clump-scale structures have a median area-equivalent radius of $\approx \! 0.45$~kpc, comparable to the imaging resolution and uncorrected for the PSF. They should therefore not be interpreted as individually resolved star clusters or single star-forming regions, as they may contain multiple unresolved components.

    \item Comparison with the major merger rates of \citet{Puskas_2025} reveals a strong dependence on stellar mass and redshift. At low stellar masses and $z\gtrsim5$, the inferred merger-driven fractions approach the observed $f_{\rm clumpy}$, suggesting that major mergers could account for many clumpy systems. In contrast, at higher stellar masses, particularly at $3 \lesssim z\lesssim5$, the observed fractions substantially exceed merger-driven expectations, even for the longest assumed observability timescales.

    \item Our results favour a gradual, mass- and redshift-dependent transition between \textit{ex situ} and \textit{in situ} clump formation. Mergers and accreted companions may contribute more prominently in low-mass galaxies at high redshift, while gravitational fragmentation in gas-rich, turbulent discs becomes increasingly important in more massive systems towards cosmic noon. These mechanisms are not mutually exclusive, as interactions may themselves trigger internal gravitational instabilities.
\end{enumerate}

Our findings support a gradual, mass- and redshift-dependent transition in the relative importance of \textit{in situ} and \textit{ex situ} clump formation. External interactions and accretion may contribute more prominently in low-mass galaxies at early epochs, while gravitational instabilities in gas-rich discs become increasingly important in more massive systems. This suggests that clumpy morphologies arise from an evolving interplay between internal and external processes, rather than a single dominant formation mechanism across cosmic time.

Important questions remain regarding the physical origins, survival, and evolution of individual clumps, as well as their contribution to galaxy growth. Future studies combining deeper and wider \textit{JWST} surveys with high-resolution imaging using adaptive optics and spatially resolved spectroscopy from next-generation 30-metre-class telescopes will help address these questions. In particular, measurements of clump kinematics, stellar populations, and chemical abundances will provide stronger constraints on their formation mechanisms and their role in the structural evolution of galaxies.

\section*{Acknowledgements}

DP acknowledges support by the Huo Family Foundation through a P.C. Ho PhD Studentship. DP, FDE, AS, and RM acknowledge support by the Science and Technology Facilities Council (STFC), by the ERC through Advanced Grant 695671 “QUENCH”, and by the UKRI Frontier Research grant RISEandFALL. ST acknowledges support by the Royal Society Research Grant G125142. FvD acknowledges support from the Herchel Smith Fellowship at the University of Cambridge. AJB acknowledges funding from the ``FirstGalaxies'' Advanced Grant from the European Research Council (ERC) under the European Union’s Horizon 2020 research and innovation programme (Grant Agreement No. 789056). CC, YZ, EE, ZJ, DJE, BDJ, MR, and BER acknowledge support by a \textit{JWST}/NIRCam contract to the University of Arizona, NAS5-02015. CC and BER also acknowledge support from the \textit{JWST} Program 3215. ALD thanks the University of Cambridge Harding Distinguished Postgraduate Scholars Programme and Technology Facilities Council (STFC) Center for Doctoral Training (CDT) in Data intensive science at the University of Cambridge (STFC grant number 2742605) for a PhD studentship. QD acknowledges support from a PhD studentship awarded by Trinity College, University of Cambridge. PIN acknowledges funding by the European Union (MSCA Doctoral Network EDUCADO, GA 101119830 and Widening Participation, ExGal-Twin, GA101158446). PIN thanks the LSST-DA Data Science Fellowship Program, which is funded by LSST-DA, the Brinson Foundation, the WoodNext Foundation, and the Research Corporation for Science Advancement Foundation; her participation in the program has benefited this work. PIN acknowledge financial support from the State Research Agency of the Spanish Ministry of Science and Innovation (AEI-MCINN) under the grants ``Galaxy Evolution with Artificial Intelligence'' with reference PGC2018-100852-A-I00 and ``BASALT'' with reference PID2021-126838NB-I00. AS also thanks the Science and Technology Facilities Council (STFC) Center for Doctoral Training (CDT) in Data Intensive Science at the University of Cambridge (STFC grant number ST/W006812/1) for a PhD studentship. H\"U acknowledges support by the Max Planck Society through the Lise Meitner Excellence Program. H\"U acknowledges funding by the European Union (ERC APEX, 101164796). Views and opinions expressed are however those of the authors only and do not necessarily reflect those of the European Union or the European Research Council Executive Agency. Neither the European Union nor the granting authority can be held responsible for them. ECL acknowledges support of an STFC Webb Fellowship (ST/W001438/1). DJE is also supported as a Simons Investigator.  The research of CCW is supported by NOIRLab, which is managed by the Association of Universities for Research in Astronomy (AURA) under a cooperative agreement with the National Science Foundation.

This work is based in part on observations made with the NASA/ESA/CSA James Webb Space Telescope. The data were obtained from the Mikulski Archive for Space Telescopes at the Space Telescope Science Institute, which is operated by the Association of Universities for Research in Astronomy, Inc., under NASA contract NAS 5-03127 for \textit{JWST}. These observations are associated with PIDs 1180, 1181, 1210, 1286, 1895, 1963, and 3215. The authors acknowledge the teams led by PIs Daniel Eisenstein (PID 3215), Christina Williams, Sandro Tacchella, Michael Maseda (JEMS; PID 1963), and Pascal Oesch (FRESCO; PID 1895) for developing their observing program with a zero-exclusive-access period.

\section*{Data Availability}

The data underlying this article will be shared at a reasonable request by the corresponding author. Fully reduced NIRCam images and NIRSpec spectra are publicly available on MAST (\url{https://archive.stsci.edu/hlsp/jades}), with \doi{10.17909/8tdj-8n28}, \doi{10.17909/z2gw-mk31}, and \doi{10.17909/fsc4-dt61} \citep{Rieke_2023, Eisenstein_2025, Bunker_2024, DEugenio_2025, Curtis-Lake_2026, Scholtz_2026, Johnson_2026, Robertson_2026}.



\bibliographystyle{mnras}
\bibliography{references.bib} 



\clearpage
\appendix

\section{Spectroscopic-redshift catalogue construction}
\label{app:specz_catalogue}

To assess the accuracy of the photometric redshifts adopted in this work, we compile a catalogue containing all the available spectroscopic redshifts of sources in the GOODS-South and GOODS-North fields. The catalogue is constructed by matching each spectroscopic catalogue to the JADES photometric catalogues in the two fields. The different catalogues are homogenised into a common format, and their reliability and coordinate-match quality flags are converted to a unified system, where the highest-quality and best-coordinate matches are assigned \texttt{flag\_z\_qual = 1} and \texttt{flag\_match = 1}, respectively.

For GOODS-South, the compilation includes the legacy spectroscopic collection in the CANDELS fields (hereafter GDS collection; N. Hathi, private communication), JADES NIRSpec observations \citep{Bunker_2024, DEugenio_2025, Curtis-Lake_2026, Scholtz_2026}, JADES Origins Field \citep[JOF; PID: 4540; PI: Eisenstein;][]{Sun_2026}, FRESCO and CONGRESS redshifts (F. Sun, private communication; see also \citealt{Meyer_2024} for H$\alpha$, \citealt{Covelo-Paz_2025} for H$\beta$+[O {\sc III}], and the survey paper by \citealt{Oesch_2023}), DarkHorse \citep{DEugenio_2026}, OASIS (T. Looser, private communication), NIRSpec Wide GTO Program (hereafter WIDE; M. Maseda, private communication), MUSE-Wide DR1 \citep{Urrutia_2019}, MUSE DR2 \citep{Bacon_2023}, ASPECS \citep{Decarli_2019}, DEIMOS \citep{Pharo_2022}, ALMA-LESS \citep{Danielson_2017}, and the DAWN \textit{JWST} Archive products associated with PID: 2198 \citep[PIs: L. Barrufet \& P. Oesch;][]{Barrufet_2025} and PID: 6541 \citep[PI: Egami;][]{DeCoursey_2025}. For GOODS-North, the compilation includes the legacy spectroscopic collection in the CANDELS field (hereafter GDN collection; N. Hathi, private communication), JADES NIRSpec observations \citep{DEugenio_2025, Curtis-Lake_2026, Scholtz_2026}, FRESCO+CONGRESS redshifts (F. Sun and X. Lin, private communication; see also \citealt{Oesch_2023, Meyer_2024, Covelo-Paz_2025}), WIDE (M. Maseda, private communication), DEIMOS \citep{Pharo_2022}, DESI Early Data Release \citep[EDR;][]{DESI_Collaboration_2024}, and the DAWN \textit{JWST} Archive products associated with PID: 2674 \citep[PI: Arrabal-Haro;][]{Arrabal_Haro_2023} and PID: 1871 \citep[PI: Chisholm;][]{Chisholm_2024}. The spectroscopic source catalogues included in the GOODS-South and GOODS-North compilations are summarised in Tables~\ref{tab:specz_sources_goods_s} and \ref{tab:specz_sources_goods_n}, respectively.

For catalogues where small systematic astrometric offsets are present, we correct the coordinates by the median RA and Dec offset measured after an initial match. In GOODS-South, this correction is applied to the GDS collection, MUSE, MUSE-Wide, ASPECS, DEIMOS, and ALMA-LESS catalogues, while the JADES NIRSpec, FRESCO, OASIS, WIDE, JOF, DarkHorse, PID: 2198, and PID: 6541 catalogues are matched without such a correction because the median offsets are negligible. In GOODS-North, the offset correction is applied to the GDN collection, WIDE, and DESI EDR catalogues, while JADES NIRSpec, FRESCO+CONGRESS, DEIMOS, PID: 2674, and PID: 1871 are matched directly. In all cases, the closest JADES photometric object is selected within the adopted matching radius, with \texttt{flag\_match = 1} corresponding to matches within $0.25$ arcsec and \texttt{flag\_match = 2} to matches between $0.25$ and $0.5$ arcsec.

The redshift quality was homogenised into a three-tier system based on the original quality flags from the individual catalogues. In the case of most \textit{JWST} spectra, we assigned the best quality, \texttt{flag\_z\_qual = 1}, to redshifts which were determined from at least one emission line detection in the medium-resolution grating or from two or more emission lines detected in low-resolution Prism/Clear spectrum. The less certain flag, \texttt{flag\_z\_qual = 2}, was assigned to clear redshifts determined from the continuum, or from the continuum and a single Prism/Clear emission line. The uncertain flag of \texttt{flag\_z\_qual = 3} was assigned to tentative redshifts determined from visual inspection. This three-tier system is based on the flags described by \citet{Bunker_2024}, and we applied it to most \textit{JWST}-based spectroscopic redshift catalogues where appropriate information was given. In the case of other instruments and legacy catalogues, we tried to categorise the redshifts into the three tiers as best as possible based on their pre-existing quality flags.

As an additional comparison, we first compare the high-quality spectroscopic redshift measurements to the photometric redshifts from \texttt{EAZY} \citep{Brammer_2008}, using the catalogue value $z_a$. These photometric redshift fits, the applied templates, and parameter assumptions are detailed in \citet{Hainline_2024} and \citet{Hainline_2026}. For the combined GOODS-South and GOODS-North sample, we find an outlier fraction of $10.98$ per cent when comparing the high-quality spectroscopic measurements to the \texttt{EAZY} photometric redshifts. For our fiducial sample we use the photometric redshifts estimated by \texttt{Prospector} (see Section~\ref{sec:Photometric and spectroscopic redshifts}).

The same spectroscopic redshift compilation is used to construct a catalogue of the best available redshift for each galaxy. Where a secure spectroscopic redshift is available, it is preferred over the photometric redshift from \texttt{Prospector}. If multiple spectroscopic redshifts are available for the same object, we first select the measurements with the best combination of redshift quality and coordinate match flags. When several measurements with the same best flags are available and at least two of them agree within $\Delta z_{\mathrm{spec}} \leq 0.1$, we adopt the redshift from the highest-priority catalogue. In GOODS-South, the priority order is JADES NIRSpec, DarkHorse, JOF, OASIS, WIDE, FRESCO, PID: 2198, PID: 6541, MUSE, MUSE-Wide, GDS collection, ASPECS, DEIMOS, and ALMA-LESS. In GOODS-North, the priority order is JADES NIRSpec, WIDE, FRESCO+CONGRESS, PID: 1871, PID: 2674, GDN collection, DESI EDR, and DEIMOS. This priority ordering corresponds to the order in which the catalogues are listed in Tables~\ref{tab:specz_sources_goods_s} and \ref{tab:specz_sources_goods_n}. If only one valid redshift is available, it is automatically assigned as the best spectroscopic redshift. This procedure ensures that the final redshift catalogue uses the most reliable available redshift while avoiding ambiguous spectroscopic identifications.

\begin{table*}
\centering
\caption{Spectroscopic redshift source catalogues included in the GOODS-South compilation, listed in the priority order used when multiple spectroscopic redshifts with the same best redshift-quality and coordinate-match flags are available for the same object.}
\label{tab:specz_sources_goods_s}
\scriptsize
\setlength{\tabcolsep}{3pt}
\renewcommand{\arraystretch}{1.18}
\begin{tabular}{p{0.21\textwidth} p{0.15\textwidth} >{\centering\arraybackslash}p{0.13\textwidth} p{0.22\textwidth} p{0.22\textwidth}}
\hline
\hline
Catalogue & Instrument & Astrometric correction & Reference & Notes \\
\hline
JADES NIRSpec & \textit{JWST}/NIRSpec & No & \citet{Bunker_2024, DEugenio_2025, Curtis-Lake_2026, Scholtz_2026} & JADES spectroscopic redshifts. \\
DarkHorse & \textit{JWST}/NIRSpec & No & \citet{DEugenio_2026} &  \\
JADES Origins Field & \textit{JWST}/NIRSpec & No & \citet{Sun_2026} & JOF; PID: 4540; PI: Eisenstein. \\
OASIS & \textit{JWST}/NIRSpec & No & T. Looser, priv. comm. &  \\
WIDE & \textit{JWST}/NIRSpec & No & M. Maseda, priv. comm. &  \\
FRESCO / CONGRESS & \textit{JWST}/NIRCam grism & No & F. Sun, priv. comm.; \citet{Oesch_2023, Meyer_2024, Covelo-Paz_2025} & FRESCO and CONGRESS grism redshifts. \\
PID: 2198; PIs: L. Barrufet and P. Oesch & \textit{JWST} & No & \citet{Barrufet_2025} & DAWN \textit{JWST} Archive product. \\
PID: 6541; PI: Egami & \textit{JWST} & No & \citet{DeCoursey_2025} & DAWN \textit{JWST} Archive product. \\
MUSE DR2 & VLT/MUSE & Yes & \citet{Bacon_2023} &  \\
MUSE-Wide DR1 & VLT/MUSE & Yes & \citet{Urrutia_2019} &  \\
GDS collection & Various & Yes & N. Hathi, priv. comm. & Legacy spectroscopic collection in the CANDELS GOODS-South field. \\
ASPECS & ALMA & Yes & \citet{Decarli_2019} & Millimetre spectroscopic redshifts. \\
DEIMOS & Keck/DEIMOS & Yes & \citet{Pharo_2022} & Ground-based optical spectroscopy. \\
ALMA-LESS & ALMA & Yes & \citet{Danielson_2017} &  \\
\hline
\end{tabular}
\end{table*}

\begin{table*}
\centering
\caption{Spectroscopic redshift source catalogues included in the GOODS-North compilation, listed in the priority order used when multiple spectroscopic redshifts with the same best redshift-quality and coordinate-match flags are available for the same object.}
\label{tab:specz_sources_goods_n}
\scriptsize
\setlength{\tabcolsep}{3pt}
\renewcommand{\arraystretch}{1.18}
\begin{tabular}{p{0.21\textwidth} p{0.15\textwidth} >{\centering\arraybackslash}p{0.13\textwidth} p{0.22\textwidth} p{0.22\textwidth}}
\hline
\hline
Catalogue & Instrument & Astrometric correction & Reference & Notes \\
\hline
JADES NIRSpec & \textit{JWST}/NIRSpec & No & \citet{DEugenio_2025, Curtis-Lake_2026, Scholtz_2026} & JADES spectroscopic redshifts. \\
WIDE & \textit{JWST}/NIRSpec & Yes & M. Maseda, priv. comm. & \\
FRESCO / CONGRESS & \textit{JWST}/NIRCam grism & No & F. Sun and X. Lin, priv. comm.; \citet{Oesch_2023, Meyer_2024, Covelo-Paz_2025} & \\
PID: 1871; PI: Chisholm & \textit{JWST} & No & \citet{Chisholm_2024} & DAWN \textit{JWST} Archive product. \\
PID: 2674; PI: Arrabal-Haro & \textit{JWST} & No & \citet{Arrabal_Haro_2023} & DAWN \textit{JWST} Archive product. \\
GDN collection & Various & Yes & N. Hathi, priv. comm. & Legacy spectroscopic collection in the CANDELS GOODS-North field. \\
DESI EDR & DESI & Yes & \citet{DESI_Collaboration_2024} & DESI Early Data Release. \\
DEIMOS & Keck/DEIMOS & No & \citet{Pharo_2022} & Ground-based optical spectroscopy. \\
\hline
\end{tabular}
\end{table*}

\section{Completeness assessment}
\label{app:completeness_assessment}

In this appendix, we show supplementary one-dimensional completeness diagnostics from the injection--recovery simulations described in Section~\ref{sec:Completeness assessment}. These figures illustrate the dependence of the recovery fraction on clump luminosity fraction and normalised galactocentric radius for the blue rest-frame wavelength, $\lambda_{\rm rest}=0.4\,\mu{\rm m}$. The luminosity-dependent completeness curves are used to derive the bin-dependent 50 per cent completeness limits listed in Table~\ref{tab:lclump_lgal_50_limits}, while the two-dimensional completeness maps are used for the probability-weighted completeness correction (see Figure~\ref{fig:2d_completeness}).

\FloatBarrier

\begin{figure*}
 \includegraphics[width=\linewidth]{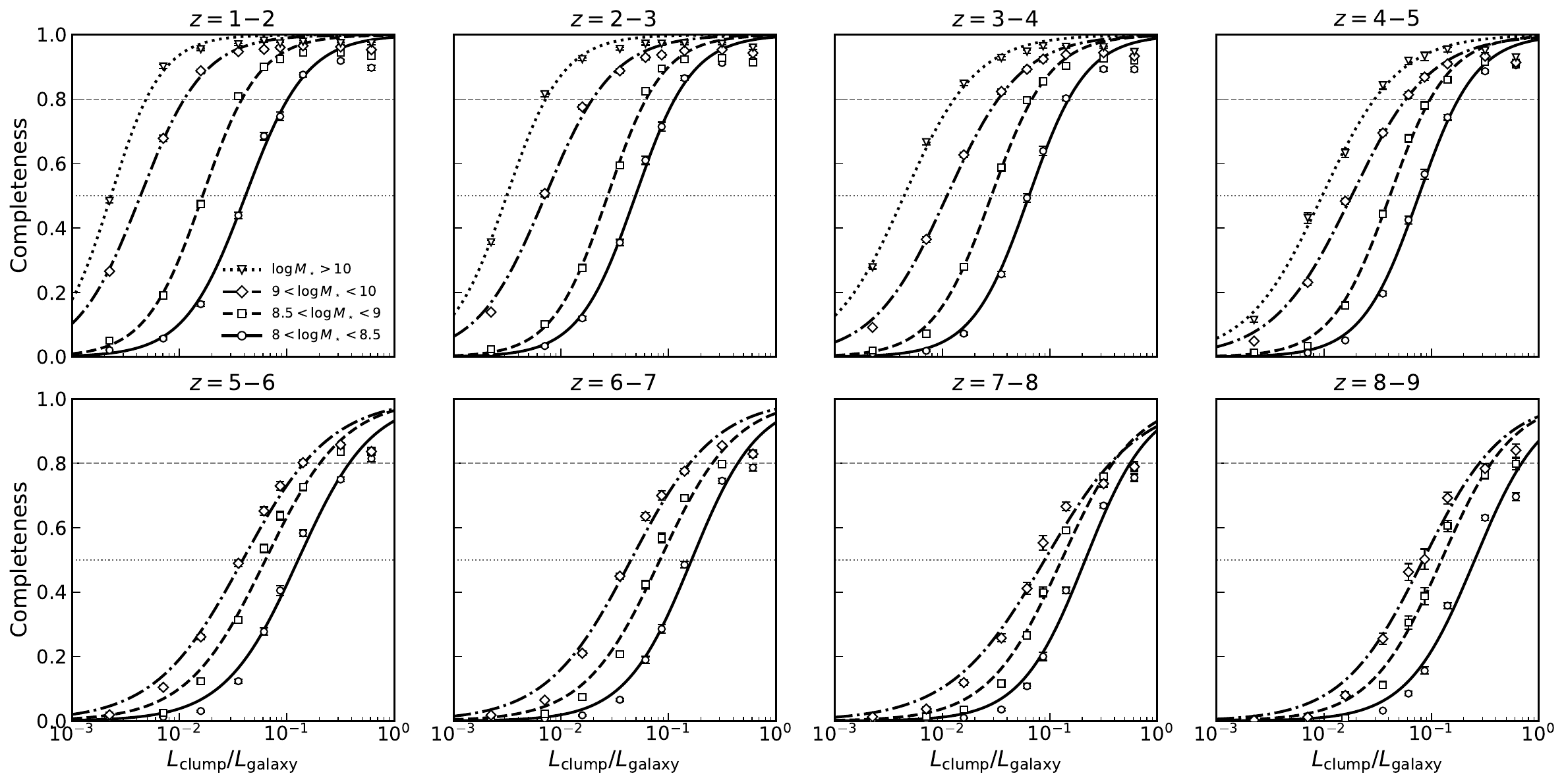}
 \caption{Completeness as a function of clump luminosity fraction, $L_{\rm clump}/L_{\rm galaxy}$, shown for different redshift and stellar mass bins at the blue rest-frame wavelength ($\lambda_{\rm rest}=0.4\,\mu{\rm m}$). Points show the measured recovery fractions from the injection-recovery simulations, with 68 per cent beta-distribution confidence intervals. Black curves show sigmoid fits to the completeness measurements. The completeness increases with clump luminosity fraction, while the luminosity required for a fixed completeness level increases towards higher redshift and lower stellar mass. The $\log_{10}(M_\star/{\rm M_\odot})>10$ bin is not shown at $z\geq5$ because of the low number of galaxies in this regime.}
 \label{fig:flux_completeness}
\end{figure*}

\begin{figure*}
 \includegraphics[width=\linewidth]{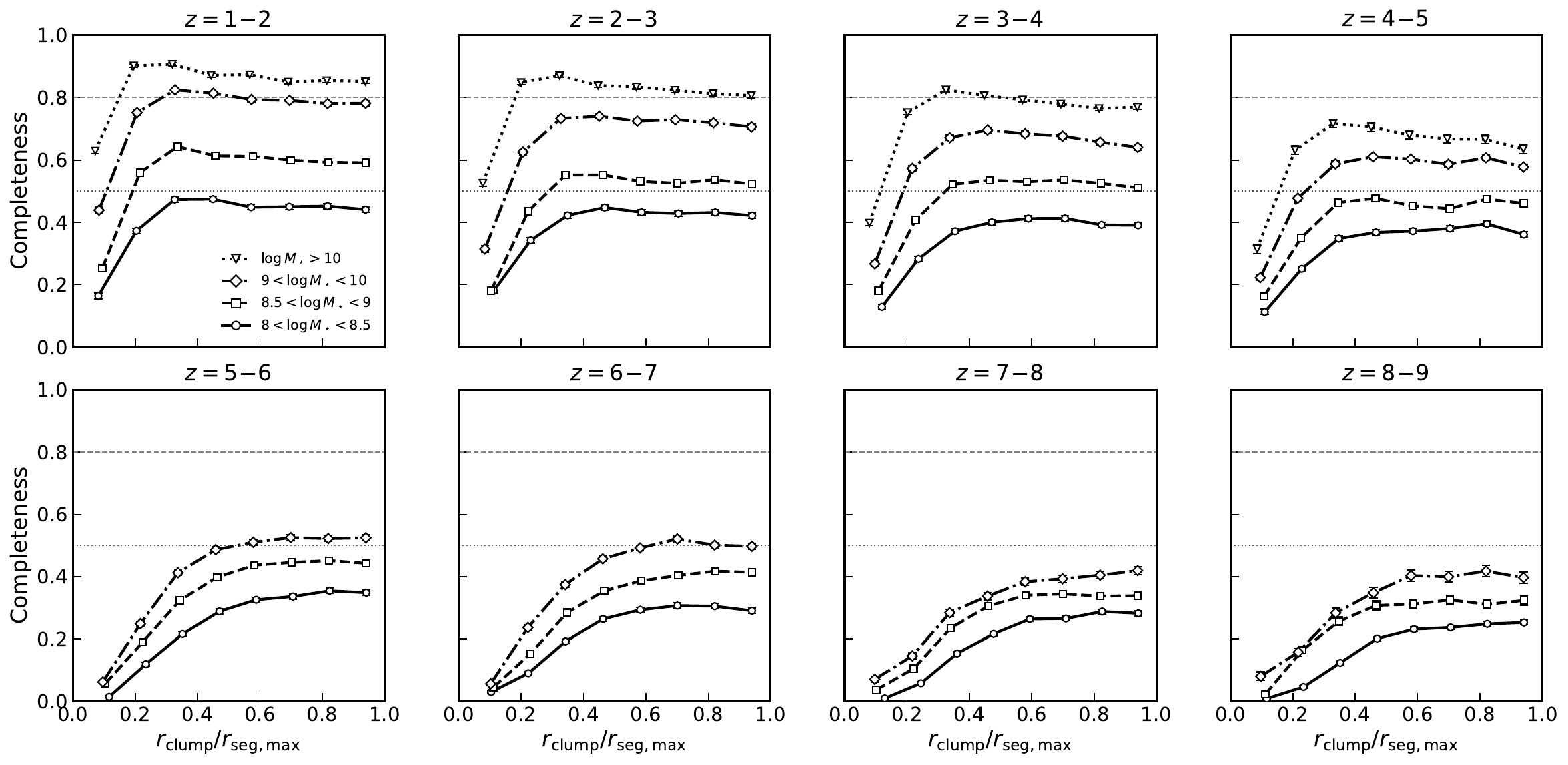}
 \caption{Completeness as a function of normalised galactocentric radius, $r_{\rm clump}/r_{\rm seg,max}$, shown for different redshift and stellar-mass bins at the blue rest-frame wavelength ($\lambda_{\rm rest}=0.4\,\mu{\rm m}$). Here $r_{\rm seg,max}$ is the maximum radial extent of the central galaxy segmentation region measured from the morphological centre. The completeness shows a weaker dependence on radius than on clump luminosity, although recovery becomes less efficient at the smallest radii due to blending with central emission and the exclusion of central detections. As in Figure~\ref{fig:flux_completeness}, the highest-mass bin is excluded at $z\geq5$ because of low number statistics.}
 \label{fig:radial_completeness}
\end{figure*}

\section{Clumpy fraction values}
\label{app:clumpy_fraction_values}

Table~\ref{tab:clumpy_fractions_all} presents the numerical values of the fiducial 50 per cent completeness-limited and probability-weighted completeness corrected clumpy galaxy fractions. The measurements are listed as a function of redshift, stellar mass, and rest-frame wavelength, together with their uncertainties.

\begin{table*}
    \centering
    \caption{Clumpy galaxy fractions as a function of redshift, stellar mass, and rest-frame wavelength. The left-hand columns show the fiducial fractions measured using the 50 per cent completeness-limited clump samples, while the right-hand columns show the probability-weighted completeness corrected fractions. All values are dimensionless fractions, and the uncertainties on $f_{\rm clumpy}^{50}$ are asymmetric, while those on $f_{\rm clumpy}^{\rm corr}$ are symmetric but are displayed using lower and upper indices for consistency. ``Total'' denotes the full stellar-mass-selected sample, $\log_{10}(M_\star/{\rm M_\odot})\geq 8$, and ellipses indicate bins without measurements.}
    \label{tab:clumpy_fractions_all}
    \resizebox{\linewidth}{!}{%
    \renewcommand{\arraystretch}{1.19}%
    \begin{tabular}{ccccc@{\hspace{1.5em}}ccccc}
        \cline{1-5}\cline{6-10}
        \noalign{\vskip 2pt}
        \multicolumn{5}{c}{$f_{\rm clumpy}^{50}$} &
        \multicolumn{5}{c}{$f_{\rm clumpy}^{\rm corr}$} \\[2pt]
        \cline{1-5}\cline{6-10}
        \noalign{\vskip 2pt}
        Redshift bin & $\log_{10}(M_\star/{\rm M_\odot})$
        & UV & Blue & Optical
        &
        Redshift bin & $\log_{10}(M_\star/{\rm M_\odot})$
        & UV & Blue & Optical \\[2pt]
        \cline{1-5}\cline{6-10}
        \noalign{\vskip 1pt}
        \cline{1-5}\cline{6-10}
        \noalign{\vskip 2pt}

        \multirow{6}{*}{$1$--$2$} & ${\rm Total}$ & $\cdots$ & $0.53^{+0.01}_{-0.01}$ & $0.53^{+0.01}_{-0.01}$
        & \multirow{6}{*}{$1$--$2$} & ${\rm Total}$ & $\cdots$ & $0.55^{+0.01}_{-0.01}$ & $0.54^{+0.01}_{-0.01}$ \\
         & $>9$ & $\cdots$ & $0.83^{+0.01}_{-0.01}$ & $0.79^{+0.01}_{-0.01}$
        &  & $>9$ & $\cdots$ & $0.80^{+0.01}_{-0.01}$ & $0.77^{+0.01}_{-0.01}$ \\
         & $8$--$8.5$ & $\cdots$ & $0.31^{+0.01}_{-0.01}$ & $0.33^{+0.01}_{-0.01}$
        &  & $8$--$8.5$ & $\cdots$ & $0.36^{+0.01}_{-0.01}$ & $0.37^{+0.01}_{-0.01}$ \\
         & $8.5$--$9$ & $\cdots$ & $0.57^{+0.01}_{-0.01}$ & $0.56^{+0.01}_{-0.01}$
        &  & $8.5$--$9$ & $\cdots$ & $0.59^{+0.01}_{-0.01}$ & $0.59^{+0.01}_{-0.01}$ \\
         & $9$--$10$ & $\cdots$ & $0.82^{+0.01}_{-0.01}$ & $0.78^{+0.01}_{-0.01}$
        &  & $9$--$10$ & $\cdots$ & $0.78^{+0.01}_{-0.01}$ & $0.78^{+0.01}_{-0.01}$ \\
         & $>10$ & $\cdots$ & $0.85^{+0.02}_{-0.02}$ & $0.79^{+0.02}_{-0.02}$
        &  & $>10$ & $\cdots$ & $0.86^{+0.02}_{-0.02}$ & $0.74^{+0.02}_{-0.02}$ \\
        \cline{1-5}\cline{6-10}

        \multirow{6}{*}{$2$--$3$} & ${\rm Total}$ & $0.29^{+0.01}_{-0.01}$ & $0.45^{+0.01}_{-0.01}$ & $0.42^{+0.01}_{-0.01}$
        & \multirow{6}{*}{$2$--$3$} & ${\rm Total}$ & $0.36^{+0.01}_{-0.01}$ & $0.48^{+0.01}_{-0.01}$ & $0.45^{+0.01}_{-0.01}$ \\
         & $>9$ & $0.60^{+0.01}_{-0.01}$ & $0.74^{+0.01}_{-0.01}$ & $0.68^{+0.01}_{-0.01}$
        &  & $>9$ & $0.64^{+0.01}_{-0.01}$ & $0.74^{+0.01}_{-0.01}$ & $0.70^{+0.01}_{-0.01}$ \\
         & $8$--$8.5$ & $0.10^{+0.01}_{-0.01}$ & $0.27^{+0.01}_{-0.01}$ & $0.25^{+0.01}_{-0.01}$
        &  & $8$--$8.5$ & $0.16^{+0.01}_{-0.01}$ & $0.29^{+0.01}_{-0.01}$ & $0.29^{+0.01}_{-0.01}$ \\
         & $8.5$--$9$ & $0.23^{+0.01}_{-0.01}$ & $0.42^{+0.01}_{-0.01}$ & $0.39^{+0.01}_{-0.01}$
        &  & $8.5$--$9$ & $0.35^{+0.01}_{-0.01}$ & $0.47^{+0.01}_{-0.01}$ & $0.44^{+0.01}_{-0.01}$ \\
         & $9$--$10$ & $0.62^{+0.01}_{-0.01}$ & $0.72^{+0.01}_{-0.01}$ & $0.68^{+0.01}_{-0.01}$
        &  & $9$--$10$ & $0.65^{+0.01}_{-0.01}$ & $0.72^{+0.01}_{-0.01}$ & $0.68^{+0.01}_{-0.01}$ \\
         & $>10$ & $0.51^{+0.03}_{-0.03}$ & $0.85^{+0.02}_{-0.03}$ & $0.74^{+0.03}_{-0.03}$
        &  & $>10$ & $0.58^{+0.03}_{-0.03}$ & $0.86^{+0.02}_{-0.02}$ & $0.83^{+0.02}_{-0.02}$ \\
        \cline{1-5}\cline{6-10}

        \multirow{6}{*}{$3$--$4$} & ${\rm Total}$ & $0.35^{+0.01}_{-0.01}$ & $0.37^{+0.01}_{-0.01}$ & $0.25^{+0.01}_{-0.01}$
        & \multirow{6}{*}{$3$--$4$} & ${\rm Total}$ & $0.41^{+0.01}_{-0.01}$ & $0.39^{+0.01}_{-0.01}$ & $0.28^{+0.01}_{-0.01}$ \\
         & $>9$ & $0.62^{+0.01}_{-0.01}$ & $0.65^{+0.01}_{-0.01}$ & $0.44^{+0.01}_{-0.01}$
        &  & $>9$ & $0.66^{+0.01}_{-0.01}$ & $0.67^{+0.01}_{-0.01}$ & $0.48^{+0.01}_{-0.01}$ \\
         & $8$--$8.5$ & $0.16^{+0.01}_{-0.01}$ & $0.20^{+0.01}_{-0.01}$ & $0.14^{+0.01}_{-0.01}$
        &  & $8$--$8.5$ & $0.24^{+0.01}_{-0.01}$ & $0.23^{+0.01}_{-0.01}$ & $0.14^{+0.01}_{-0.01}$ \\
         & $8.5$--$9$ & $0.41^{+0.01}_{-0.01}$ & $0.39^{+0.01}_{-0.01}$ & $0.26^{+0.01}_{-0.01}$
        &  & $8.5$--$9$ & $0.46^{+0.01}_{-0.01}$ & $0.38^{+0.01}_{-0.01}$ & $0.31^{+0.01}_{-0.01}$ \\
         & $9$--$10$ & $0.62^{+0.01}_{-0.01}$ & $0.64^{+0.01}_{-0.01}$ & $0.43^{+0.01}_{-0.01}$
        &  & $9$--$10$ & $0.66^{+0.01}_{-0.01}$ & $0.65^{+0.01}_{-0.01}$ & $0.47^{+0.01}_{-0.01}$ \\
         & $>10$ & $0.56^{+0.04}_{-0.04}$ & $0.75^{+0.04}_{-0.04}$ & $0.56^{+0.04}_{-0.04}$
        &  & $>10$ & $0.59^{+0.04}_{-0.04}$ & $0.81^{+0.03}_{-0.03}$ & $0.58^{+0.04}_{-0.04}$ \\
        \cline{1-5}\cline{6-10}

        \multirow{6}{*}{$4$--$5$} & ${\rm Total}$ & $0.33^{+0.01}_{-0.01}$ & $0.29^{+0.01}_{-0.01}$ & $0.14^{+0.01}_{-0.01}$
        & \multirow{6}{*}{$4$--$5$} & ${\rm Total}$ & $0.35^{+0.01}_{-0.01}$ & $0.31^{+0.01}_{-0.01}$ & $0.13^{+0.00}_{-0.00}$ \\
         & $>9$ & $0.55^{+0.02}_{-0.02}$ & $0.51^{+0.02}_{-0.02}$ & $0.28^{+0.02}_{-0.02}$
        &  & $>9$ & $0.57^{+0.02}_{-0.02}$ & $0.53^{+0.02}_{-0.02}$ & $0.28^{+0.02}_{-0.02}$ \\
         & $8$--$8.5$ & $0.19^{+0.01}_{-0.01}$ & $0.17^{+0.01}_{-0.01}$ & $0.07^{+0.01}_{-0.01}$
        &  & $8$--$8.5$ & $0.23^{+0.01}_{-0.01}$ & $0.21^{+0.01}_{-0.01}$ & $0.07^{+0.01}_{-0.01}$ \\
         & $8.5$--$9$ & $0.41^{+0.01}_{-0.01}$ & $0.34^{+0.01}_{-0.01}$ & $0.16^{+0.01}_{-0.01}$
        &  & $8.5$--$9$ & $0.41^{+0.01}_{-0.01}$ & $0.32^{+0.01}_{-0.01}$ & $0.13^{+0.01}_{-0.01}$ \\
         & $9$--$10$ & $0.55^{+0.02}_{-0.02}$ & $0.51^{+0.02}_{-0.02}$ & $0.27^{+0.02}_{-0.02}$
        &  & $9$--$10$ & $0.57^{+0.02}_{-0.02}$ & $0.53^{+0.02}_{-0.02}$ & $0.28^{+0.02}_{-0.02}$ \\
         & $>10$ & $0.58^{+0.08}_{-0.08}$ & $0.66^{+0.07}_{-0.08}$ & $0.45^{+0.08}_{-0.08}$
        &  & $>10$ & $0.57^{+0.07}_{-0.07}$ & $0.63^{+0.07}_{-0.07}$ & $0.39^{+0.08}_{-0.08}$ \\
        \cline{1-5}\cline{6-10}

        \multirow{6}{*}{$5$--$6$} & ${\rm Total}$ & $0.24^{+0.01}_{-0.01}$ & $0.09^{+0.01}_{-0.01}$ & $0.05^{+0.01}_{-0.01}$
        & \multirow{6}{*}{$5$--$6$} & ${\rm Total}$ & $0.29^{+0.01}_{-0.01}$ & $0.13^{+0.01}_{-0.01}$ & $0.06^{+0.01}_{-0.01}$ \\
         & $>9$ & $0.43^{+0.03}_{-0.03}$ & $0.17^{+0.02}_{-0.02}$ & $0.11^{+0.02}_{-0.02}$
        &  & $>9$ & $0.47^{+0.03}_{-0.03}$ & $0.22^{+0.02}_{-0.02}$ & $0.13^{+0.02}_{-0.02}$ \\
         & $8$--$8.5$ & $0.14^{+0.01}_{-0.01}$ & $0.05^{+0.01}_{-0.01}$ & $0.03^{+0.01}_{-0.01}$
        &  & $8$--$8.5$ & $0.19^{+0.01}_{-0.01}$ & $0.08^{+0.01}_{-0.01}$ & $0.03^{+0.01}_{-0.01}$ \\
         & $8.5$--$9$ & $0.28^{+0.02}_{-0.02}$ & $0.11^{+0.01}_{-0.01}$ & $0.05^{+0.01}_{-0.01}$
        &  & $8.5$--$9$ & $0.36^{+0.02}_{-0.02}$ & $0.16^{+0.02}_{-0.02}$ & $0.08^{+0.01}_{-0.01}$ \\
         & $9$--$10$ & $0.43^{+0.03}_{-0.03}$ & $0.17^{+0.02}_{-0.02}$ & $0.11^{+0.02}_{-0.02}$
        &  & $9$--$10$ & $0.47^{+0.03}_{-0.03}$ & $0.22^{+0.02}_{-0.02}$ & $0.13^{+0.02}_{-0.02}$ \\
         & $>10$ & $\cdots$ & $\cdots$ & $\cdots$
        &  & $>10$ & $\cdots$ & $\cdots$ & $\cdots$ \\
        \cline{1-5}\cline{6-10}

        \multirow{6}{*}{$6$--$7$} & ${\rm Total}$ & $0.19^{+0.01}_{-0.01}$ & $0.08^{+0.01}_{-0.01}$ & $0.05^{+0.01}_{-0.01}$
        & \multirow{6}{*}{$6$--$7$} & ${\rm Total}$ & $0.22^{+0.01}_{-0.01}$ & $0.08^{+0.01}_{-0.01}$ & $0.02^{+0.01}_{-0.01}$ \\
         & $>9$ & $0.46^{+0.05}_{-0.05}$ & $0.17^{+0.04}_{-0.03}$ & $0.09^{+0.03}_{-0.02}$
        &  & $>9$ & $0.43^{+0.04}_{-0.04}$ & $0.22^{+0.04}_{-0.04}$ & $0.05^{+0.02}_{-0.02}$ \\
         & $8$--$8.5$ & $0.08^{+0.01}_{-0.01}$ & $0.04^{+0.01}_{-0.01}$ & $0.02^{+0.01}_{-0.01}$
        &  & $8$--$8.5$ & $0.15^{+0.02}_{-0.02}$ & $0.05^{+0.01}_{-0.01}$ & $0.02^{+0.01}_{-0.01}$ \\
         & $8.5$--$9$ & $0.26^{+0.03}_{-0.03}$ & $0.11^{+0.02}_{-0.02}$ & $0.07^{+0.02}_{-0.02}$
        &  & $8.5$--$9$ & $0.26^{+0.03}_{-0.03}$ & $0.08^{+0.02}_{-0.02}$ & $0.02^{+0.01}_{-0.01}$ \\
         & $9$--$10$ & $0.46^{+0.05}_{-0.05}$ & $0.17^{+0.04}_{-0.03}$ & $0.09^{+0.03}_{-0.02}$
        &  & $9$--$10$ & $0.43^{+0.04}_{-0.04}$ & $0.22^{+0.04}_{-0.04}$ & $0.05^{+0.02}_{-0.02}$ \\
         & $>10$ & $\cdots$ & $\cdots$ & $\cdots$
        &  & $>10$ & $\cdots$ & $\cdots$ & $\cdots$ \\
        \cline{1-5}\cline{6-10}

        \multirow{6}{*}{$7$--$8$} & ${\rm Total}$ & $0.15^{+0.02}_{-0.02}$ & $0.04^{+0.01}_{-0.01}$ & $0.02^{+0.01}_{-0.01}$
        & \multirow{6}{*}{$7$--$8$} & ${\rm Total}$ & $0.21^{+0.02}_{-0.02}$ & $0.05^{+0.01}_{-0.01}$ & $0.03^{+0.01}_{-0.01}$ \\
         & $>9$ & $0.16^{+0.05}_{-0.05}$ & $0.06^{+0.04}_{-0.03}$ & $0.06^{+0.04}_{-0.03}$
        &  & $>9$ & $0.24^{+0.06}_{-0.06}$ & $0.07^{+0.03}_{-0.03}$ & $0.09^{+0.04}_{-0.04}$ \\
         & $8$--$8.5$ & $0.14^{+0.02}_{-0.02}$ & $0.03^{+0.01}_{-0.01}$ & $0.00^{+0.01}_{-0.00}$
        &  & $8$--$8.5$ & $0.18^{+0.02}_{-0.02}$ & $0.05^{+0.01}_{-0.01}$ & $0.02^{+0.01}_{-0.01}$ \\
         & $8.5$--$9$ & $0.18^{+0.04}_{-0.03}$ & $0.07^{+0.03}_{-0.02}$ & $0.02^{+0.02}_{-0.01}$
        &  & $8.5$--$9$ & $0.28^{+0.04}_{-0.04}$ & $0.04^{+0.02}_{-0.02}$ & $0.03^{+0.02}_{-0.02}$ \\
         & $9$--$10$ & $0.16^{+0.05}_{-0.05}$ & $0.06^{+0.04}_{-0.03}$ & $0.06^{+0.04}_{-0.03}$
        &  & $9$--$10$ & $0.24^{+0.06}_{-0.06}$ & $0.07^{+0.03}_{-0.03}$ & $0.09^{+0.04}_{-0.04}$ \\
         & $>10$ & $\cdots$ & $\cdots$ & $\cdots$
        &  & $>10$ & $\cdots$ & $\cdots$ & $\cdots$ \\
        \cline{1-5}\cline{6-10}

        \multirow{6}{*}{$8$--$9$} & ${\rm Total}$ & $0.12^{+0.03}_{-0.02}$ & $0.04^{+0.02}_{-0.01}$ & $0.02^{+0.01}_{-0.01}$
        & \multirow{6}{*}{$8$--$9$} & ${\rm Total}$ & $0.16^{+0.03}_{-0.03}$ & $0.02^{+0.01}_{-0.01}$ & $0.01^{+0.01}_{-0.01}$ \\
         & $>9$ & $0.26^{+0.10}_{-0.08}$ & $0.09^{+0.07}_{-0.05}$ & $0.01^{+0.03}_{-0.01}$
        &  & $>9$ & $0.24^{+0.09}_{-0.09}$ & $0.08^{+0.06}_{-0.06}$ & $0.04^{+0.04}_{-0.04}$ \\
         & $8$--$8.5$ & $0.09^{+0.03}_{-0.02}$ & $0.02^{+0.02}_{-0.01}$ & $0.01^{+0.01}_{-0.01}$
        &  & $8$--$8.5$ & $0.13^{+0.03}_{-0.03}$ & $0.01^{+0.01}_{-0.01}$ & $0.00^{+0.00}_{-0.00}$ \\
         & $8.5$--$9$ & $0.15^{+0.07}_{-0.05}$ & $0.09^{+0.06}_{-0.04}$ & $0.06^{+0.05}_{-0.03}$
        &  & $8.5$--$9$ & $0.22^{+0.07}_{-0.07}$ & $0.00^{+0.00}_{-0.00}$ & $0.03^{+0.03}_{-0.03}$ \\
         & $9$--$10$ & $0.26^{+0.10}_{-0.08}$ & $0.09^{+0.07}_{-0.05}$ & $0.01^{+0.03}_{-0.01}$
        &  & $9$--$10$ & $0.24^{+0.09}_{-0.09}$ & $0.08^{+0.06}_{-0.06}$ & $0.04^{+0.04}_{-0.04}$ \\
         & $>10$ & $\cdots$ & $\cdots$ & $\cdots$
        &  & $>10$ & $\cdots$ & $\cdots$ & $\cdots$ \\
        \cline{1-5}\cline{6-10}

    \end{tabular}%
    }
\end{table*}

\section{Clump multiplicity and SFR enhancement}
\label{app:clump_multiplicity_sfms}

\begin{figure*}
 \includegraphics[width=\linewidth]{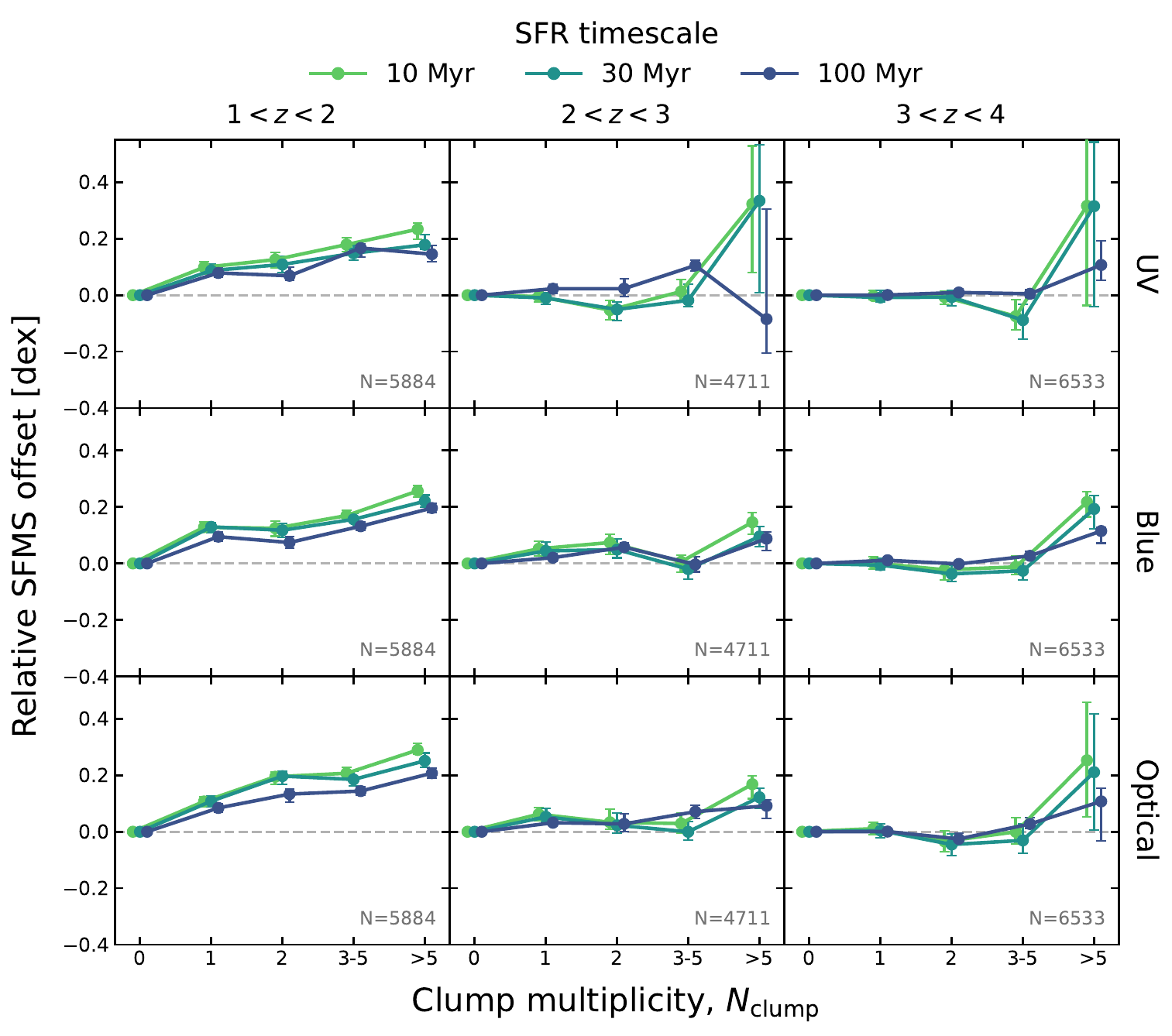}
 \caption{Relative SFMS offset as a function of clump multiplicity, redshift, and rest-frame wavelength, for SFRs averaged over 10, 30, and 100 Myr. Columns correspond to redshift bins and rows to UV, blue, and optical measurements. Each point shows the median offset from the SFMS of \citet{Simmonds_2025}, relative to galaxies with $N_{\rm clump}=0$ in the same redshift and wavelength bin. Error bars indicate the 16th--84th percentile bootstrap intervals, while the dashed line marks zero enhancement. Panel annotations indicate the number of galaxies.}
 \label{fig:clump_multiplicity_sfms_enhancement}
\end{figure*}

To investigate whether the association between clump multiplicity and enhanced SFR depends on the averaging timescale, we repeat the analysis using SFRs averaged over 10, 30, and 100 Myr. We calculate each galaxy's offset from the corresponding SFMS of \citet{Simmonds_2025} at its stellar mass and redshift, and subtract the median offset of non-clumpy galaxies in the same redshift and rest-frame wavelength bin. Figure~\ref{fig:clump_multiplicity_sfms_enhancement} shows a generally increasing SFR enhancement with multiplicity at $1<z<2$ across all three wavelengths. For $N_{\rm clump}>5$, the enhancements range from $0.15-0.23$ dex in the UV, $0.20-0.26$ dex in the blue, and $0.21-0.29$ dex in the optical across the three timescales. The enhancement persists over 100 Myr but generally weakens with increasing averaging timescale, suggesting a stronger connection to recent star formation. At $2<z<4$, the trends are less pronounced, with stronger enhancements largely restricted to the highest-multiplicity systems, particularly in the blue and optical. The UV measurements at these redshifts show greater scatter and uncertainties, indicating that the relationship between multiplicity and SFR enhancement is not universal across redshift and wavelength. We restrict this comparison to $z<4$ due to the limited number of high-multiplicity galaxies at higher redshifts.

\section{Clump multiplicity and axis ratio}
\label{app:multiplicity_axis_ratio}

\begin{figure*}
 \includegraphics[width=\linewidth]{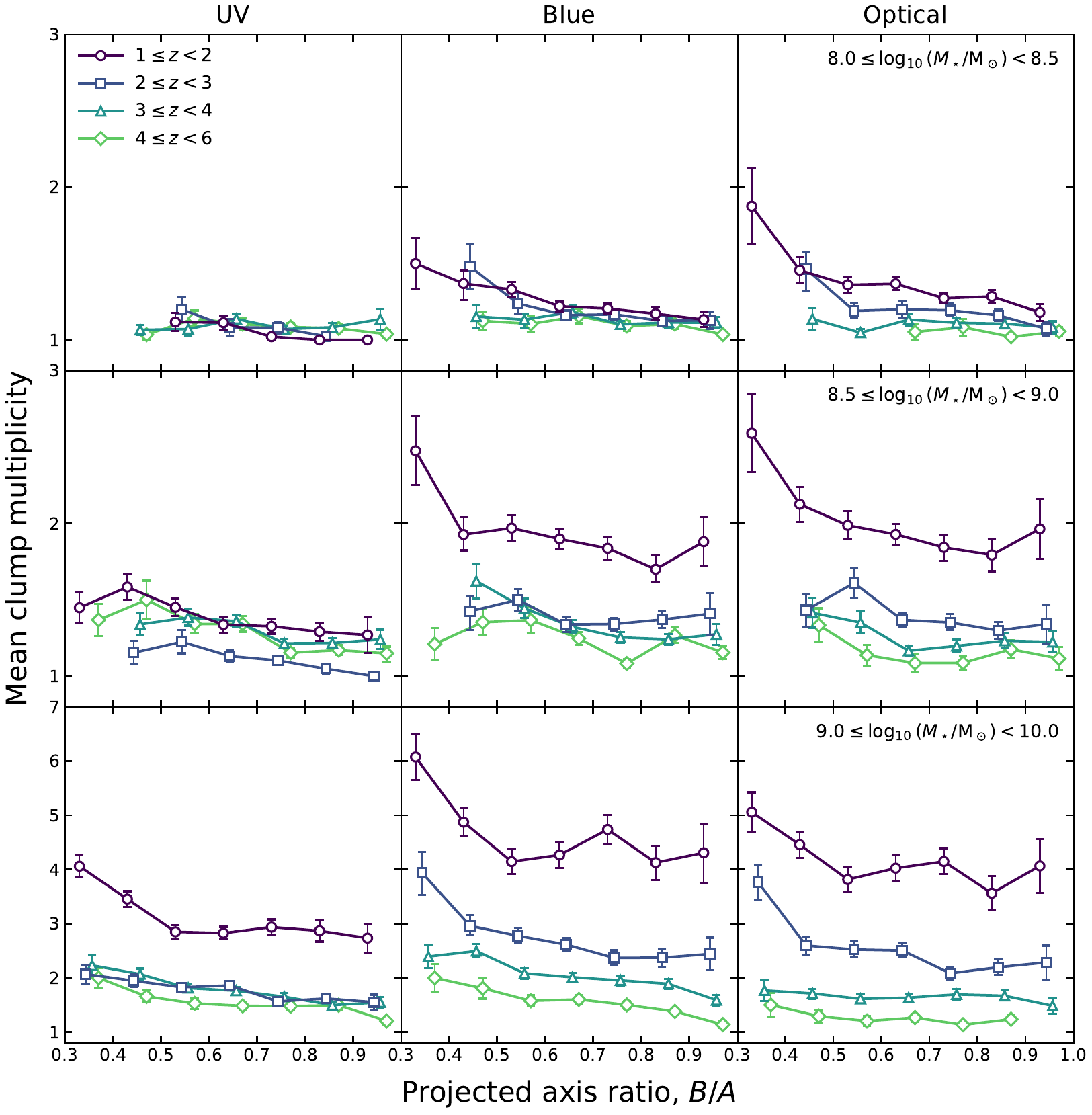}
 \caption{Mean clump multiplicity as a function of projected galaxy axis ratio ($B/A$), stellar mass, redshift, and rest-frame wavelength. Columns correspond to the UV, blue, and optical measurements, while rows show three stellar-mass bins. Different colours and symbols indicate the redshift intervals. The mean multiplicity is calculated among galaxies hosting at least one detected non-central clump, using axis-ratio bins of width 0.1. Error bars indicate the 16$^{\rm th}$--84$^{\rm th}$ percentile bootstrap uncertainties, and bins containing fewer than 15 galaxies are excluded.}
 \label{fig:clump_multiplicity_axis_ratio}
\end{figure*}

We investigate the dependence of clump multiplicity on the projected galaxy axis ratio ($B/A$), calculated from the major- and minor-axis measurements ($A$ and $B$) provided in the JADES DR5 photometric catalogues \citep{Robertson_2026}. Figure~\ref{fig:clump_multiplicity_axis_ratio} shows that clump multiplicity generally decreases with increasing $B/A$, particularly in the blue and optical measurements at lower redshifts ($z<3$). The anticorrelation is strongest in the highest stellar-mass bin considered ($9\leq\log_{10}(M_*/M_\odot)<10$), where at $1\leq z<2$ the mean blue multiplicity decreases from approximately 6 to 4 between $B/A\approx0.35$ and 0.85, with a similar but weaker trend in the optical. At lower stellar masses and higher redshifts, the dependence becomes weaker, with mean multiplicities approaching unity. The UV measurements show a less pronounced dependence on axis ratio, including at $1\leq z<2$. As discussed in Section~\ref{sec:Clump multiplicity}, this trend may reflect a contribution from morphologically disturbed or interacting systems, although projection and detection effects cannot be excluded.

\section{Estimated clump radii distributions}
\label{app:clump_radii}

\begin{figure*}
 \includegraphics[width=0.9\linewidth]{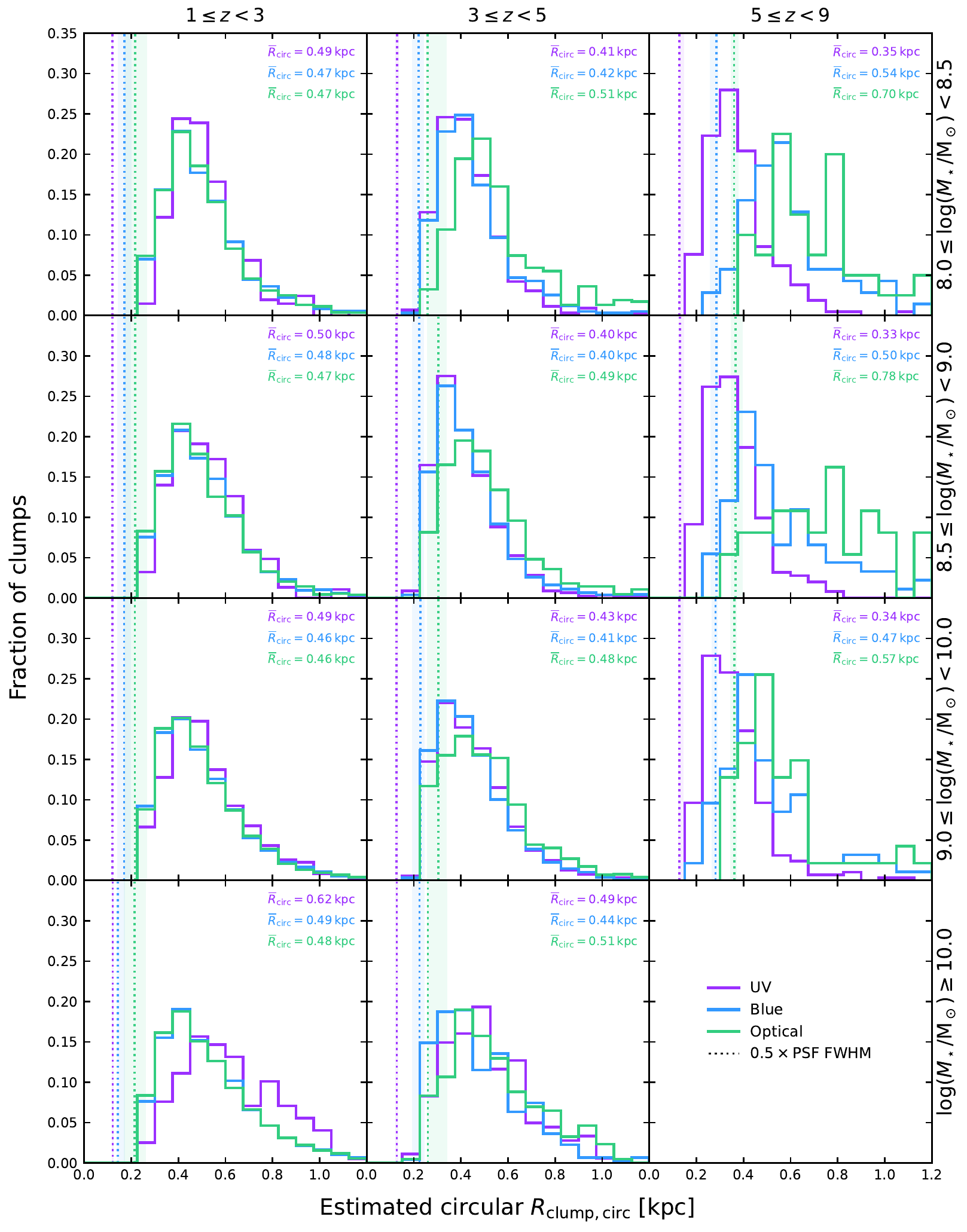}
 \caption{Distributions of the estimated circular radii of non-central clumps in the 50 per cent completeness-limited catalogue, divided by host-galaxy stellar mass (rows) and redshift (columns). The UV, blue, and optical samples are shown in purple, blue, and green, respectively, with each distribution normalised by the number of clumps in that subsample. Coloured values indicate the median clump radii. Dotted vertical lines mark the median physical scale corresponding to $0.5\times{\rm FWHM}_{\rm PSF}$, with shaded regions showing the 16--84 percentile ranges. The reported radii are derived from the dendrogram segmentation areas and are not corrected for PSF broadening.}
 \label{fig:clump_circular_radius_mass_redshift}
\end{figure*}

Figure~\ref{fig:clump_circular_radius_mass_redshift} shows the clump-radius distributions divided by stellar mass and redshift. At $1 \leq z < 3$, the distributions are broadly similar across wavelengths and stellar masses, except for somewhat larger UV clumps in the most massive galaxies. At higher redshifts, the largest median radii are found mainly in the optical samples of low- and intermediate-mass galaxies. This trend is qualitatively compatible with a larger contribution from extended merger-related structures or merger-induced star formation, as discussed in Section~\ref{sec:Clump sizes}. However, wavelength-dependent resolution, blending, and small-number statistics prevent a definitive interpretation of these differences.

\bsp	
\label{lastpage}
\end{document}